%% file: arxiv-acm.tex
\documentclass[acmsmall, authorversion, screen]{acmart}
\setcopyright{rightsretained}
\acmPrice{}
\acmDOI{10.1145/3371083}
\acmYear{2020}
\copyrightyear{2020}
\acmJournal{PACMPL}
\acmVolume{4}
\acmNumber{POPL}
\acmArticle{15}
\acmMonth{1}

\usepackage[
    type={CC},
    modifier={by},
    version={4.0},
]{doclicense}
\usepackage{import}
\import{./}{packages}
\import{./}{macros}

\includeversion{arxiv}
\excludeversion{popl}

\begin{document}

\title{Recurrence Extraction for Functional Programs through
Call-by-Push-Value (Extended Version)}

\author{G. A. Kavvos}
\authornote{Current affiliation: Department of Computer Science, Aarhus University}
\orcid{0000-0001-7953-7975}
\email{g.a.kavvos@gmail.com}
\affiliation{
  \department{Department of Mathematics and Computer Science}
  \institution{Wesleyan University}
  \streetaddress{265 Church Street}
  \city{Middletown}
  \state{CT}
  \postcode{06459}
  \country{United States of America}
}

\author{Edward Morehouse}
\email{emorehouse@wesleyan.edu}
\affiliation{
  \department{Department of Mathematics and Computer Science}
  \institution{Wesleyan University}
  \streetaddress{265 Church Street}
  \city{Middletown}
  \state{CT}
  \postcode{06459}
  \country{United States of America}
}

\author{Daniel R. Licata}
\orcid{0000-0003-0697-7405}             
\email{dlicata@wesleyan.edu}
\affiliation{
  \department{Department of Mathematics and Computer Science}
  \institution{Wesleyan University}
  \streetaddress{265 Church Street}
  \city{Middletown}
  \state{CT}
  \postcode{06459}
  \country{United States of America}
}

\author{Norman Danner}
\email{ndanner@wesleyan.edu}

\affiliation{
  \department{Department of Mathematics and Computer Science}
  \institution{Wesleyan University}
  \streetaddress{265 Church Street}
  \city{Middletown}
  \state{CT}
  \postcode{06459}
  \country{United States of America}
}

\authorsaddresses{Authors' address: G. A. Kavvos,
g.a.kavvos@gmail.com; Edward Morehouse, emorehouse@wesleyan.edu;
Daniel R. Licata, dlicata@wesleyan.edu; Norman Danner,
ndanner@wesleyan.edu, Department of Mathematics and Computer
Science, Wesleyan University, 265 Church Street, Middletown, CT,
06459, United States of America.}

\import{./}{abstract}


\maketitle

\import{./}{recextr-cbpv}

\begin{acks}                            
\import{./}{acks}

\doclicenseThis

\end{acks}

\input{popl2020.bbl}

\appendix

\import{./}{supp}

\end{document}

%% file: packages.tex
\usepackage{derivation}
\usepackage{enumitem}
\usepackage{multirow}
\usepackage{prooftree}
\usepackage{mathtools}
\usepackage{suffix}
\usepackage{tikz}
\import{./}{tikzstyle}
\usepackage{versions}

%% file: tikzstyle.tex
\usetikzlibrary
{
	calc ,
	matrix ,
	arrows ,
	angles ,
	quotes ,
	intersections ,
	positioning ,
	fit ,
	shapes.geometric ,
	shapes.misc ,
	decorations ,
	decorations.markings ,
	decorations.pathmorphing ,
	decorations.pathreplacing ,
	decorations.text 
}

\usepgflibrary
{
	arrows.meta
}


\usepackage{xcolor}

\tikzset
{
	diagram/.style =
	{
		line cap = butt ,
		line join = bevel ,
		arrows = -> ,
		> = angle 60 ,
		auto = left ,
		text height = 1.5ex , 
		text depth = 0.25ex , 
		align = center ,
	} ,
	code node/.append style =
	{
		every node/.append style =
		{
			execute at begin node = \begin{texttt} ,
			execute at end node = \end{texttt}
		}
	} ,
	code diagram/.style =
	{
		diagram ,
		code node
	} ,
	math node/.append style =
	{
		every node/.append style =
		{
			execute at begin node = \begin{math} ,
			execute at end node = \end{math}
		}
	} ,
	math diagram/.style =
	{
		diagram ,
		math node
	} ,
	string diagram/.style =
	{
		math diagram ,
		arrows = - ,
	} ,
	online/.style =
	{
		shape = rectangle ,
		rounded corners = 2mm ,
		inner sep = 1.5pt ,
		fill = white ,
		opacity = 1.0 ,
		anchor = center	
	} ,
	double arrow/.style =
	{
		double distance between line centers = 2pt ,
		-implies 
	} ,
	equals/.style =
	{
		double distance between line centers = 2pt ,
		-implies , 
		arrows = -
	} ,
	mapto/.append style =
	{
		arrows = |-> ,
	} ,
	paph/.append style =
	{
		decorate ,
		decoration = {snake , segment length = 5mm , amplitude = 0.5mm} ,
	} ,
	includel/.append style =
	{
		arrows = Hooks[right]-> ,
	} ,
	includer/.append style =
	{
		arrows = Hooks[left]-> ,
	} ,
	monomorphism/.append style =
	{
		arrows = >-> ,
	} ,
	epimorphism/.append style =
	{
		arrows = ->> ,
	} ,
	cofibration/.append style =
	{
		arrows = >-> ,
	} ,
	fibration/.append style =
	{
		arrows = ->> ,
	} ,
	equivalence/.append style =
	{
		decoration = {markings , mark = at position 1/2 with {\node[online , transform shape] {∼};}} ,
		postaction = {decorate} ,
	} ,
	proarrow/.append style =
	{
		decoration = {markings , mark = at position 1/2 with {\draw [solid , -] (0 , -0.6ex) to (0 , 0.6ex);}} ,
		postaction = {decorate} ,
	} ,
	string/.append style =
	{
		arrows = - ,
	} ,
	cofibration string/.append style =
	{
		decoration = {markings , mark = at position 1/4 with {\arrow{>}} , mark = at position 3/4 with {\arrow{>}}} ,
		postaction = {decorate} ,
	} ,
	fibration string/.append style =
	{
		decoration = {markings , mark = at position 4/8 with {\arrow{>}} , mark = at position 5/8 with {\arrow{>}}} ,
		postaction = {decorate} ,
	} ,
	overcross/.append style =
	{
		preaction = {draw = white , - , line width = 5pt}
	} ,
	label/.append style =
	{
		font = \scriptsize
	} ,
	incoming/.append style =
	{
		pos = 0 ,
		anchor = south
	} ,
	outgoing/.append style =
	{
		pos = 1 ,
		anchor = north
	} ,
	outcoming/.append style =
	{
		pos = 0 ,
		anchor = north
	} ,
	ingoing/.append style =
	{
		pos = 1 ,
		anchor = south
	} ,
	fromabove/.append style =
	{
		pos = 0 ,
		anchor = south
	} ,
	tobelow/.append style =
	{
		pos = 1 ,
		anchor = north
	} ,
	frombelow/.append style =
	{
		pos = 0 ,
		anchor = north
	} ,
	toabove/.append style =
	{
		pos = 1 ,
		anchor = south
	} ,
	fromleft/.append style =
	{
		pos = 0 ,
		anchor = east
	} ,
	toright/.append style =
	{
		pos = 1 ,
		anchor = west
	} ,
	fromright/.append style =
	{
		pos = 0 ,
		anchor = west
	} ,
	toleft/.append style =
	{
		pos = 1 ,
		anchor = east
	} ,
	forall/.append style =
	{
		line width = 0.5pt ,
	} ,
	exists/.append style =
	{
		densely dashed
	} ,
	structural/.append style =
	{
		opacity = 2/3 ,
	} ,
	0-cell/.style =
	{
		shape = rectangle ,
		rounded corners = 2mm ,
		inner sep = 2pt
	} ,
	1-cell/.style =
	{
		shape = rectangle ,
		rounded corners = 2mm ,
		inner sep = 1.5pt ,
		fill = white ,
		opacity = 1.0 ,
		anchor = center	
	} ,
	2-cell/.style =
	{
		draw = black!75 ,
		fill = white ,
		opacity = 0.9 ,
		inner sep = 0.75mm ,
		minimum size = 4.0mm
	} ,
	bead/.style =
	{
		2-cell ,
		shape = rectangle ,
		rounded corners = 2.0mm
	} ,
	diamond/.style =
	{
		2-cell ,
		shape = diamond
	} ,
	bra/.style =
	{
		2-cell ,
		shape = isosceles triangle ,
		isosceles triangle apex angle = 70 ,
		shape border rotate = -90 ,
		minimum size = 5mm
	} ,
	ket/.style =
	{
		2-cell ,
		shape = isosceles triangle ,
		isosceles triangle apex angle = 70 ,
		shape border rotate = 90 ,
		minimum size = 6mm
	} ,
	box/.style =
	{
		2-cell ,
		shape = rectangle
	} ,
	white dot/.style =
	{
		2-cell ,
		shape = circle ,
		minimum size = 1.5mm ,
		fill = white
	} ,
	black dot/.style =
	{
		2-cell ,
		shape = circle ,
		minimum size = 1.5mm ,
		fill = gray
	} ,
	dot/.style =
	{
		2-cell ,
		shape = circle ,
		inner sep = 0mm ,
		minimum size = 0.5mm ,
		fill = black
	} ,
	sheet/.style =
	{
		draw = black!75 ,
		fill = gray!10 ,
		opacity = 0.9 ,
		fill opacity = 0.5
	} ,
	torn/.append style =
	{
		decoration = {random steps , segment length = 2pt , amplitude = 1pt}
	} ,
	edge annotation/.append style =
	{
		anchor = west
	} ,
	callout/.append style =
	{
		minimum height = 2mm ,
		minimum width = 4mm ,
		draw ,
		draw opacity = 0.25
	} ,
	callout-pre/.append style =
	{
		callout ,
		inner sep = 1.2mm ,
		solid
	} ,
	callout-post/.append style =
	{
		callout ,
		inner sep = 0.8mm ,
		dashed
	} ,
	pullback/.pic =
	{
		\draw [semithick , arrows = -] (-0.1 , 0.1) -- (0.1 , 0.1) -- (0.1 , -0.1) ;
	} ,
	pushout/.pic =
	{
		\draw [semithick , arrows = -] (-0.1 , 0.1) -- (-0.1 , -0.1) -- (0.1 , -0.1) ;
	} ,
}

\pgfdeclaredecoration{simple line}{start}
{
  \state{start}[width = +0pt,
                next state=step]{
    \pgfpathmoveto{\pgfpoint{0pt}{0pt}}
  }
  \state{step}[auto end on length    = 3pt,
               auto corner on length = 3pt,               
               width=+1pt]
  {
    \pgfpathlineto{\pgfpoint{1pt}{0pt}}
  }
  \state{final}
  {}
}

%% file: abstract.tex
\begin{abstract}
The main way of analyzing the complexity of a program is that of
extracting and solving a recurrence that expresses its running
time in terms of the size of its input. We develop a method that
automatically extracts such recurrences from the syntax of
higher-order recursive functional programs. The resulting
recurrences, which are programs in a call-by-name language with
recursion, explicitly compute the running time in terms of the
size of the input. In order to achieve this in a uniform way that
covers both call-by-name and call-by-value evaluation strategies,
we use Call-by-Push-Value (CBPV) as an intermediate language.
Finally, we use domain theory to develop a denotational cost
semantics for the resulting recurrences.
\end{abstract}

%% file: recextr-cbpv.tex
\section{Introduction}

Functional programmers typically analyze time, space, or other resource
usage of their programs using the \emph{extract-and-solve} method.
First, we \emph{extract a recurrence} from the program. In this context,
a recurrence is a mathematical object---usually an inequality---that
expresses an upper bound for the running time of a program in terms of
the \emph{size} of its input. Depending on the task at hand, this notion
of size may vary. For example, if the input is a tree, we may define its
size to be its number of nodes, its depth, or some more complicated
expression. The second step consists of \emph{solving} this recurrence:
mathematical methods are used to express it (or a suitably looser
version of it) in a non-recursive \emph{closed form} and big-$O$ bound.
While this method is taught in introductory textbooks (e.g. \citet[\S
  7]{Bird2014}), there is no \emph{formal} connection between the
program and the extracted recurrence.  In this work we concentrate on
the first of those steps: we seek a method to \emph{automatically
  extract a recurrence} from the syntax of a recursive functional
program, in such a way that we can prove a formal \emph{bounding
  theorem} relating the extracted recurrence to the program's
operational cost.

\citet{Danner2015} present such a method for a call-by-value terminating
language with inductive types and structural recursion. Their method
consists of the following steps:
\begin{itemize}
  \item Given a program $M$ in the \emph{source language}---i.e. the
    language that we wish to analyse---a \emph{syntactic recurrence}
    $\complexity M$ is extracted. This recurrence is expressed in an
    appropriate \emph{recurrence language}, which includes primitives
    for expressing cost. This step is close in spirit to a monadic
    translation of the program into the writer monad.  This approach
    is able to accommodate higher-order functions, assigning to
    them a higher-order recurrence expressing their cost in terms
    of a recurrence for their input.

  \item A \emph{bounding relation} between source programs and syntactic
    recurrences is defined by induction on the types of the source
    language, i.e. as a logical relation. Intuitively, a source program
    is bounded by a recurrence if the components of the recurrence
    express \emph{upper bounds} for the attributes of the source
    program, e.g. evaluation cost, size, etc.

  \item Following that, a \emph{bounding theorem} is proved. This
  shows that every source program $M$ is bounded by the extracted
  recurrence $\complexity M$.

  \item Finally, a denotational semantics is provided for the
  recurrence language. Depending on the intended application, this
  semantics abstracts inductive data types to some notion of size.
  For example, to consider binary trees up to their height we
  might interpret them as natural numbers, with the node
  constructor interpreted as the maximum.  Alternatively, the node
  constructor may be interpreted as addition, thus yielding the
  number of nodes.  
\end{itemize}

\begin{figure}
  \import{diagrams/}{cbx_recurrence_extraction}
  \caption{Recurrence extraction}
  \label{fig:recextr1}
\end{figure}
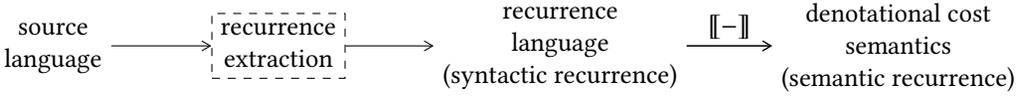

\noindent This strategy is shown schematically in Figure
\ref{fig:recextr1}.  

In this previous work, the \emph{recurrence language} was taken to
be a call-by-name language, so as to be as `close to mathematics'
as possible. Composing the extraction with the semantic
interpretation then yields what one might call a \emph{semantic
recurrence}, which is intended to match the recurrence we would
informally write when teaching undergraduate students---at least
in the context of first-order programs. Thus the entire process
gives a formal account and justification for informal cost
analysis techniques.  Factoring this into a syntactic and a
semantic step is a useful tool for obtaining a simplifying
account, which is additionally modular in the different notions of
size for inductive data types.  Nevertheless, it is still somewhat
rigid with respect to changes to the \emph{source language}: each
source language requires a new and complex logical relations proof.

\begin{figure}
  \import{diagrams/}{cbpv_recurrence_extraction}
  \caption{Recurrence extraction through CBPV}
  \label{fig:recextr2}
\end{figure}
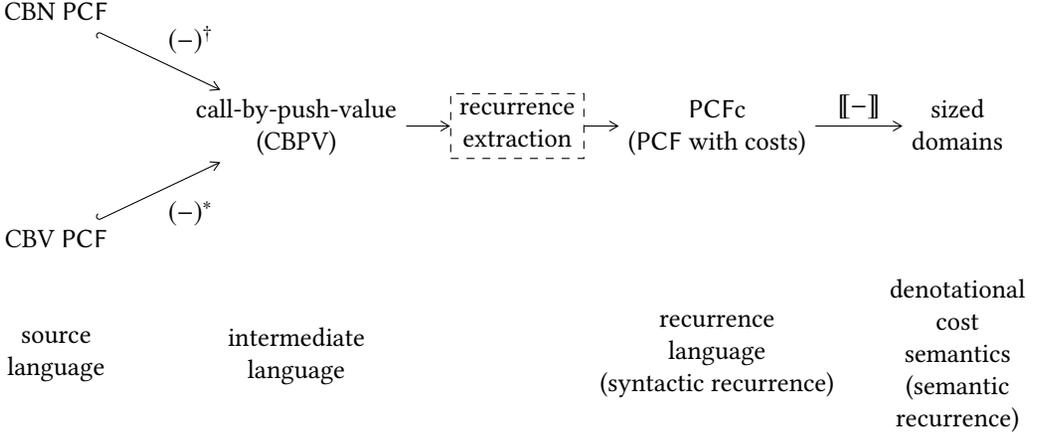

In this paper, we improve upon the above approach in several ways:
\begin{enumerate}
\item We factor the syntactic phase into a cost-preserving embedding
  of the source language in an \emph{intermediate language}, followed by
  a recurrence extraction for the intermediate language in
  the style of \citet{Danner2015}. The bounding theorem is proved
  once for the intermediate language, so for each source language we
  need only prove that the cost-preserving embeddings are correct.  This
  strategy is illustrated Figure \ref{fig:recextr2}.

\item In order to support a wide variety of languages, we choose as an
  intermediate language Call-by-Push-Value (CBPV)~\citep{Levy2003},
  which embeds both call-by-value and call-by-name evaluation.  We
  define recurrence extraction and prove a bounding theorem for CBPV,
  and obtain recurrence extraction and bounding theorems for both
  call-by-value (CBV) and call-by-name (CBN) source languages by
  embedding them in CBPV.
  
\item Our intermediate language is a version of CBPV with general
  recursion. Thus, we obtain a new recurrence extraction and
  bounding theorem result for CBV with general recursion, and a
  simpler recurrence extraction and bounding theorem for CBN than
  \citet{Kim2016}. General recursion is important in our setting
  for writing divide-and-conquer algorithms in the standard way:
  in order to give a formal analogue of the standard informal
  techniques, we should not require the programmer to embed a
  termination metric in the source program.

\item We generalize the denotational semantics from sets equipped with a
  size ordering to domains equipped with a size ordering.  This
  involves some subtle questions regarding the interaction of the
  size ordering with the information order of the domains. We
  develop a semantics that identifies the \emph{bottom} element of
  the domains (non-termination) with the \emph{maximum} element of
  the size ordering (infinity), and more generally
  reverse-includes the information order in the size order---a
  more defined recurrence is a more precise bound.  The key
  features of this semantics are encoded in a syntactic recurrence
  language \pcfc.
\end{enumerate}

In more detail, our technical development proceeds as follows.  We
first define three variants of \pcf{}, starting in Section \ref{section:source} with relatively
standard CBV and CBN source languages. To isolate the issues involved in
considering general recursion, we choose source languages with
only natural numbers as data (treated as a flat base type). Next,
in Section \ref{section:recurrence} we define the recurrence
language $\pcfc{}$, a version of call-by-name~\pcf{} with an
additional type of costs. The key axiomatic component of $\pcfc{}$
is its \emph{size order} (really, a pre-order relation).
The size order codifies the minimum requirements
necessary to prove the bounding theorem, while subsequent denotational 
interpretations of $\pcfc{}$ can refine this to specific notions of cost and
size.   The primary conceptual
difficulty in the size order is understanding how it interacts
with the implicit information order that arises from models
of~$\pcf{}$, which is necessary for verifying the bounding theorem
for recurrences extracted from recursively-defined functions. Our
approach takes more
defined recurrences to be \emph{smaller} in the size order, i.e. to be better
bounds. This is encoded in two of the rules that axiomatize the
size order: one that asserts that rational chains are decreasing
in the size order, and one that asserts that if all the
approximants to a fixed point are a bound, then so is the fixed
point itself.

Next, we describe the ``end-user'' recurrence extraction functions and
corresponding bounding theorems for CBV and CBN source languages in
Section~\ref{section:extraction-cbx}.  The statements of these
theorems for the two source languages are independent of
the intermediate language, but we will prove them by factoring
them through it, as indicated in Figure~\ref{fig:recextr2}.
These two examples illustrate the flexibility of our approach:
even though they originate from the same general theorem, the
the two bounding relations are very different in spirit.

The rest of the paper is largely devoted to proving that the extraction
functions in Section~\ref{section:extraction-cbx} are correct using the
approach of Figure~\ref{fig:recextr2}.  We define the version of Levy's
\emph{Call-by-Push-Value} (\cbpv{}) \citep{Levy2003} that we use as an
intermediate language in Section~\ref{section:cbpv}. \cbpv{} is a
polarised $\lambda$-calculus that is structured around a fundamental
dichotomy between \emph{values} and \emph{computations}.  The primitives
of $\cbpv{}$ provide logical mechanisms that control evaluation.
In terms of expressivity, these mechanisms are strong enough to
allow a faithful embedding of both CBN and CBV \pcf{} in \cbpv{}.
The structure of \cbpv{} ensures
compatibility with computational effects such as printing, memory, etc.
For the purposes of this paper, we will consider cost to be an effect:
the embeddings of CBN and CBV into \cbpv{} will explicitly mention a
\emph{command} that incurs evaluation cost whenever some should be incurred
according to the source language operational semantics.  This makes the
embeddings essentially parametric in the evaluation costs: even if we
decide to change the costs---e.g. by charging twice as much for a
function call---we only need to slightly alter the embedding into
\cbpv{} and the accompanying cost-preservation proof.  We thus
demonstrate that \cbpv{} is an extremely flexible and general
intermediate language for recurrence extraction.

The extraction function and bounding relation for \cbpv\ is defined in
Section~\ref{section:extraction-cbpv}.  It is here that we reap the
benefits of the distinction between values and computations, for it is
clear that they require different notions of bounding.  This in turn
explicates the differences in the bounding relations for our original
call-by-value and call-by-name versions of \pcf.  The cost-preserving
embeddings of the two languages that we give in
Section~\ref{section:embed} are fundamentally different: following
\citet{Levy2003}, CBN types are translated to \cbpv\ computation types,
whereas CBV types are translated to \cbpv\ value types.

We define a denotational semantics for the recurrence language
\pcfc{} in Section~\ref{section:sized-domains} by introducing
\emph{sized domains}, which combine size and information orders.
We observe that the standard model of \pcf{} extends to a model in
sized domains, in which the natural numbers are given their usual
order. In Section~\ref{section:exponentiation-example}, we use
this model to give an end-to-end example of recurrence extraction
for a non-structural recursion, in order to show that the
recurrence we obtain indeed matches the one we would expect from
an informal analysis. Finally, in Section~\ref{section:inductive}
we sketch an extension of our work to a call-by-value source
language with inductive types, following the approach of
\cite{Danner2015}, and use it to extract a recurrence from a
program implementing merge sort. Both this and the exponentiation
example validate our claim that our technique formalizes the usual
informal extract-and-solve approach to cost analysis.

\section{The Source Languages: PCF}
  \label{section:source}

Our objective in this paper is to present a completely formal and
provably correct way to extract recurrences from general-recursive
functional programs. For the sake of clarity, we target a small
formal calculus that illustrates the core features of this style
of programming. Perhaps the most well-understood calculus of this
sort is \pcf{} \citep{Plotkin1977}. \pcf{} essentially consists of
a simply-typed $\lambda$-calculus, a base type of natural numbers,
and some mechanism for obtaining fixed points. Almost every
introductory book on programming language semantics includes a
wealth of material on it: see e.g. \cite{Gunter1992, Mitchell1996,
Streicher2006}. As discussed in the introduction, we would like to
extract recurrences from both a CBN and a CBV variant of \pcf{}.

\begin{figure}
  \input{pcf-types}
  \caption{PCF}
  \label{fig:pcf-types}
\end{figure}

The types, terms, and typing judgments of \pcf{} are defined in
Figure \ref{fig:pcf-types}. The version we use comes with flat
natural numbers, product, and function types. We write
$\mathcal{T}^{\pcf{}}_{A}$ for the (open and closed) terms of
\pcf{} of type $A$, and $\mathcal{V}^{\pcf{}}_{A}$ for the CBV
canonical forms of type $A$.

Our natural numbers are introduced by constants $\num{0}, \num{1},
\dots$ and correspond to eager, rather than lazy, natural numbers.
Instead of the usual predecessor and successor functions, we
introduce five arithmetic operations on natural numbers, as well
as the zero test $\pif{N}{P}{Q}$, which behaves like $P$ if $N$ is
$\num{0}$, and like $Q$ otherwise. This choice of primitives gives
them the flavor of \emph{machine words}. We sometimes refer to
these as ``flat'' natural numbers, as they admit a domain
interpretation with a flat information order (bottom below all
numerals, and no other relations).

The terms of the CBN and the CBV variants differ only on fixed
points. In the case of CBN we may take fixed points at every type:
if we have $M : A$ in terms of $x : A$, we may construct
$\fix{x}{M} : A$. However, in CBV we are only allowed to take
fixed points at function types, i.e. to make recursive function
definitions: if we have $M : B$ in terms of an argument $x : A$
and a recursive variable $f : A \rightarrow B$, then we can obtain
a recursively defined function $\rec{f}{x}{M} : A \rightarrow B$.
This formulation of CBV PCF is commonly found in the literature;
see e.g. \citep[\S 7]{Levy2006}, \citep[\S 1.2.5]{Fiore1996}, or
\citep{Plotkin2001}.

\begin{figure}
  \centering
  \input{pcf-bigstep}
  \caption{Big-step semantics for PCF}
  \label{fig:pcf-bigstep}
\end{figure}

The big-step semantics of the two variants of \pcf{} are defined in
Figure \ref{fig:pcf-bigstep}. Judgments of the form $\evalpcf{M}{V}{n}$
are read as ``$M$ evaluates to canonical form $V$ incurring a cost $n$
in the process.'' Except the presence of the cost superscripts, these
rules are standard in the literature on \pcf{}, while the cost
annotations follow \citet{blelloch-greiner:fpca95}.  Each rule sums the costs
incurred in the premises and adds additional cost for the evaluation
step represented by that rule.  For simplicity, we choose to charge one
unit of cost for pair projections and (possibly recursive) function
applications, and no cost otherwise. This could easily be adjusted to
give any constant cost for each reduction, but---as recurrence
extraction is syntax-directed---each reduction for the same term
constructor must incur the same cost (e.g. we cannot charge $1$ for
reducing an $\textsf{if}$ to the $0$ branch and $2$ for reducing it to the
other branch). We have that
\begin{theorem}[Determinacy]
  \label{theorem:PCF-determinacy}
  There is at most one pair $n$, $V$ such that $\evalpcf{M}{V}{n}$.
\end{theorem} 

\noindent We let $\bounded{M} \defiff \exists n, V.\
\evalpcf{M}{V}{n}$ which we pronounce as ``$M$ is bounded,'' and
write $\unbounded{M}$ to mean $\lnot\left(\bounded{M}\right)$.

\section{The Recurrence Language: PCF with Costs}
  \label{section:recurrence}

\begin{figure}
  \centering
  \input{pcf-reclang}
  \caption{PCF with costs}
  \label{fig:pcf-reclang}
\end{figure}

The recurrences we aim to extract from \pcf{} terms will
themselves be expressed in an appropriate \emph{recurrence
language}, which will also be a version of \pcf{}, and which we
call \emph{\pcf{} with costs} (\pcfc{}). As mentioned before, we
would like the recurrence language to behave in a way that is as
close to mathematics as possible. Hence, we choose \pcfc{} to be
a CBN language.

In addition to standard constructs, \pcfc{} also sports an
additional type of costs, which we denote by $\cost$. We assume as
little as possible about costs: there are constants for no cost
and unit cost, denoted $\zeroc$ and $\unitc$. Furthermore, \pcfc{}
comes with an operator that adds arbitrary costs together: if $M :
\cost$ and $N : \cost$ we have $M \plusc N : \cost$. We define
$\numc{n}$ to be the right-associated sum $\unitc \plusc \dots
\plusc \zeroc$ of $n$ unit costs.  We write
$\mathcal{T}^{\pcfc{}}_{A}$ for the (open and closed) terms of
\pcfc{} of type $A$.

The operational semantics of \pcfc{} consists of a judgment
$\evalpcf{M}{V}{}$, which is read as ``term $M$ evaluates to
canonical form $V$.'' Notice that this judgment does not record
any evaluation costs, as we are not interested in the running time
of recurrences. The rules for $\evalpcf{M}{V}{}$ consist of all
the CBN rules of Figure \ref{fig:pcf-bigstep}---with the cost
annotations erased---along with two additional rules that handle
the evaluation of costs: one which allows $\zeroc$ and $\unitc$ to
evaluate to themselves, and one which handles the summation of
costs in the expected way. These facts are summarised in Figure
\ref{fig:pcf-reclang}.

However, the operational semantics of \pcfc{} play a very limited
r\^{o}le: they are only used to define the predicate $\bounded{M}$
for \pcfc{}. Instead, the central device used alongside \pcfc{} is
the \emph{size order}. It consists of judgments of the form
$\Gamma \vdash M \costleq N : A$, which we read as ``$M$ is bounded
above by $N$ at type $A$ in context $\Gamma$.'' The size order is
used to compare the costs denoted by recurrences, as well as the
sizes of values.

\begin{figure}
  \centering
  \input{pcf-inequality}
  \caption{Monotone contexts and the size order for \pcfc{}}
  \label{fig:pcf-inequality}
\end{figure}

The full inductive definition of the size order is introduced in
Figure \ref{fig:pcf-inequality}. To explain its definition, we
first need to discuss three auxiliary notions.

\paragraph{\bfseries Syntactic Unfolding}

The \emph{$n$th syntactic unfolding} of a fixed point term,
denoted by $\fixn{x}{n}{M}$, is in a sense the ``$n$th
approximation'' to the fixed point $\fix{x}{M}$. Intuitively,
$\fixn{x}{n}{M}$ is the term that runs for up to $n$ recursive
calls, and in the next call decides to diverge.  It is defined by
\begin{align*}
  \fixn{x}{0}{M}   &\defeq \fix{x}{x} \\
  \fixn{x}{n+1}{M} &\defeq M[\fixn{x}{n}{M}/x]
\end{align*} It is easy to see that \begin{proposition}
  If $\Gamma, x : A \vdash M : A$ then $\Gamma \vdash
  \fixn{x}{n}{M} : A$ for any $n \geq 0$.
\end{proposition}

\paragraph{\bfseries Monotone contexts}

As is usual, \pcf{} term contexts are defined in a manner similar to \pcf{}
terms. The difference is that \pcf{} contexts have had one of their
subterms replaced by a \emph{hole}, denoted by $[]$. For example, the
context $\ectx{C} \defeq \lambda x. M([] + 3)$ is obtained by replacing the
term $N$ in $\lambda x. M(N + 3)$ by a hole $[]$. We may recover the
original term $\lambda x. M(N + 3)$ by \emph{filling} the hole with $N$,
which we write as $\ectx{C}[N]$ and define in the usual way.  Recall
also that---unlike substitution---filling a hole may result in variable
capture. For example, the variable $x$ is no longer free in $\ectx{C}[x]
\equiv \lambda x.  M(x + 3)$.

Of the contexts, we choose certain ones to be \emph{monotone},
which means that they preserve the size order
$\costleq$.  In order to prove the bounding theorem about
recurrence extraction into \pcfc{}, we will need that the monotone
contexts include at least the principal positions of elimination
forms, as well as introduction forms for negative types (i.e. the
bodies of functions and pairs), as stated in Figure
\ref{fig:pcf-inequality}.  It is always permissible, and sometimes
desirable, to make more contexts monotone (e.g. all of them).
However,
this places more requirements on semantic models of \pcfc, so in
this figure we only include those that are necessary for
proving the bounding theorem.  Monotone contexts are introduced by
a typing judgment $\mctxtt{\mctx{C}}{\Gamma}{A}{\Delta}{B}$. The
meaning of this judgment is this: the hole in $\mctx{C}$ is
assumed to represent a term $\Gamma \vdash M : A$, i.e. a term of
type $A$ in context $\Gamma$. When we fill the hole with a $\Gamma
\vdash M : A$, the resultant term $\mctx{C}[M]$ has type $B$ in
context $\Delta$.  Note that the context may change, as $\mctx{C}$
might bind some free variables. 

\begin{proposition}[Filling Typing]
  $\Gamma \vdash M : A\
    \land\
  \mctxtt{\mctx{C}}{\Gamma}{A}{\Delta}{B}\
    \Longrightarrow\
  \Delta \vdash \mctx{C}[M] : B$
\end{proposition}

\paragraph{\bfseries Eliminative contexts}

Of the monotone contexts, the (negative) \emph{eliminative contexts}
consist of a series of applications and projections. They are
defined by \[
  \ectx{E}\ ::=\
         []
    \mid \pi_i(\ectx{E})
    \mid \ectx{E} M
\] Eliminative contexts are strict in effects, and in particular
they preserve divergence:

\begin{lemma}[Preservation of Infinity]
  \label{lemma:preservation-of-infinity}
  For an eliminative context $\ectx{E}$, $\unbounded{M}$ implies
  $\unbounded{\ectx{E}[M]}$.
\end{lemma}

\paragraph{\bfseries The rules of the size order}

The rules of the size order given in Figure
\ref{fig:pcf-inequality} consist of exactly those that are needed
in order to prove the bounding theorem of
\S\ref{section:extraction-cbpv}. For example, we do not assume
that $\num{n} \costleq \num{n+1} : \natt$ or $\zeroc \costleq
\unitc : \cost$, so that we can consider models where $\costleq$
is actually equality, in addition to models where these
inequalities do hold (such as the one given in Section
\ref{section:sized-domains}).

The rules for $\costleq$ can be naturally organised in three
groups. The first group is the one that makes $\costleq$ a
preorder---i.e. (\hyperlink{refl}{\textsf{refl}}),
(\hyperlink{trans}{\textsf{trans}})---and preserved by monotone
contexts (\hyperlink{ctx}{\textsf{ctx}}). The second group
contains rules that encode small-step $\beta$-reduction: these
ensure that $N \costleq M$ whenever we would have had $M
\longrightarrow_\beta N$. This group contains the rules
(\hyperlink{zero}{\textsf{zero}}) and
(\hyperlink{assoc}{\textsf{assoc}}), which ensure that $\numc{n +
m} \costleq \numc{n} + \numc{m}$ for all $n, m \in \mathbb{N}$, as
a simple induction shows. 

Finally, the third and most interesting group consists of the two
rules that concern fixed points. Both of these rules deal with
sequences of terms that we call \emph{rational chains}, i.e.
sequences of the form $\left(\fixn{x}{i}{M}\right)_{i \in
\omega}$, which semantically correspond to chains of the form
$\left(f^i(\bot)\right)_{i \in \omega}$. Rational chains are
ubiquitous in syntactic and intensional models of \pcf{}: see
e.g. \cite{Milner1977, Abramsky1996, Pitts1997, Escardo2009}.

The rule (\hyperlink{rat}{\textsf{rat}}) ensures that rational
chains are \emph{decreasing} in the size order, i.e.
that
\[
  \dots \costleq \fixn{x}{2}{M}
        \costleq \fixn{x}{1}{M} 
        \costleq \fixn{x}{0}{M} \defeq \infty
\] Recall that the $(n+1)$th syntactic unfolding may make more
recursive calls than the $n$th, so it is a more defined term.  Hence,
this rule intuitively states that \emph{a more defined recurrence is a
  tighter bound}. Note that as $\fixn{x}{n+1}{M}$ is defined to be
$M[\fixn{x}{n}{M}/x]$, this rule is analogous to $\beta$ for
$\textsf{fix}$.\footnote{We do not require $\beta$ for $\textsf{fix}$
  itself as a size order axiom in order prove the bounding theorem,
  essentially because in the case of the bounding theorem for fixed
  points, we will apply a fixed point induction principle to instead
  reason about $\textsf{fix}_n$.  The $\beta$ axioms are generally used
  in the proof of the bounding theorem to head expand, showing that a
  redex is in the relation if its reduct is, but this particular
  reduction does not come up.}

Finally, the rule (\hyperlink{cpind}{\textsf{cpind}}) is a form of
\emph{computational induction} for \pcfc{}. It ensures that this
process of iterative tightening of bounds does not `overshoot': if
we can prove that $\fixn{x}{n}{E}$ is an upper bound for $M$ for
every $n \geq 0$, then so is $\fix{x}{E}$. Moreover, this rule
ensures that computational induction can take place under any
eliminative context $\ectx{E}$.

We immediately obtain the following results.

\begin{lemma}[Infinite Loop is a Top Element (Infinity)]
  $\Gamma \vdash M \costleq \fix{x}{x} : A$ for all $\Gamma
  \vdash M : A$.
\end{lemma}
\begin{proof}
  Use the rule $(\textsf{\hyperlink{rat}{\textsf{rat}}})$, $n
  = 0$ for a fresh variable $x \not\in \textsc{Vars}(\Gamma)$.
\end{proof}

\begin{lemma}[Bounded terms are a lower set]
  \label{lemma:bounded}
  $
    \bounded{M}\ 
      \land\ 
    N \costleq M : A\ 
      \Longrightarrow\ 
    \bounded{N}
  $
\end{lemma}

\begin{proof}
  By induction on $N \costleq M : A$.
\end{proof}

\section{Recurrence Extraction for Call-by-Name and Call-by-Value}
  \label{section:extraction-cbx}

We now have enough details in place to show how to extract
recurrences for both call-by-name and call-by-value \pcf{}.

\subsection{Call-by-Value}

\begin{figure}
  \centering
  \input{cbv-extraction}
  \caption{Recurrence extraction for CBV \pcf{}}
  \label{fig:cbv-extraction}
\end{figure}

The extraction procedure for CBV is very similar to that of
\cite{Danner2013,Danner2015}, and is displayed in Figure
\ref{fig:cbv-extraction}. To begin, we map each type $A$ of CBV
\pcf{} to a type \[
  \cbvrec{A} \defeq \cost \times \cbvpot{A} \\
\] of \pcfc{}, where $\cbvpot{A}$ is defined by induction on $A$.
We call this the type of \emph{complexities for $A$}. A complexity
for $A$ consists of a pair: its first component is the \emph{cost}
of evaluating a term of type $A$ to a value, and its second
component is the \emph{potential} of the resultant value. If $E :
\cbvrec{A}$ we write $E_c \defeq \pi_1(E) : \cost$ and $E_p \defeq
\pi_2(E) : \cbvpot{A}$ to refer to these two components
respectively.

The notion of potential is crucial in CBV extraction.  There are varying
interpretations about the nature of potentials: in one reading,
they---directly or indirectly---encode information about the \emph{size}
of values; in another, they represent the future cost of using that
value, to which we often refer as \emph{use-cost}. The definition of
$\cbvpot{A}$ is also given in Figure \ref{fig:cbv-extraction}.  Perhaps
the most interesting clause is $\cbvpot{A \rightarrow B} \defeq
\cbvpot{A} \rightarrow \cbvrec{B}$.  We may read this as follows. The
cost of using a value of function type---i.e. a
$\lambda$-expression---is expressed as a function itself. This function
that maps the size/use-cost of a value of type $A$ to a pair of an
evaluation cost (the cost of evaluating the application of that
$\lambda$-expression to a value with that use-cost) and the use-cost of
a value of type $B$.  This is consistent with the idea that in CBV
\emph{variables represent values}.

Following that, we define a map that extracts a recurrence from
each term of CBV \pcf{}, which we also denote by $\cbvrec{-}$ (see
\citet{Danner2015} for an explanation of most of the cases).
Arithmetic operations $M \pcfop N$ are chosen to have zero cost in
addition to the cost of evaluating their inputs (since for big-$O$
bounds we mainly count recursive calls), but for their potential
we distinguish some cases, guided by two constraints needed by our
proof of the bounding theorem: the potential must be an upper
bound on the value of the operation, and must be monotone in the
potentials of the operands wherever mathematically possible
Addition/multiplication
are monotone in both $M$ and $N$, so we use
addition/multiplication to combine the potentials of $M$ and $N$,
which will be an upper bound on $M \{+,*\} N$.  In general,
subtraction and division are monotone in $M$ but antimonotone in
$N$, so we take the potential to be that of $M$, which is an upper
bound on $M \{-,\div\} N$. However, for subtraction or division
with $N$ a numeric constant there will be no monotonicity
obligation for $N$ (because it does not vary), so we can perform
the $\{-,\div\}$ in the potential. This precision is useful for
analyzing algorithms that subtract/divide by a constant in
recursive calls.  Finally, $\bmod$ is not monotone in either
position, so we cannot use $\bmod$ in the potential; but it is
bounded by both $M$ and $N-1$, and we somewhat arbitrarily choose
the later.  For algorithms (e.g. GCD) that recur on $x
\{-,\div,\bmod\} y$, we would need more sophisticated tracking of
monotone and antimonotone positions in the recurrence language.
The other new case is for recursive functions, and interprets them
using fixed points in \pcfc.


Extending $\cbvpot{-}$ to contexts pointwise, we have
\begin{theorem}
  If $\Gamma \vdash M : A$ then $\cbvpot{\Gamma} \vdash \cbvrec{M}
  : \cbvrec{A}$.
\end{theorem}

It remains to state a theorem to the effect that each extracted
recurrence $\cbvrec{M} : \cbvrec{A}$ encodes an upper bound for the
evaluation of $M$. Because of the presence of function types, this is
stated as a logical relation:

\begin{theorem}[CBV extraction]
  \label{theorem:cbv-extraction}
  There exist relations \[
  {(M : \mathcal{T}^{\pcf,\textrm{closed}}_{A}) \logrel_A
   (E : \mathcal{T}^{\pcfc,\textrm{closed}}_{\cbvrec{A}})}
      \quad\text{ and }\quad
    {(V : \mathcal{V}^{\pcf,\textrm{closed}}_{A} ) \valrel_A (E :
    \mathcal{T}^{\pcfc,\textrm{closed}}_{\cbvpot{A}})}  
  \] such that \begin{alignat*}{3}
    & M \logrel_{A} E
      &&\quad\Longrightarrow\quad
      &&\text{if } \bounded{E_c} \text{ then }
          \exists n, V.\ 
            \begin{cases}
              \evalpcf{M}{V}{n} \\
              \numc{n} \costleq E_c \\
              V \valrel_{A} E_p
            \end{cases} \\
    &\num{n} \valrel_{\natt} E
      &&\quad\Longrightarrow\quad
      && \num{n} \costleq E \\
    &\tuple{V_1}{V_2} \valrel_{A_1 \times A_2} E
      &&\quad\Longrightarrow\quad
      && \begin{cases}
            V_1 \valrel_{A_1} \pi_1(E) \\
            V_2 \valrel_{A_2} \pi_2(E)
          \end{cases} \\
    &\lambda x. M \valrel_{A \rightarrow B}  E
      &&\quad\Longrightarrow\quad
      &&\forall(N \valrel_{A} X).\ M[N/x] \logrel_{B} E\,X \\
    &\rec{f}{x}{P} \valrel_{A \rightarrow B} E
      &&\quad\Longrightarrow\quad
      &&\forall(N \valrel_{A} X).\ P[\rec{f}{x}{P}/f, N/x] \logrel_{B} E\,X
  \end{alignat*} and, moreover, 
  \begin{enumerate}
    \item
      $V \valrel_{A} \cbnrec{V}_p$ for any CBV \pcf{} value $\cdot \vdash V : A$,
    \item
      $M \logrel_{A} \cbnrec{M}$ for any closed CBV \pcf{} term $\cdot \vdash M : A$.
  \end{enumerate}
\end{theorem}

We prove this as a corollary of recurrence extraction for CBPV
below.  Intuitively, the \emph{expression bounding relation} $M
\logrel_{A} E$ says that the cost and value of a CBV PCF program
$M$ are predicted by $E$ in the following sense: if the cost
component of $E$ terminates, then so does $M$, and the evaluation
cost of $M$ is bounded by the cost component of $E$ according to
the size order. Moreover, the value is bounded by the
potential component of $E$, according to the value bounding
relation $\valrel_A$. The value bounding relation is
type-directed: at $\natt$, it says that $n$ is bounded by $E$
according to the size order; for pairs, it says that the
components are bounded; and for functions it says that the
applications are bounded.

Relative to the bounding relation in \citet{Danner2015}, the key change
here for supporting general recursion involves the quantifiers in
expression bounding.  In that work, expression bounding was defined as
``if $\evalpcf{M}{V}{n}$ then $\numc{n} \costleq E_c$ and $V \valrel_{A}
E_p$'' (if the source program evaluates, then the recurrence's
prediction is correct), though because all programs in the language
considered there terminate, this is equivalent to ``$\evalpcf{M}{V}{n}$
and $\numc{n} \costleq E_c$ and $V \valrel_{A} E_p$.''  Here, we assert
this same guarantee \emph{only if} the cost component of the recurrence
itself terminates.  We view a recurrence whose cost diverges as
predicting an infinite cost for the program, so the expression bounding
relation is vacuously true in this case, expressing that any program
meets this bound.  

\subsection{Call-by-Name}

The extraction procedure for CBN \pcf{} is very different
to the one for CBV. We will discover the precise reasons for that
through \cbpv{} in \S\ref{section:embed}, but for now we will
satisfy ourselves with the following intuitions. When studying
sundry versions of \pcf{} we often speak of certain types
as being \emph{observable}, in the sense that the successful
normalization of terms at those types provides some manifest
information to the user. For example, every term $M : \natt$
either diverges or is observed to converge to a constant
$\num{n}$. However, in CBN we are unable to observe a function:
the only action we may perform is that of applying it to an
argument. In contrast, in CBV \emph{every type is
observable}:\footnote{There is a third paradigm which is close to
CBN, but where termination at function type is observable. This is
usually called the \emph{lazy} paradigm; see \cite{Abramsky1990b},
\cite{Riecke1993}, and \cite[\S 1.7.3]{Levy2003}.} for example,
when evaluating a term at function type we either expect
divergence or convergence to a $\lambda$-expression.

In the case of CBN we only have one observable type, i.e. that of
natural numbers. This is to say that we expect the end user to
only evaluate terms of type $\natt$. It follows that only
complexities for those terms should contain cost components.
Complexities at any other type will need to come with some
mechanism for ``pushing costs down to $\natt$.'' Thus, a
complexity for a CBN type will not merely be a type of \pcfc{},
but will also come with some kind of algebra structure that
enables us to do that at every type.

\begin{definition}[Cost algebra]
  A (\pcfc{}) \emph{cost algebra} $A = (\carrier{A}, \alpha_A)$
  consists of a type $\carrier{A}$ and a \pcfc{} term $ c : \cost,
  x : \carrier{A} \vdash \alpha_A(c, x) : \carrier{A} $ such that
  \[
    c : \cost, x : \carrier{A} 
      \vdash 
        \alpha_A(c_1 \plusc c_2, x) 
          \costleq 
        \alpha_A(c_1, \alpha_A(c_2, x)) : \carrier{A}
  \]
\end{definition}

A \pcfc{} cost algebra\footnote{The analogous equation $x : \carrier{A}
  \vdash x \costleq \alpha_A(\zeroc, x)$ is true if $\costleq$
  additionally includes certain $\eta$-contractions, but this
  property is not used here so we omit these $\eta$ rules for maximum
  generality.} $A = (\carrier{A}, \alpha_A)$ consists of a type
$\carrier{A}$, which we call its \emph{carrier}, and a \emph{structure
  map} $\alpha_A(c, x)$, which is expressed as a term in two free
variables. The structure map allows one to ``add costs'' to any term of
carrier type.  From this point onwards we will abuse notation by not
making a formal distinction between algebras $A = (\carrier{A},
\alpha_A)$ and their carriers $\carrier{A}$.

\begin{figure}
  \centering
  \input{cbn-extraction}
  \caption{Recurrence extraction for CBN \pcf{}}
  \label{fig:cbn-extraction}
\end{figure}

The extraction procedure for CBN may be found in Figure
\ref{fig:cbn-extraction}. We begin by defining an algebra
$(\cbnrec{A}, \alpha_A)$ for each type $A$ of CBN \pcf{}.  In
particular, $\natt$ is mapped to the \emph{``free algebra''}
$(\cost \times \natt, \alpha_\natt)$ on $\natt$. For simplicity we
use the notation $L \algp{A} M \defeq \alpha_A(L, M)$ whenever $L
: \cost$ and $M : \cbnrec{A}$.

We then inductively define a map $\cbnrec{-}$ from \emph{typed} terms of
\pcf{} to terms of \pcfc{}. This definition uses the aforementioned
algebras, and so the output depends on the type of each input
term.\footnote{Formally, the function is defined on typing
  derivations/intrinsically typed syntax $\cbnrec{\Gamma \vdash M : A}$,
  but we elide the annotations for brevity.}  This definition uses the
algebras to push the cost of evaluation towards the ground
types. Indeed, costs here are not added at the elimination rules, but
\emph{at the introduction rules}.  Consider the lazy product type $\natt
\times \natt$ as an example.  A pair $\tuple{M}{N}$ of this type is not
directly observable, but contains two observable computations $M$ and
$N$. A computation that uses this pair costs one unit more than a
computation that uses either $M$ or $N$ directly, because it must do the
product projection.  

This translation is well-typed:

\begin{theorem}
  If $\Gamma \vdash M : A$ then $\cbnrec{\Gamma} \vdash \cbnrec{M}
  : \cbnrec{A}$.
\end{theorem}

\noindent Furthermore, it satisfies a bounding theorem given as a
logical relation:

\begin{theorem}[CBN extraction]
  \label{theorem:cbn-extraction}
  There exists a relation \[
    (M : \mathcal{T}^{\pcf,\textrm{closed}}_A)\
      \logrel\
    (E : \mathcal{T}^{\pcfc,\textrm{closed}}_{\cbnrec{A}} )
  \] such that \begin{alignat*}{3}
    &M \logrel_{\natt} E
      &&\quad\Longrightarrow\quad
      &&\text{if } \bounded{E_c} \text{ then }
          \exists n, V.\ 
            \begin{cases}
              \evalpcf{M}{V}{n} \\
              \numc{n} \costleq E_c : \cost \\
              V \costleq E_p : \natt
            \end{cases} \\
    &M \logrel_{A_1 \times A_2} E
      &&\quad\Longrightarrow\quad
      && \begin{cases}
            \pi_1(M) \logrel_{A_1} \pi_1(E) \\
            \pi_2(M) \logrel_{A_2} \pi_2(E)
          \end{cases} \\
    &M \logrel_{A \rightarrow B}  E
      &&\quad\Longrightarrow\quad
      &&\forall(N \logrel_{A} X).\ M\,N \logrel_{B} E\,X
  \end{alignat*} and, moreover, $M \logrel_{A} \cbvrec{M}$ for any
  \pcf{} term $\cdot \vdash M : A$.
\end{theorem}

Unlike in CBV, we do not distinguish between value and expression
relations.  The CBN relation for $\natt$ (or other observable
types, if we had them) is analogous to the CBV expression
relation, while the clauses for negative types simply say that the
elimination forms preserve relatedness. Again, rather than proving
this bounding theorem directly, we will obtain it as a corollary
of a general theorem about CBPV.

\section{The Intermediate Language: Levy's Call-by-Push-Value (CBPV)}
  \label{section:cbpv}

We aim to prove both theorems of Section
\ref{section:extraction-cbx} at once. In order to do so, we will
embed both CBN and CBV \pcf{} in an \emph{intermediate language},
and we will do so in a cost-preserving manner. Finally, we will
define a recurrence extraction function for this intermediate
language, and prove that it is correct. The composite of these two
processes will prove the results of Section
\ref{section:extraction-cbx}.

\begin{figure}
  \centering
  \input{cbpv-types}
  \caption{Call-by-Push-Value (CBPV) with natural numbers and recursion}
  \label{fig:cbpv-types}
\end{figure}

Our intermediate language of choice of Levy's
\emph{Call-by-Push-Value} (\cbpv{}) \cite{Levy2003}. \cbpv{} is a
polarised $\lambda$-calculus whose type structure provides ways of
controlling evaluation. The typing judgments of the version of
\cbpv{} that we will use in this paper---which includes natural
numbers and recursion---may be found in Figure
\ref{fig:cbpv-types}.

\cbpv{} is structured around a fundamental dichotomy between
values and computations. In particular, it comes with two sorts of
types: \emph{value types}, usually denoted by $A$, and
\emph{computation types}, usually denoted by $\ct{B}$.  Value
types often include certain ground types, \emph{positive}
products, and \emph{thunks}. Computation types include \emph{free}
computation types, \emph{negative} product types, and function
types. Like in CBV, variables in \cbpv{} may only denote values,
and are thus assigned value types.

The terms at value types are exactly values: the introduction
rules encode all the possible ways of constructing them. For
example, if $V : A_1$ and $W : A_2$, we have $(V, W) : A_1 \times
A_2$. However, we may \emph{not} project the components of that
pair back into value types. But if we have a term $N : \ct{B}$ at
some computation type with two free variables $x_1 : A_1$ and $x_2
: A_2$, then the elimination rule allows us to split a pair of
values into $N$, yielding $\splitprod{(V, W)}{x_1}{x_2}{N} :
\ct{B}$. It is in this way that the invariant of value types is
maintained.

As their name suggests, the terms at computation type may
demonstrate some non-trivial computational behaviour, including
effects (e.g. printing, output, etc). In the words of Paul Levy,
\emph{``a value is, a computation does.''} The most basic such
type is the \emph{free} (computation) type $FA$, where $A$ is a
value type. A term $M : FA$ is a \emph{returner}, which may engage
in some effectful behaviour before returning some value by
normalizing to the form $\return{V}$ where $V : A$. Given $M : FA$
and as $x : A \vdash N : \ct{B}$ we may form $\bind{x}{M}{N} :
\ct{B}$. This term will first run $M$, effecting some changes and
returning a value which will then be substituted for $x$.
Computation types also include \emph{negative products}, which we
write as $\ct{B}_1 \with \ct{B}_2$, and which come with the usual
projections.  Finally, they include function types: given a
computation $M : \ct{B}$ in a free value variable $x : A$, we may
form $\lambda x.\ M : A \rightarrow \ct{B}$. Notice that $A
\rightarrow \ct{B}$ is a `mixed type,' which mirrors the fact that
variables may only stand for values.

Only one thing remains, and that is to overcome the restriction of
variables to value types. This is achieved by introducing a value
type $U\ct{B}$, the so-called type of \emph{thunks} of $\ct{B}$,
for every computation type $\ct{B}$.  Each computation $M :
\ct{B}$ may be \emph{thunked} as $\thunk{M} : U\ct{B}$, and each
$N : U\ct{B}$ may be \emph{forced} into a computation $\force{M} :
\ct{B}$. Thunk types are crucial in simulating CBN. They are also
necessary in the \cbpv{} fixed point rule, which assumes that the
recursive call is given in the form of a thunk.

Our version of \cbpv{} also comes with `flat' natural numbers
as a value type. As in \pcf{}, these are introduced by an
infinite number of constants $\numl{0}, \numl{1}, \dots$. We may
use natural numbers in either of two ways. First, using the term
$\ifz{\numl{n}}{P}{Q}$ we may query them on whether they are zero,
following the computation $P$ if so, and $Q$ otherwise. Second,
using the construct $\calc{v}{\numl{m} \cbpvop \numl{n}}{N}$ we may
\emph{calculate} one of four numerical operations on them, and
substitute the result into some variable $v : \natt$ that is free
in the `continuation' $N$.

Finally, we equip our version of \cbpv{} with a single effect,
namely the one that incurs a unit cost: for each computation $M :
\ct{B}$ we may form a computation $\charge{M} : \ct{B}$. In
section \ref{section:embed} we will use this effect in our
embedding of CBN and CBV into \cbpv{} to record the places where
the +1's appear in the operational semantics of \pcf{}.

\begin{figure}
  \centering
  \input{cbpv-bigstep}
  \caption{Big-step semantics for CBPV}
  \label{fig:cbpv-bigstep}
\end{figure}

Figure \ref{fig:cbpv-bigstep} introduces big-step semantics for
CBPV. The judgment $\eval{M}{T}{n}$ means that the closed \cbpv{} term
$M$ evaluates to the \emph{terminal computation} $T$, incurring a cost
$n$ in the process. Terminal computations include $\return{V}$, pairs, and
$\lambda$-expressions. Unlike the operational semantics of \pcf{}, costs
are incurred here only when they are explicitly mentioned as effects:
with the exception of the rule for $\charge{M}$, all the other rules sum
the costs of their premises.

\section{Recurrence Extraction for CBPV \& the Bounding Theorem}
  \label{section:extraction-cbpv}

Having introduced our intermediate language, we will now show how
to extract recurrences from its terms. Furthermore, we will prove the
central result of this paper, the \emph{Bounding Theorem}, which
ensures that extracted recurrences express upper bounds for the
running times of \cbpv{} terms.

Compared with the results of \cite{Danner2015}, our results are a
significant technical improvement. First, the distinction between
\emph{potentials} and \emph{complexities} is now formally
motivated: our techniques demonstrate that \emph{values have
potentials}, whereas \emph{computations have complexities}. We
thus get a very clear conceptual basis for higher-order
recurrences. Second, the fact that our version of \cbpv{} comes
with an explicit cost effect makes this result reusable and
extensible: our theorem applies not just to a specific source
language, but to any source language that can be faithfully
embedded in \cbpv{}. All we need to do is encode its terms in
\cbpv{} in a way that explicitly records where evaluation costs
are incurred, and then invoke the bounding theorem. 

\begin{figure}
  \centering
  \input{cbpv-extraction}
  \caption{Recurrence extraction for \cbpv{}}
  \label{fig:cbpv-extraction}
\end{figure}

\subsection{CBPV Recurrence Extraction}

The extraction procedure for \cbpv{} incorporates elements of the
extraction procedures for both CBV and CBN given in
\S\ref{section:extraction-cbx}. It is rather close to the monad/algebra
categorical semantics of \cbpv{} given in \citep[\S 3]{Levy2006} in the
special case of the writer monad ($\cost \times -$).

To begin, we translate \cbpv\ types to \pcfc\ types as follows:
a value type $A$ is translated to a PCF type
$\potential{A}$, its so-called type of \emph{potentials}, and
a computation type $\ct{B}$ is translated to a
\emph{cost algebra} $\complexity{\ct{B}} \defeq
\left(\carrier{\complexity{\ct{B}}}, \alpha_{\ct{B}}\right)$.
Again, the idea is that values only have potential
(future, indirect use-cost), whereas computations may be evaluated
now, hence incurring (immediate, direct) costs. Because of the
interdependence between value and computation types introduced by
$F(-)$ and $U(-)$, the translation---which may be found in Figure
\ref{fig:cbpv-extraction}---is defined in a mutually recursive
manner. For the sake of precision we have maintained the formal
distinction between algebras $\complexity{\ct{B}}$ and their
carriers $\carrier{\complexity{\ct{B}}}$. As before, we abbreviate
$\alpha_{\ct{B}}(L, E)$ by $L \algp{\ct{B}} E$.

Next, we give a translation of \cbpv{} terms to \pcfc{} terms.  In
Figure \ref{fig:cbpv-extraction} we introduce the judgments
$\cbpvtv{\Gamma}{V}{P}{A}$ and $\cbpvtc{\Gamma}{M}{Q}{\ct{B}}$.
The first one is read as ``the \cbpv{} term $V$ of value type $A$
translates to the \pcfc{} term $P$ in context $\Gamma$.'' The
interpretation of the second judgment is similar, but involves
terms of computation type. We have that

\begin{theorem}[\cbpv{}-to-\pcfc{} translation] \hfill
  \begin{enumerate}
    \item 
      If $\Gamma \vturn V : A$ then there is a unique \pcfc{} term
      $P$ such that $\cbpvtv{\Gamma}{V}{P}{A}$, and
      $\potential{\Gamma} \vdash P : \potential{A}$.
    \item 
      If $\Gamma \cturn M : \ct{B}$ then there is a unique \pcfc{}
      term $Q$ such that $\cbpvtc{\Gamma}{M}{Q}{\ct{B}}$, and
      $\potential{\Gamma} \vdash Q :
      \carrier{\complexity{\ct{B}}}$.
  \end{enumerate}
\end{theorem} Hence, whenever $\cbpvtv{\Gamma}{V}{P}{A}$ we know
that this $P$ is unique. We write $\potential{V}$ for it, so that
$\potential{\Gamma} \vdash \potential{V} : \potential{A}$. We also
write $\complexity{M}$ for the unique $Q$ such that
$\cbpvtc{\Gamma}{M}{Q}{\ct{B}}$. In that case we have
$\potential{\Gamma} \vdash \complexity{M} :
\carrier{\complexity{\ct{B}}}$. From this point onwards we again
confuse algebras with their carriers.

The main ideas of this recurrence extraction are: implementing the
effect $\charge{M}$ in CBPV as adding unit cost using the algebra on the
computation type $\ct B$ of $M$, and undoing some of the finer
operational distinctions that are made in CBPV but not in \pcfc. For
example, both eager and lazy products in CBPV are interpreted as
products in \pcfc{}, and both $\mathsf{split}$ and projections in CBPV
are interpreted as projections in \pcfc.  Because of the latter, it may
be surprising that we can obtain a bounding theorem that works for the
CBV features of CBPV.  A precedent for this kind of translation of CBV
into a CBN-like setting is the adequacy theorem of \cite{Plotkin2001}
for CBV \pcf{} with algebraic effects: the main requirement in that
result is a strong monad with a \emph{strict} strength, and the writer
monad $\cost \times -$ is such a monad.  

\subsection{The Bounding Theorem}

\begin{figure}
  \centering
  \input{cbpv-logrel}
  \caption{Bounding relations for CBPV}
  \label{fig:cbpv-logrel}
\end{figure}

We now wish to argue that the extraction procedure for \cbpv{} is
correct via a logical relation that generalizes the CBV and CBN
ones. We define two relations \[
  (V : \mathcal{T}^{\cbpv,\textrm{closed}}_{A}) \vrel{}{}{A} (E
  :\mathcal{T}^{\pcfc,\textrm{closed}}_{\potential{A}})
  \qquad
  (M : \mathcal{T}^{\cbpv,\textrm{closed}}_{\ct{B}})
  \crel{}{}{\ct{B}} (E : \mathcal{T}^{\pcfc,\textrm{closed}}_{\complexity{\ct{B}}})
\]
by induction on types; the full definitions are shown
in Figure \ref{fig:cbpv-logrel}. $\vrel{V}{E}{A}$ is read as ``the
potential of the value $V : A$ is bounded above by $E$,'' and
$\crel{M}{E}{\ct{B}}$ is read as ``the complexity of the computation $M
: \ct{B}$ is bounded above by $E$.''  The value relation is similar to
the CBV value bounding relation, with the clause for $U$-types
corresponding to the switch from value bounding to expression bounding
in function types. The computation relation is similar to the CBN
bounding relation, with the clause for $F$-types corresponding to the
CBV expression relation.

We then establish that this relation holds at the diagonal,
i.e. that \begin{theorem}[Bounding theorem] \hfill
  \label{theorem:cbpv-bounding-theorem}
  \begin{enumerate}
    \item
      If $\cdot \vdash V : A$ then $\vrel{V}{\potential{V}}{A}$.
    \item
      If $\cdot \vdash M : \ct{B}$ then $\crel{M}{\complexity{M}}{\ct{B}}$.
  \end{enumerate}
\end{theorem} 

While technical, the proof of this theorem is unsurprising, and
may be found 
\processifversion{popl}{in an extended version of this paper.}%
\processifversion{arxiv}{in Appendix~\ref{sec:bounding-theorem-proof}.}
Amongst other
things, it requires an auxiliary notion of \emph{costless weak
head reduction}. It is denoted by $M \wh{0} N$, and encodes all
weak head reductions, plus a `commuting conversion' that makes
negative eliminators commute with the unit cost construct, e.g.
$\pi_i(\charge{M}) \wh{0} \charge{\pi_i(M)}$. The main lemmas used are: 

\begin{lemma}[Algebra Monotonicity]
  \label{lemma:algebra-monotone}
    $E \costleq E' : \cost
      \Longrightarrow
    E \algp{\ct{B}} M \costleq E' \algp{\ct{B}} M$
\end{lemma}

\begin{lemma}[Bound weakening] \hfill
  \label{lemma:bound}
  \begin{enumerate}
    \item
      $
      \vrel{V}{E}{A}\ \land\ E \costleq E' : \potential{A}\
        \Longrightarrow\
      \vrel{V}{E'}{A}
      $
    \item
      $
      \crel{M}{E}{\ct{B}}\ \land\ E \costleq E' : \complexity{\ct{B}}\
        \Longrightarrow\
      \crel{M}{E'}{\ct{B}}
      $
  \end{enumerate}
\end{lemma}

\begin{lemma}[Head expansion/reduction] \hfill
  \label{lemma:headexp}
\begin{enumerate}
\item
  $
    \crel{M'}{E}{\ct{B}}\
      \land\
    M \wh{0} M'
      \Longrightarrow
    \crel{M}{E}{\ct{B}}
  $

  \item
      $
    \crel{M}{E}{\ct{B}}\
      \land\
    M \wh{0} M'
      \Longrightarrow
    \crel{M'}{E}{\ct{B}}
  $
\end{enumerate}
\end{lemma}

\begin{lemma}[Unit charge]
  \label{lemma:unit}
  $
    \crel{M}{E}{\ct{B}}\
      \Longrightarrow
    \crel{\charge{M}}{\unitc +_{\ct{B}} E}{\ct{B}}
  $
\end{lemma}

\begin{theorem}[Compactness]
  \label{theorem:compactness}
  $
  \bounded{\ectx{E}[\fix{x}{M}]}
    \Longrightarrow
  \exists n \geq 0. \ \bounded{\ectx{E}[\fixn{x}{n}{M}]}
  $
\end{theorem}

\begin{lemma}[Bounds for Recursion] \hfill
  \label{lemma:recbounds}
  \begin{enumerate}
    \item (Infinity)
      $\unbounded{E} \Longrightarrow \crel{M}{E}{\ct{B}}$.
    \item (Infinity-Algebra)
      $\unbounded{L} \Longrightarrow \crel{M}{L \algp{\ct{B}}
      E}{\ct{B}}$.
    \item (Fixed Point Induction)
      $\left(\forall n \geq 0.\
            \crel{M}{\fixn{x}{n}{E}}{\ct{B}}\right)
        \Longrightarrow
          \crel{M}{\fix{x}{E}}{\ct{B}}$
  \end{enumerate}
\end{lemma}

Compactness is known to hold for \pcf{}. It can be obtained
through a standard denotational semantics---see e.g. \citet[\S
10.9]{Pitts1994}---or through syntactic methods---see \cite[\S
4]{Pitts1997}.
\begin{arxiv}
Proofs of these lemmas can be found in Appendix~\ref{sec:bounding-relation-proofs}.
\end{arxiv}

\section{Embedding PCF into CBPV}
  \label{section:embed}

The only piece of the puzzle that remains is to show how both CBN and
CBV \pcf{} can be embedded in \cbpv{} in a cost-preserving way. At a
high level, we follow \cite[\S 2.7]{Levy2003} in defining a term
translation, but insert $\mathsf{charge}$ expressions wherever the
operational semantics of CBV or CBN PCF incurs a cost. We then prove
that this translation preserves and reflects the operational semantics,
including the cost.

\subsection{CBN}

In order to embed CBN into \cbpv{}, we must recall that
\begin{enumerate}
  \item 
    the only \emph{observable} type of CBN \pcf{} is $\natt$, and
  \item 
    the variables of CBN terms represent \emph{thunks}, as we may
    substitute them with arbitrary terms.
\end{enumerate} These two facts directly lead us to a translation
of any CBN type $A$ to a \cbpv{} \emph{computation} type
$\cbnt{A}$. As $\natt$ is observable, it is mapped to
$F\left(\natt\right)$ (it returns a value). Products are
compositionally mapped to \emph{negative} products. Finally, as
variables stand for thunks, we must thunk the domain when
translating the function type.

\begin{figure}
  \centering
  \input{cbn-embedding}
  \caption{Call-by-name \pcf{} to \cbpv{} translation}
  \label{fig:cbn-embedding}
\end{figure}

The translation of both types and terms is defined in Figure
\ref{fig:cbn-embedding}. Notice that computation steps where
explicit evaluation to canonical form is necessary---e.g. for
testing whether a number is zero---are punctuated by the
appearance of the $\bind{x}{(-)}{(-)}$ construct. Furthermore,
notice that a variable represents a thunk, so it must be forced.
Finally, notice the appearance of $\charge{(-)}$ whenever the
operational semantics of \pcf{} would impose a unit charge.

Extending $\cbnt{(-)}$ to contexts pointwise, we have that

\begin{proposition}
  If $\Gamma \vdash M : A$ in call-by-name \pcf{} then
  $\cbnt{\Gamma} \cturn \cbnt{M} : \cbnt{A}$.
\end{proposition}

To prove that the operational semantics are preserved and
reflected under this translation we follow the technique of
\cite[\S 7]{Levy2006}. We inductively define a relation
${\cbnbisim}$ between terms of \pcf{} and terms of $\cbpv{}$.  The
relation includes one rule for each line of the table of Figure
\ref{fig:cbn-embedding}, e.g. \[
  \begin{prooftree}
      M \cbnbisim M'
    \quad
      N \cbnbisim N'
    \justifies
      M\,N \cbnbisim M'(\thunk{N'})
  \end{prooftree}
\] Moreover, the definition also includes the rule \[
  \begin{prooftree}
      M \cbnbisim M'
    \justifies
      M \cbnbisim \force{\thunk{M'}}
  \end{prooftree}
\] which---for technical reasons---inserts enough occurrences of
$\force{\thunk{(-)}}$. Evidently, \begin{proposition}
  \label{prop:bisim-dag}
  $M \cbnbisim M^\dag$
\end{proposition} 

\noindent We can then show that

\begin{theorem}[CBN bisimulation] \hfill
  \begin{enumerate}
    \item
      If $\evalpcf{M}{V}{n}$ and $M \cbnbisim M'$, then
      $\eval{M'}{T'}{n}$ for some $T'$ with $V \cbnbisim T'$.
    \item
      If $\eval{M'}{T'}{n}$ and $M \cbnbisim M'$, then
      $\evalpcf{M}{V}{n}$ for some $V$ with $V \cbnbisim T'$.
  \end{enumerate}
\end{theorem}
\begin{proof}
  We first show that $M \cbnbisim M'$ and $N \cbnbisim N'$ implies
  $M[N/x] \cbnbisim M'[\thunk{N'}/x]$, by an easy induction on $M
  \cbnbisim M'$. We then show the main claims. In the first case,
  we proceed by induction on $\evalpcf{M}{V}{n}$, followed by
  local inductions on $M \cbnbisim M'$. In the second case, we
  proceed by induction on $\eval{M'}{T'}{n}$, followed by local
  inductions on $M \cbnbisim M'$.
\end{proof}

\begin{corollary}[CBN preservation] \hfill
  \label{cor:cbnpres}
  \begin{enumerate}
    \item
      If $\evalpcf{M}{V}{n}$ then $\eval{M^\dag}{T}{n}$ with $V
      \cbnbisim T$.
    \item
      If $\eval{M^\dag}{T}{n}$ then $\eval{M}{V}{n}$ with $V
      \cbnbisim T$.
  \end{enumerate}
\end{corollary}

\subsection{CBV}

In order to embed CBV into \cbpv{}, we must recall that
\begin{enumerate}
  \item 
    in CBV, termination at every type is observable, and returns a
    \emph{value}
  \item 
    the variables of CBV terms represent \emph{values}
\end{enumerate}
In short: \emph{all terms of CBV are returners}.
Thus, we aim to translate each type CBV type $A$ to a \cbpv{}
value type $\cbvt{A}$, and each term $x : A \vdash M : B$ to a
term $x : \cbvt{A} \cturn \cbvt{M} : F\left(\cbvt{B}\right)$ of
\cbpv{}.

\begin{figure}
  \centering
  \input{cbv-embedding}
  \caption{Call-by-value \pcf{} to \cbpv{} translation}
  \label{fig:cbv-embedding}
\end{figure}

Translating types is then fairly evident: $\natt$ is translated to
itself, and products are compositionally mapped to \emph{positive}
products. Finally, $A \rightarrow B$ is mapped to $\cbvt{A}
\rightarrow F\left(\cbvt{B}\right)$, which is then thunked in
order to become a value type. 

On the level of terms there are two translations: one takes CBV
\pcf{} terms to \cbpv{} terms of $F(-)$ type, and the other one
takes CBV \pcf{} values to \cbpv{} values. They are both defined
in Figure \ref{fig:cbv-embedding}.  Notice once again that
computation steps where explicit evaluation to canonical form is
necessary---e.g. for function application, or construction of
pairs---necessitate a use of the $\bind{x}{(-)}{(-)}$ construct,
which forces evaluation. Furthermore, notice that a variable
represents a value, so it must be returned.  Finally, notice the
appearance of $\charge{(-)}$ whenever the operational semantics of
\pcf{} would impose a unit charge.

Extending $\cbvt{(-)}$ to contexts pointwise, we have that

\begin{proposition} \hfill
  \begin{enumerate}
    \item
      If $\Gamma \vdash V : A$ is a \pcf{} value, then
      $\cbvt{\Gamma} \vturn \val{V} : \cbvt{A}$.
    \item
      If $\Gamma \vdash M : A$ in call-by-value \pcf{}, then
      $\cbvt{\Gamma} \cturn \cbvt{M} : F\left(\cbvt{A}\right)$.
  \end{enumerate}
\end{proposition}

It is even easier than CBN to establish that the operational
semantics of CBV are preserved and reflected under this
translation.

\begin{theorem}[CBV preservation] \hfill
  \begin{enumerate}
    \item
      If $\evalpcf{M}{V}{n}$ then $\eval{\cbvt{M}}{\val{V}}{n}$.
    \item
      If $\eval{\cbvt{M}}{\return{W}}{n}$ then there exists $V$
      such that $\evalpcf{M}{V}{n}$ and $W \aequiv \val{V}$.
  \end{enumerate}
\end{theorem}
\begin{proof}
  First, we need to establish two lemmas: \begin{itemize}
    \item[-]
      $\cbvt{V} \aequiv \return{\val{V}}$ for a call-by-value
      \pcf{} value $V$
    \item[-]
      $\cbvt{(M[V/x])} \aequiv \cbvt{M}[\val{V}/x]$
  \end{itemize} These follow straightforwardly by induction on
  values and terms respectively. The first claim then follows by
  induction on $\evalpcf{M}{V}{n}$, and the second by induction on
  $\eval{\cbvt{M}}{\return{W}}{n}$.
\end{proof}

\subsection{Completing the Circle}

Combining the bounding theorem of Section
\ref{section:extraction-cbpv} and these results, we can prove the
main theorems of \S\ref{section:extraction-cbx}. For Theorem
\ref{theorem:cbv-extraction} we will make use of head reduction,
viz. Lemma \ref{lemma:headexp}. We then define \begin{align*}
  M \valrel_A E
    &\defiff
  \vrel{\val{V}}{E}{\cbvt{A}}
    &
  M \logrel_A E
    &\defiff
  \crel{\cbvt{M}}{E}{F\left(\cbvt{A}\right)}
\end{align*} 
and calculate that the desired conditions hold. We then do the
same for Theorem \ref{theorem:cbn-extraction} and \[
  M \logrel_A E
    \quad\defiff\quad
  \crel{\cbnt{M}}{E}{\cbnt{A}}
\]


\section{Denotational Semantics for PCF with Costs}
  \label{section:sized-domains}

This section is concerned with the development of a denotational
semantics for \pcfc{}. Having completed the step of recurrence
extraction, we would now like to interpret our \emph{syntactic
recurrences}---expressed in the formal language of \pcf{}---into
\emph{semantic recurrences}. These should be as close as possible
to good old-fashioned recurrences found in textbooks on
algorithms.

Because of the presence of recursion, this endeavour requires a
modicum of \emph{domain theory}. Standard references to domain
theory include \cite{Abramsky1994, Stoltenberg-Hansen1994},
whereas references on the denotational semantics of \pcf{} in
domains may be found in \S\ref{section:source}. We assume knowledge
of the standard definition of \emph{complete partial order (cpo)},
i.e.  a partial order $(D, \sqsubseteq)$ with least upper bounds
of directed subsets, and of Scott-continuous functions. We also
recall the notion of an $\omega$-chain $\left(x_i\right)_{i \in
\omega}$ in a cpo, i.e. an increasing sequence of elements $x_1
\sqsubseteq x_2 \sqsubseteq \dots$.

\subsection{Sized Domains}

The relation usually modelled by the denotational semantics of
\pcf{} is that of \emph{equality}: the semantics of \pcf{} terms
is \emph{sound}, in that it is stable under evaluation. However,
our main notion here is that of the axiomatic size order, and in
order to model it we introduce the following kind of domain.

\begin{definition}
  \label{definition:sized-domain}
  A \emph{sized domain} (with joins) $(D, \sqsubseteq, \semleq,
  \lor, \bot, \zeroc, \unitc)$ consists of \begin{itemize}
    \item[-]
      a set $D$,
    \item[-]
      a partial order $\sqsubseteq$, which we call the \emph{information
      order on $D$}, and
    \item[-]
      a preorder $\semleq$, which we call the \emph{size order on $D$}, and
    \item[-]
      an operation $\lor : D \times D \rightarrow D$
  \end{itemize} such that \begin{enumerate}
    \item
      $(D, \sqsubseteq)$ is a cpo with least element $\bot$,
    \item
      $(D, \semleq)$ has chosen binary \emph{joins} (maximums/least upper bounds) given by $\lor
      : D \times D \rightarrow D$ and least element $\zeroc$,
    \item
      $\lor : D \times D \rightarrow D$ is continuous with respect
      to $(D, \sqsubseteq)$,
    \item
      $x \sqsubseteq y$ implies $y \semleq x$, and
    \item
      if $(x_i)_{i \in \omega}$ is a $\omega$-chain in $D$
      and $\forall i \in \omega.\ z \semleq x_i$, then 
      $z \semleq \bigsqcup_{i \in \omega} x_i$.
  \end{enumerate}
\end{definition}

\noindent We immediately know that 

\begin{proposition}
  \label{proposition:bot-is-greatest}
  $\bot$ is the greatest element (up to iso) with respect to $(D,
  \semleq)$.
\end{proposition}
\begin{proof}
  For any $x \in D$ we have that $\bot \sqsubseteq x$, and hence
  $x \semleq \bot$ by axiom (4). Thus, $\bot$ is a greatest
  element in $(D, \semleq)$. However, as $(D, \semleq)$ is not a
  partial order, it is only unique up to isomorphism.
\end{proof}

The most interesting axioms in this definition are (4) and (5).
The first one formalizes the idea that \emph{a more defined bound
is a better bound}. Consider the function $f = \{ 1 \mapsto 2, x
\mapsto \bot \}$ on the flat domain of natural numbers. When
interpreted 
as a recurrence, $f$ gives us an upper bound for the input $1$,
but no information about any other input. In fact, Proposition
\ref{proposition:bot-is-greatest} tells us that $\bot$ is the
greatest element in the size order: it represents the
\emph{infinite value}. Thus, the recurrence $f$ maps
all elements other than $1$ to the trivial upper bound of
infinity. The recurrence $g = \{ 1 \mapsto 2, 2 \mapsto 2, x
\mapsto \bot \}$ is more well-defined than $f$, as it is also
defined (i.e. not $\bot$) at input $2$: we have $f \sqsubseteq g$.
As $\bot$ represents infinity, this is indeed a \emph{tighter}
bound, and axiom (4) stipulates that we should be able to treat it
as such: we should have $g \semleq f$. Axiom (4) is closely
related to rule (\hyperlink{rat}{\textsf{rat}}) of the axiomatic
size order of Figure \ref{fig:pcf-inequality}, and is indeed used
to verify its soundness.

Axiom (5) expresses the idea that \emph{a recursively defined
recurrence is an upper bound if all its approximants are}. Recall
that fixed points in \pcf{} are interpreted as the least upper
bound of a chain $\left(f^i(\bot)\right)_{i \in \omega}$ that is
increasing in the information order $\sqsubseteq$. Intuitively,
each element of this chain is more well-defined than its
predecessor.  Axiom (5) requires that if every extra step of
definition $f^i(\bot)$ of a recurrence is indeed an upper bound
for an element, then so must be the limit $\bigsqcup_{i \in
\omega} x_i$.  Axiom (5) is very close to the rule
(\hyperlink{cpind}{\textsf{cpind}}), and is used to verify its
soundness.

We define the category $\mathbf{SizeDom}$ to consist of sized domains
and Scott-continuous functions. The morphisms need not preserve any
aspect of the size order; for example, we do not requiring \emph{size
  monotonicity} in general, because only the \pcfc~monotone contexts
$\mctx{C}$ are required to be interpreted as size-monotonic
functions. It is also easy to see that
this category is cartesian closed, as it inherits products and
exponentials from the underlying category $\mathbf{Cpo}$ of complete
partial orders and Scott-continuous functions: we only need to define
the size order and the chosen joins---which are in both cases defined
pointwise---and show that the axioms of Definition
\ref{definition:sized-domain} follow.

\subsection{Interpreting PCF with Costs into Sized Domains}

Defining a semantic interpretation of \pcfc{} into sized domains
is reasonably straightforward. To each type $A$ of PCF we
associate a sized domain $\sem{A}{} = (D_A, \sqsubseteq_{A},
\semleq_A, \lor_A, \bot_A, 0_A, 1_A)$. We interpret both the
natural numbers $\natt$ and the type $\cost$ of costs by the
\emph{flat sized domain of natural numbers}, which is defined to
be $ \mathbb{N}_\bot \defeq (\mathbb{N} \cup \{\bot\},
\sqsubseteq, \semleq, \lor, \bot, 0, 1)$ where $\bot \not\in
\mathbb{N}$. Its underlying set consists of the natural numbers
augmented with a `fresh' element $\bot$, which simultaneously
models non-termination and infinity. Writing $\leq_\mathbb{N}$,
for the usual order on natural numbers, we define 
\[
  x \sqsubseteq y
    \defiff
      x = \bot \text{ or } x = y, \qquad
  x \semleq y
    \defiff
      y = \bot \text{ or } x \leq_\mathbb{N} y, \qquad
  x \lor y
    \defeq
        \begin{cases}
          \bot         \text{ if } x = \bot \text{ or } y = \bot \\
          \maxn{x}{y}  \text{ otherwise }
        \end{cases} \\
\]
It is easy to see that the requisite axioms are
satisfied. The rest of the types are interpreted by
\[
  \sem{A_1 \times A_2}{} \defeq \sem{A_1}{} \times \sem{A_2}{} 
  \qquad
  \sem{A \rightarrow B}{} \defeq
    \left[\sem{A}{} \rightarrow \sem{B}{}\right]
\]
where $\left[X \rightarrow Y\right]$ denotes the set of
Scott-continuous functions from a cpo $(X, \sqsubseteq_X)$ to a
cpo $(Y, \sqsubseteq_Y)$, and where both the information and size
orders are given pointwise.

To each judgment $\Gamma \vdash M : A$, where $\Gamma \equiv x_1 : A_1, \dots, x_n : A_n$,  we associate a function
$\sem{\Gamma \vdash M : A}{} : \sem{\Gamma}{} \rightarrow \sem{A}{}$ where
$\sem{\Gamma}{} \defeq \sem{A_1}{} \times \dots \times \sem{A_n}{}$. This definition is mostly standard---see e.g.
\citep[\S 3.2]{Streicher2006}. However, there are two unexpected
elements, which arise because the $\pcfc$ monotone contexts $\mctx{C}$ must be
interpreted as 
size-monotone functions in order to interpret the syntactic size order axiom
$(\hyperlink{ctx}{\textsf{ctx}})$, and because we interpret $\natt$
with the usual $0 \le 1 \le 2 \ldots \le \infty$ size order
to facilitate reasoning about semantic recurrences.  
First, because $\pif{\mctx{C}}{P}{Q}$ is a monotone context,
we must ``monotonize'' the
conditional.
We do this by sending $\infty$ to $\infty$, and using the chosen joins to make the
interpretation for $\ge 1$ dominate\footnote{The standard definition of the conditional
along with $\num{0} \costleq \num{1}$ and
(\hyperlink{ctx}{\textsf{ctx}}) would imply $\sem{P} =
\sem{\pif{\num{0}}{P}{Q}} \costleq \sem{\pif{\num{1}}{P}{Q}} =
\sem{Q}$ for arbitrary terms $P$ and $Q$.} the answer for $0$:
\[
  \sem{\Gamma \vdash \pif{N}{P}{Q} : A}(\vec{d}) 
    \defeq 
      \begin{cases}
        \bot_A & \text{ if } \sem{\Gamma \vdash N : \natt}(\vec{d}) = \bot \\
        \sem{P}(\vec{d}) & \text{ if } \sem{\Gamma \vdash N : \natt}(\vec{d}) = 0 \\
        \sem{P}(\vec{d}) \lor_A \sem{Q}(\vec{d}) & \text{ otherwise}
      \end{cases}
      \]
Second, because the operands of an arithmetic operation are monotone contexts,
we must interpret the operations by a monotone 
upper bound of their actual numerical value.
While addition and multiplication are already monotone, for
subtraction, division and modulo, we use the following definitions, 
where $\pcfopa \in \{-, \div\}$:\footnote{
  We could in fact remove
  $\{-,\div\}$ with non-numerals and $\bmod$ from \pcfc, because
  they are never used as the target of 
  the recurrence extraction in Section \ref{section:extraction-cbpv}.
  In general, there is a choice about whether to perform this monotonization in recurrence extraction or in the semantic interpretation.
  The above recurrence extraction from \cbpv\/ to \pcfc\/ already does monotonization,
  eliminating all mods and $\{-,\div\}$ with non-numerals.
  However, we prefer to leave these operations in \pcfc\/ and interpret them similarly here
  to also support the other style of recurrence extraction into \pcfc.  
}
\begin{align*}
  \sem{\Gamma \vdash M \pcfopa \num{n} : \natt}(\vec{d})
    &\defeq
  \sem{\Gamma \vdash M : \natt}(\vec{d}) \pcfopa n\\
  \sem{\Gamma \vdash M \pcfopa N : \natt}(\vec{d})
    &\defeq
  \sem{\Gamma \vdash M : \natt}(\vec{d})\\
  \sem{\Gamma \vdash M \bmod N : \natt}(\vec{d})
    &\defeq
  \sem{\Gamma \vdash N : \natt}(\vec{d})-1
\end{align*} The `new' clauses interpret $\sem{\zeroc}(\vec{d})
\defeq 0$, $\sem{\unitc}(\vec{d}) \defeq 1$, and \[
  \sem{\Gamma \vdash M \plusc N : \cost}(\vec{d})
    \defeq 
    \begin{cases}
      \bot & \text{ if } \sem{M}{}(\vec{d}) = \bot
             \text{ or } \sem{N}{}(\vec{d}) = \bot \\
      \sem{M}{}(\vec{d}) +_\mathbb{N} \sem{N}{}(\vec{d}) & \text{ otherwise}
    \end{cases}
\] We have that
\begin{theorem}[Soundness]
  $
  \Gamma \vdash M \costleq N : A\
    \Longrightarrow\
  \forall \vec{d} \in \sem{\Gamma}{}.\
  \sem{M}{}(\vec{d}) \semleq_A \sem{N}{}(\vec{d})
  $
\end{theorem}

\noindent Moreover, we can use standard logical relations
techniques to show that this interpretation is \emph{adequate} in
an appropriately relaxed sense.

\begin{theorem}[Adequacy] \hfill
  \label{theorem:pcf-adequacy}
  \begin{enumerate}
    \item
      If $M : \natt$, then $\sem{M}{} = m$ implies
      $\evalpcf{M}{\num{k}}{}$ for some $k \leq_\mathbb{N} m$.
    \item
      If $M : \cost$, then $\sem{M}{} = m$ implies
      $\evalpcf{M}{\numc{k}}{}$ for some $k \leq_\mathbb{N} m$.
  \end{enumerate}
\end{theorem}

We anticipate the existence of other models of $\pcfc$ in sized
domains, which we expect to obtain by varying the orders on the
interpretation of $\natt$, or even $\times$ and $\to$.  For
example, there should be a more precise discrete model, where all
size orders are interpreted by equality, all cost bounds
are exact, and the potentials include the programs themselves.

\section{Example: Exponentiation}
  \label{section:exponentiation-example}

\begin{figure}
\input{exp}
\caption{Divide-and-conquer exponentiation and its extracted recurrence}
\label{fig:d-c-exp}
\end{figure}

Next, we show how the above technique works on an example of a
non-structurally-recursive function, which is not supported by
\citet{Danner2015}.  
Consider the recursive specification for ``exponentiation-by-squaring'':
$$
	2^n =
	\begin{cases}
		(2^{n/2})^2 & \text{if $n$ is even} \\
		(2^{(n-1)/2})^2 * 2 & \text{otherwise}
	\end{cases}
$$
We implement this in CBV \pcf\ in Figure~\ref{fig:d-c-exp}, using
$\textsf{let}$ expressions as syntactic sugar for $\beta$-expansions
and recalling that we are performing floor division so that $n / 2 = (n-1) / 2$ when $n$ is odd.
We extract a recurrence in \pcfc\
using the rules of Figure \ref{fig:cbv-extraction} and show the result in the
same figure.  In
order to keep the term from becoming unwieldy, we assume a few
additional size order axioms on \pcfc{} terms beyond
those of Figure \ref{fig:pcf-inequality}, all of which are true in
the semantics of sized domains (we write equality for the size
order $\costleq$ in both directions):
\begin{itemize}
	\item[-]
    the opposite direction of rules
    (\hyperlink{assoc}{\textsf{assoc}}) and
    (\hyperlink{zero}{\textsf{zero}}), so that $(\plusc, \zeroc)$
    forms a monoid,
	\item[-]
    the opposite direction of the $\beta$ rules, as well as $\eta$
    equality rules for $\times$ and $\rightarrow$ types,
	\item[-]
    the `commuting conversion' $\pi_1(\pif{N}{M_1}{M_2}) =
    \pif{N}{\pi_1(M_1)}{\pi_1(M_2)}$, and
  \item[-]
    $\pif{\num{k}}{N}{N} = N$ 
\end{itemize}


It is $\complexity{\prog{exp}}_p$ that is the desired recurrence, and we observe that
if $T(n) = \pi_1(\potentialof{\complexity{\prog{exp}}}\,n)$, then
\begin{align*}
  T(0) &= 0 & T(n) &= 3 + T(n/2)
\end{align*} 
which is the expected recurrence from an informal analysis.

When $n$ is even, the potential is an over-approximation of the actual
value: we get $2 * (\potentialof{(\prog{exp}(n/2))})^2$ whether $n$ is even or odd because
the recurrence extraction sends $n \bmod 2$ to $1$ in order to be monotone.
We leave a more precise treatment of non-monotone operations to future work.

\section{Inductive Types}
  \label{section:inductive}

Our previous work~\citep{Danner2015} only permitted call-by-value
functional programs with strictly positive inductive datatypes
and structural recursion. The present paper extends these
techniques to general non-structural recursion, which often leads
to significantly more concise and efficient code. It is thus
interesting to examine the combination of this presentation with
the datatypes supported by the previous techniques of
\citet{Danner2015}. We believe that this combination is
achievable. In this section we illustrate it by a careful
treatment of the case of lists in CBV \pcf{}, and we briefly
discuss the general case. In future work we plan to extend our
techniques to \emph{general recursive types}, which subsume
lazy/CBN \pcf{} lists as well as other coinductive types, though
we expect that defining the bounding relation will be somewhat
more challenging.

\begin{figure}
  \input{ind-types}
  \caption{Extensions for handling call-by-value lists}
  \label{fig:ind-types}
\end{figure}

Figure \ref{fig:ind-types} introduces the necessary additions to the
syntax of CBV \pcf{} and \cbpv{}, and also extends the cost-preserving
embedding, recurrence extraction and bounding relation to the new
constructs. Intuitively, we choose here to represent a list by its
length. This is useful for analyzing many algorithms whose running time
does not depend on the list elements; if the cost does depend on list
elements, one should instead represent a list by its length and maximum
element, or some other more precise information.  We could perform this
abstraction of lists as lengths in the semantics, as in
\citet{Danner2015}, but this would introduce quite a bit of notational
overhead.  For the sake of simplicity, we instead translate a list
to its length in the recurrence extraction phase, defining the type of
potentials of lists to be $\natt$.  This way we do not need to add lists
to $\pcfc$.

The only essentially non-trivial clause is the recurrence
extraction for $\cbpvlcase V {M_{\cbpvnil}} x {xs}
{M_{\cbpvcons}}$. Intuitively, this should map to a \pcfc{} zero
test $\pif{W}{P}{Q}$, where $V$, $M_\cbpvnil$, and $M_\cbpvcons$
extract to $W$, $P$, and $Q$ respectively. However, $Q$ has two
free variables, $x$ and $xs$, which must be substituted for. These
correspond to the potentials of the head and tail of $V$. The
answer for $xs$ is immediate, as its potential is bounded above by
$W - \num{1}$. As we have taken the potential of a list to be its
length, we do not have any information about the size of the list
elements. Hence, we choose the potential of $x$ to be $\infty$
($\fix x x$). This gives an upper bound in all cases, and is
useful when the cost of the function to be analyzed does not
depend on the size of the list elements: in that case the $\infty$
substituted for~$x$ will drop out of the recurrence at some point,
so we will still obtain a finite bound.  To complete the
development, we extend the proof of the bounding theorem by adding
a straightforward case for each new term construct. To do so we
need the size order axioms $M \costleq (M + \num 1) - \num 1$,
$\num 0 \costleq \num 1$ and $\fix x x \costleq \ectx{E}[\fix x
x]$ for eliminative contexts $\ectx{E}$, all of which are valid in
the standard semantics.

Though we have not written out all of the details, we expect that we
can extend the above to allow potentials other than list
length (e.g. length and maximum element), and to all strictly positive inductive types.  Briefly, we need
to add the corresponding inductive types to \pcfc{}, add new size order
axioms, extract case expressions as case expressions, and then modify
the notion of sized domains to provide an interpretation of lists and
other inductive types, as in \citet{Danner2015}.  By doing so, the
syntactic recurrence becomes a cost-annotated version of the original
\cbpv\ program that preserves a maximal amount of
information about the inductive type values in the original program.
Different sized domains can then be used to give different abstractions
of size for inductive types, e.g. length for lists, height for
(labelled) trees, or more complicated metrics, such as one that records
both the size of the tree along with the maximum label.  These different
abstractions in turn allow for more or less detailed cost analyses of
the original programs.

\begin{figure}
  \input{merge-sort}
  \caption{The merge sort function in CBV \pcf\ and its extracted recurrence}
  \label{fig:merge-sort}
\end{figure}

\subsection{Merge Sort Example}

We end with an example, viz. recurrence extraction for merge sort
in CBV \pcf{}. We assume the existence of two functions
$\prog{divide}: \pcflist A \to \pcflist A \times \pcflist A$ and
$\prog{merge}: \pcflist A \times \pcflist A \to \pcflist A$, which
perform the usual tasks of splitting a list into two (nearly)
equal-sized pieces and merging two sorted lists into a single
sorted list.  The translation into \cbpv{} is extremely verbose,
so we directly give the extracted recurrence in Figure
\ref{fig:merge-sort}. Corresponding to $\prog{divide}$ and
$\prog{merge}$ are \cbpv{} programs $\prog{divide} : U(\cbpvlist
\cbvt{A} \to F(\cbpvlist \cbvt{A} \times \cbpvlist \cbvt{A}))$ and
$\prog{merge} : U(\cbpvlist \cbvt{A} \times \cbpvlist \cbvt{A} \to
F(\cbpvlist \cbvt{A}))$, and thus \pcfc{} recurrences
$\prog{divide} : \natt \rightarrow \cost \times \natt \times
\natt$ and $\prog{merge} : \natt\times\natt \to \cost \times
\natt$. The recurrence for $\prog{divide}$ expresses the cost of
the CBV \pcf{} $\prog{divide}$ and the lengths of the two returned
lists in terms of the length of its input. The recurrence
extraction is, as with exponentiation, a tedious unwinding of
definitions which uses a few additional size order axioms that are
valid in the standard semantics of \S\ref{section:sized-domains}.

We can then analyze the semantic recurrence---i.e. the denotation
of $\complexity{\prog{sort}}$---and the two projections for
potential and cost. For this discussion, we will write syntactic
expressions but manipulate them as though they are the
corresponding denotations. If $S(n) =
\pi_2(\complexity{\prog{sort}}_p n)$ and $T(n) =
\pi_1(\complexity{\prog{sort}}_c n)$, then the Bounding Theorem
tells us that $S(n)$ and $T(n)$ are upper bounds on the length and
cost of $\prog{sort}(xs)$ respectively, when $xs$ has length~$n$.
We will assume that \[
  \prog{divide}\ k = \tuple{ck}{\ceil{k/2}, \floor{k/2}}
    \qquad
  \prog{merge}\ \tuple k \ell = \tuple{d(k+l)}{k+l}
\] for some fixed constants $c$ and~$d$. This is one way to
formalize the assumption that the complexity of $\prog{sort}$ does
not depend on the potentials of the actual list elements:
while $\prog{merge}$ depends on a comparison function, 
we are asserting that its cost and the length of the result depend
only on the lengths of its arguments.  Of course, this is exactly
what we typically do for an informal analysis of merge sort
on lists of constant-size elements (e.g. machine words), or on lists of arbitrary
data under the assumption that the comparison function takes constant time.
Moving the abstraction of lists to numbers from the recurrence
extraction to the denotational semantics
would permit an analysis that takes into account
the complexity of the comparison function.  

We can now write out a more `traditional' recurrence for $S$: 
\begin{align*}
S(0) &= 0 & S(1) &= 1
  &S(n) &= (E((\prog{divide}(n))_p))_p = (E(\tuple{\ceil{n/2}}{\floor{n/2}}))_p \\
       &&&&&= (\prog{merge}\ \tuple{S(\ceil{n/2})}{S(\floor{n/2})})_p \\
       &&&&&= S(\ceil{n/2}) + S(\floor{n/2})
\end{align*}
The standard proof tells us that $S(n) = n$.  Next we
write out a more traditional recurrence for~$T(n)$ when $n$ is a
power of~$2$, which mimics the usual approach for analyzing such
algorithms: 
\begin{align*}
  T(1) &= 0 
  &T(n) &= (2 + (\prog{divide}(n))_c) + (E((\prog{divide}(n))_p)_c \\
       &&&= (2 + cn) + (E(\tuple{n/2}{n/2}))_c \\
       &&&= (2 + cn) + (5 + T(n/2) + T(n/2)) + (\prog{merge}\ \tuple{S(n/2)}{S(n/2)})_c \\
       &&&= 7 + cn + 2T(n/2) + (\prog{merge}\ \tuple{n/2}{n/2})_c \\
       &&&= 7 + (c+d)n + 2T(n/2).
\end{align*} 
This is precisely the recurrence we expect.
To proceed further with this example, we could define appropriate $\prog{divide}$ and
$\prog{merge}$ functions and prove that the extracted recurrences
satisfy the aforementioned assumptions.

\section{Related Work}
  \label{section:related-work}

There is a long history of techniques for extracting cost information from
programs, probably starting with \citet{wegbreit:cacm75}, who computes closed
bounds for Lisp programs.  Much of the work has been done for imperative
languages, as exemplified by the COSTA project for Java bytecode analysis
\citep{albert-et-al:tcs12:cost-analysis,albert-et-al:tocl13:inference} and
SACO for parallel cost \citep{albert-et-al:tocl18}.  We cannot hope to review
the entire field here, so concentrate on recent work on
cost analysis for functional programs.

The idea that the potential of a function is itself a function is taken from
\citet{Danner2015}, who in turn adapt it from
\citet{danner-royer:ats-lmcs}.
\citet{danielsson:popl08} focuses on amortized cost of lazy programs and makes
use of what is essentially the writer monad~$\cost\times-$, though he requires
the user to explicitly annotate programs with ticks ($\textsf{charge}$s).
The Resource Aware ML (RAML) project has achieved great results in automating
the cost analysis of higher-order functional programs using automatic
amortized resource analysis (AARA).  
This approach reduces the establishment of cost
bounds on a program to type inference in a resource-aware type system, and
that in turn is reduced to a linear programming problem.  It has been used to
verify the cost of significant portions of the OCaml libraries 
\citep{hoffman-et-al:popl17} and for
analyzing space usage in the presence of garbage collection
\citep{niu-hoffman:lpar18}, to name just a
couple of recent achievements.
However, the version of RAML that is current as of the time of this writing is
restricted to deriving polynomial bounds, and hence it
derives a quadratic bound for merge-sort; we simply
derive a recurrence, and if the user of the system can prove a tighter bound
on it than quadratic (e.g., by using the Master Theorem), more power to her.
Another type-based approach is described by
\citet{avanzini-dal-lago-popl17}, where a form of size types with index
polymorphism is used.  As with AARA, cost analysis comes down to type
inference.  This work also makes use of ticks ($\textsf{charge}$s); an
interesting aspect of it is that the clock itself becomes part of the program
to which type inference is applied, and so cost comes down to an analysis of
size.  Somewhat earlier work takes a program transformation approach,
defunctionalizing higher-order programs in a cost-preserving way, and then
applying first-order analysis techniques to the result
\citep{avanzini-et-al:icfp15}.

The accomplishments of these approaches in terms of automatically
computing cost bounds is impressive, and not something we claim our
approach as described here does.  It is possible that after deploying
our recurrence extraction technique, recurrence solvers such as OCRS
\citep{kincaid-et-al:popl18:ocrs} could be used, similarly to how RAML
uses an off-the-shelf LP solver (of course, since our recurrences are
higher-order functions, this would presumably require some form of
defunctionalization, but perhaps
\citeauthor{benzinger:tcs04}'s~\citeyearpar{benzinger:tcs04} work would be
applicable).  The main contrast between our work and all of the
above is that our work puts the standard informal approaches to cost
analysis via recurrence extraction on firm mathematical footing, while
the automatic techniques use different methods than what we teach
students in introductory classes or what programmers do in their heads.

\section{Conclusions and Future Work}

We have improved upon the results of \citet{Danner2015} by giving
a uniform recurrence extraction method for a variety of source
languages. Instead of formulating extraction directly for each
source language, we embed them in an intermediate
language---namely \cbpv---and perform recurrence extraction for
that language instead. Our method uniformly handles general
recursion irrespective of the evaluation strategy of the source
language. We have shown this strategy in action by showing how to
extract recurrences from both call-by-value and call-by-name
\pcf{} programs.

The natural next step would be to increase the expressiveness of
the source language types.
For example, it would be very interesting to examine whether our
strategy can be extended to cover \emph{recursive types} \citep[\S
20]{Pierce2002}. This would lead us to an extraction function for
call-by-value and call-by-name versions of Plotkin's \emph{Fixed
Point Calculus (FPC)} \citep{Plotkin1985}, \citep[\S
7.4]{Gunter1992}, thereby enabling the consideration of a large
number of very expressive programming languages. On the other
hand, such a programme would present significant technical
challenges: amongst other things, it would require the solution of
sized domain equations.

In either case, the combination of inductive or recursive data
types and recursion would also aid us in obtaining a handle on the
complexity of \emph{coinductive data}. For example, in a
call-by-value setting with inductive types and recursion we can
express \emph{streams} of natural numbers by the type $\mu X.
\mathbf{1} \rightarrow (\mathbf{1} + \natt \times X)$. General
recursion then allows us to define non-trivial streams and
functions that compute on them. Recurrence extraction in this
setting would guide us towards various notions of \emph{stream
complexity}.  Moreover, this would be a special case of
\emph{functional reactive programming (FRP)} \citep{Elliott1997},
and extending recurrence extraction to FRP would naturally
lead us to interesting notions of complexity for functional
programs that intentionally do not terminate.

In another direction, deterministic programs with interesting
average-case behavior, such as those implementing Quicksort,
typically use general recursion. It should be the case that the
notions of probabilistic recurrences \citep{karp:jacm94} used to
informally analyze such algorithms can be formalized in terms of
appropriate semantic models in which the size of an argument is
interpreted as an appropriate random variable.  Finally,
formalizing amortized analysis is another direction of significant
interest. The current approach analyzes composition by composing
worst-case bounds, which do not always yield a tight result.  The goal would
not be the automated analysis of AARA, but a formalization of informal
techniques such as
\citeauthor{tarjan:amortized-complexity}'s~\citeyearpar{tarjan:amortized-complexity}
banker's and physicist's methods.

%
%
%
%

%% file: diagrams/cbx_recurrence_extraction.tex
\tikzset
{
		line cap = butt ,
		line join = bevel ,
		arrows = -> ,
		> = angle 60 ,
		auto = left ,
		text depth = 0.25ex , 
		align = center ,
}

\begin{tikzpicture}[x = {(45mm , 0mm)} , y = {(0mm , 15mm)}]
	\node (source) at (0 , 0) {source \\ language} ;
	\node (syntactic recurrence) at (3/2 , 0) {recurrence \\ language \\ (syntactic recurrence)} ;
	\node (semantic recurrence) at (5/2 , 0) {denotational  cost \\ semantics \\ (semantic recurrence)} ;
	\node [draw , dashed] (extraction) at ($ (source) + (4/6 , 0) $) {recurrence \\ extraction} ;
	\draw (source) to (extraction) ;
	\draw (extraction) to (syntactic recurrence) ;
	\draw (syntactic recurrence) to (semantic recurrence) ;
	\draw (syntactic recurrence) to node [auto] {$\sem{-}$} (semantic recurrence) ;
\end{tikzpicture}

%% file: diagrams/cbpv_recurrence_extraction.tex
\tikzset
{
		line cap = butt ,
		line join = bevel ,
		arrows = -> ,
		> = angle 60 ,
		auto = left ,
		text depth = 0.25ex , 
		align = center ,
}

\begin{tikzpicture}[x = {(40mm , 0mm)} , y = {(0mm , 15mm)}]
	\node (source) at (0 , 0) {source \\ language} ;
	\node (intermediate) at (4/5 , 0) {intermediate \\ language} ;
	\node (recurrence) at (11/5 , 0) {recurrence \\ language \\ (syntactic
    recurrence)} ;
	\node (cost) at (3 , 0) {denotational \\ cost \\ semantics \\ (semantic \\
    recurrence)} ;
	\node (sized domains) at ($ (cost) + (0 , 2) $) {sized \\ domains} ;
	\node (PCFc) at (recurrence |- sized domains) {\pcfc{} \\ (\pcf{} with costs)} ;
	\node (CBPV) at (intermediate |- sized domains) {call-by-push-value \\ (CBPV)} ;
	\node [draw , dashed] (recurrence extraction) at ($(CBPV) ! 0.525 ! (PCFc)$) {recurrence \\ extraction} ;
	\node (CBN PCF) at ($ (source |- sized domains) + (0 , 1) $) {CBN \pcf{}} ;
	\node (CBV PCF) at ($ (source |- sized domains) + (0 , -1) $) {CBV \pcf{}} ;
	\draw [includer] (CBN PCF) to node [auto] {$\cbnt{(-)}$} (CBPV) ;
	\draw [includel] (CBV PCF) to node [swap] {$\cbvt{(-)}$} (CBPV) ;
	\draw (CBPV) to (recurrence extraction) ;
	\draw (recurrence extraction) to (PCFc) ;
	\draw (PCFc) to node [auto] {$\sem{-}$} (sized domains) ;
\end{tikzpicture}

%% file: pcf-types.tex
\begin{small}
  \begin{align*}
  \textbf{Types} \qquad
    &A, B
      && ::= 
            &&\natt
      \mid  A \times B
      \mid  A \rightarrow B
  \\[1ex]
  \textbf{Contexts} \qquad
    &\Gamma
      && ::=
            &&\cdot
      \mid  \Gamma, x : A
  \\[1ex]
  \textbf{Numerical operations} \qquad
    &\pcfop
      && ::= 
            && +
          \mid \ast
          \mid -
          \mid \div
          \mid \bmod
  \\[1ex]
  \textbf{Canonical forms (CBN)} \qquad
    &V
      && ::= 
            && \num{n}
      \mid  \tuple{M}{N}
      \mid  \lambda x.\ M
  \\[1ex]
  \textbf{Canonical forms (CBV)} \qquad
    &V, W, Z
      && ::= 
            && \num{n}
      \mid  \tuple{V}{W}
      \mid  \lambda x.\ M
      \mid  \rec{f}{x}{M}
\end{align*}

\begin{tabular}{c}
\\
  $
  \begin{prooftree}
      \phantom{\Gamma \vdash N : \natt}
    \justifies
      \Gamma, x : A, \Gamma' \vdash x : A
  \end{prooftree}
  $

  \qquad

  $
  \begin{prooftree}
      \phantom{\Gamma \vdash N : \natt}
    \justifies
      \Gamma \vdash \num{n} : \natt
  \end{prooftree}
  $

  \qquad

  $
  \begin{prooftree}
      \Gamma \vdash N : \natt
        \quad
      \Gamma \vdash P, Q : A
    \justifies
      \Gamma \vdash \pif{N}{P}{Q} : A
  \end{prooftree}
  $

  \qquad
  
  $
  \begin{prooftree}
      \Gamma \vdash M : \natt
        \quad
      \Gamma \vdash N : \natt
    \justifies
      \Gamma \vdash M \pcfop N : \natt
  \end{prooftree}
  $
  
  \\\\

  $
  \begin{prooftree}
      \Gamma, x : A \vdash M : A
    \justifies
      \Gamma \vdash \fix{x}{M} : A
  \end{prooftree}
  $

  \qquad

  $
  \begin{prooftree}
      \Gamma \vdash M : A_1
        \quad
      \Gamma \vdash N : A_2
    \justifies
      \Gamma \vdash \tuple{M}{N} : A_1 \times A_2
  \end{prooftree}
  $

  \qquad

  $
  \begin{prooftree}
      \Gamma \vdash P : A_1 \times A_2
    \justifies
      \Gamma \vdash \pi_i(P) : A_i
  \end{prooftree}
  $

  \\\\

  $
  \begin{prooftree}
      \Gamma, f : A \rightarrow B, x : A  \vdash M : B
    \justifies
      \Gamma \vdash \rec{f}{x}{M} : A \rightarrow B
  \end{prooftree}
  $

  \qquad

  $
  \begin{prooftree}
      \Gamma, x : A \vdash M : B
    \justifies
      \Gamma \vdash \lambda x.\ M : A \rightarrow B
  \end{prooftree}
  $

  \qquad

  $
  \begin{prooftree}
      \Gamma \vdash M : A \rightarrow B
        \quad
      \Gamma \vdash N : A
    \justifies
      \Gamma \vdash M\,N : B
  \end{prooftree}
  $

\end{tabular}

\end{small}

%% file: pcf-bigstep.tex
\vspace{2ex}
    
\begin{small}
\begin{tabular}{c}

  \textbf{Rules for all variants}

  \\[2ex]

  $
  \begin{prooftree}
      \phantom{no premise}
    \justifies
      \evalpcf{\num{n}}{\num{n}}{0}
  \end{prooftree}
  $

  \quad

  \begin{prooftree}
      \evalpcf{M}{\num{m}}{a}
        \quad
      \evalpcf{N}{\num{n}}{b}
    \justifies
      \evalpcf{M \pcfop N}{\num{m \pcfop n}}{a + b}
  \end{prooftree}

  \quad

  $
  \begin{prooftree}
      \phantom{no premise}
    \justifies
      \evalpcf{\lambda x.\ M}{\lambda x.\ M}{0}
  \end{prooftree}
  $

  \\\\

  $
  \begin{prooftree}
      \evalpcf{N}{\num{0}}{a}
        \quad
      \evalpcf{P}{V}{b}
    \justifies
      \evalpcf{\pif{N}{P}{Q}}{V}{a+b}
  \end{prooftree}
  $

  \quad

  $
  \begin{prooftree}
      \evalpcf{N}{\num{n+1}}{a}
        \quad
      \evalpcf{Q}{V}{b}
    \justifies
      \evalpcf{\pif{N}{P}{Q}}{V}{a+b}
  \end{prooftree}
  $

  \\\\

  {\textbf{Call-by-name PCF}}
  \\[2ex]

  $
  \begin{prooftree}
      \phantom{no premise}
    \justifies
      \evalpcf{\tuple{M}{N}}{\tuple{M}{N}}{0}
  \end{prooftree}
  $

  \quad

  $
  \begin{prooftree}
      \evalpcf{P}{\tuple{M_1}{M_2}}{a}
        \quad
      \evalpcf{M_i}{V_i}{b}
    \justifies
      \evalpcf{\pi_i(P)}{V_i}{a + b + 1}
  \end{prooftree}
  $

  \quad

  $
  \begin{prooftree}
      \evalpcf{M}{\lambda x.\ P}{m}
        \quad
      \evalpcf{P[N/x]}{V}{n}
    \justifies
      \evalpcf{M N}{V}{m + n + 1}
  \end{prooftree}
  $

  \quad

  $
  \begin{prooftree}
      \evalpcf{M[\fix{x}{M}/x]}{V}{n}
    \justifies
      \evalpcf{\fix{x}{M}}{V}{n + 1}
  \end{prooftree}
  $

  \\\\

  {\textbf{Call-by-value PCF}}

  \\[2ex]

  $
  \begin{prooftree}
      \phantom{\evalpcf{H}{V}{n}}
    \justifies
      \evalpcf{\rec{f}{x}{P}}{\rec{f}{x}{P}}{0}
  \end{prooftree}
  $

  \quad

  $
  \begin{prooftree}
      \evalpcf{M}{V}{a}
        \quad
      \evalpcf{N}{W}{b}
    \justifies
      \evalpcf{\tuple{M}{N}}{\tuple{V}{W}}{a+b}
  \end{prooftree}
  $

  \quad

  $
  \begin{prooftree}
      \evalpcf{P}{\tuple{V_1}{V_2}}{n}
    \justifies
      \evalpcf{\pi_i(P)}{V_i}{n + 1}
  \end{prooftree}
  $

  \\\\

  $
  \begin{prooftree}
      \evalpcf{M}{\rec{f}{x}{P}}{m}
        \quad
      \evalpcf{N}{W}{n}
        \quad
      \evalpcf{P[\rec{f}{x}{P}/f, W/x]}{V}{k}
    \justifies
      \evalpcf{M N}{V}{m + n + k + 1}
  \end{prooftree}
  $

  \quad

  $
  \begin{prooftree}
      \evalpcf{M}{\lambda x.\ P}{m}
        \quad
      \evalpcf{N}{W}{n}
        \quad
      \evalpcf{P[W/x]}{V}{k}
    \justifies
      \evalpcf{M N}{V}{m + n + k + 1}
  \end{prooftree}
  $
\end{tabular}
\end{small}

%% file: pcf-reclang.tex
\begin{minipage}{0.40\textwidth}
  \begin{align*}
    \numc{0}   &\defeq \zeroc \\
    \numc{n+1} &\defeq \unitc \plusc \numc{n}
  \end{align*}
\end{minipage} \begin{minipage}{0.59\textwidth}
  \begin{alignat*}{3}
    \textbf{Types} \qquad
      &A, B &&::=\ \dots \mid \cost
    \\[1ex]
    \textbf{Canonical forms (CBN only)} \qquad
      &V    &&::=\  \dots \mid \numc{n}
  \end{alignat*}
\end{minipage}

\vspace{2ex}

\textbf{Typing rules}: The CBN rules of Figure
\ref{fig:pcf-types}, plus $
  \quad
    \begin{prooftree}
        \numc{n} \in \{\zeroc, \unitc\}
      \justifies
        \Gamma \vdash \numc{n} : \cost
    \end{prooftree}
  \quad 
$ and $
  \quad
    \begin{prooftree}
        \Gamma \vdash M : \cost
          \quad
        \Gamma \vdash N : \cost
      \justifies
        \Gamma \vdash M \plusc N : \cost
    \end{prooftree}
  \quad 
$

\vspace{3ex}

\textbf{Big-step semantics}: The CBN rules of Figure
\ref{fig:pcf-bigstep}, plus $
  \quad
    \begin{prooftree}
        \numc{n} \in \{\zeroc, \unitc\}
      \justifies
        \evalpcf{\numc{n}}{\numc{n}}{}
    \end{prooftree}
  \quad
$ and $
  \quad
    \begin{prooftree}
        \evalpcf{M}{\numc{m}}{}
          \qquad
        \evalpcf{N}{\numc{n}}{}
      \justifies
        \evalpcf{M \plusc N}{\numc{m+n}}{}
    \end{prooftree}
  \quad
$

%% file: pcf-inequality.tex
\renewcommand{\arraystretch}{3}

\begin{small}
  \makebox[0pt]{
\begin{tabular}{c c}


  $
  \begin{prooftree}
      \phantom{\mctxtt{\mctx{C}}{\Gamma}{A}{\Delta}{\natt}}
    \justifies
      \mctxtt{[]}{\Gamma}{A}{\Gamma}{A}
  \end{prooftree}
  $

  &

  $
  \begin{prooftree}
      \mctxtt{\mctx{C}}{\Gamma}{A}{\Delta}{\natt}
    \quad
      \Delta \vdash M, N : C
    \justifies
      \mctxtt{\pif{\mctx{C}}{M}{N}}
            {\Gamma}{A}
            {\Delta}{C}
  \end{prooftree}
  $

  \\

  $
  \begin{prooftree}
      \Delta \vdash M : \cost
    \quad
      \mctxtt{\mctx{C}}{\Gamma}{A}{\Delta}{\cost}
    \justifies
      \mctxtt{M \plusc \mctx{C}}{\Gamma}{A}{\Delta}{\cost}
  \end{prooftree}
  $

  &

  $
  \begin{prooftree}
      \mctxtt{\mctx{C}}{\Gamma}{A}{\Delta}{\cost}
    \quad
      \Delta \vdash N : \cost
    \justifies
      \mctxtt{\mctx{C} \plusc N}{\Gamma}{A}{\Delta}{\cost}
  \end{prooftree}
  $

  \\

  $
  \begin{prooftree}
      \mctxtt{\mctx{C}}
            {\Gamma}{A}
            {\Delta, x : B}{C}
    \justifies
      \mctxtt{\lambda x.\ \mctx{C}}
            {\Gamma}{A}
            {\Delta}{B \rightarrow C}
  \end{prooftree}
  $

  &

  $
  \begin{prooftree}
      \mctxtt{\mctx{C}}
            {\Gamma}{A}
            {\Delta}{B \rightarrow C}
    \quad
      \Delta \vdash N : B
    \justifies
      \mctxtt{\mctx{C} N}
            {\Gamma}{A}
            {\Delta}{C}
  \end{prooftree}
  $

  \\

  $
  \begin{prooftree}
      \Delta \vdash M : B_1
    \quad
      \mctxtt{\mctx{C}}
            {\Gamma}{A}
            {\Delta}{B_2}
    \justifies
      \mctxtt{\tuple{M}{\mctx{C}}}
            {\Gamma}{A}
            {\Delta}{B_1 \times B_2}
  \end{prooftree}
  $

  &

  $
  \begin{prooftree}
      \mctxtt{\mctx{C}}
            {\Gamma}{A}
            {\Delta}{B_1}
    \quad
      \Delta \vdash N : B_2
    \justifies
      \mctxtt{\tuple{\mctx{C}}{N}}
            {\Gamma}{A}
            {\Delta}{B_1 \times B_2}
  \end{prooftree}
  $

  \\
  
  $
  \begin{prooftree}
      \mctxtt{\mctx{C}}
            {\Gamma}{A}
            {\Delta}{A_1 \times A_2}
    \justifies
      \mctxtt{\pi_i(\mctx{C})}
            {\Gamma}{A}
            {\Delta}{A_i}
  \end{prooftree}
  $

  &

  $
  \begin{prooftree}
      \mctxtt{\mctx{C}}
             {\Gamma}{A}
             {\Delta}{\natt}
    \quad
      \Delta \vdash N : \natt
    \justifies
      \mctxtt{\mctx{C} \pcfop N}
             {\Gamma}{A}
             {\Delta}{\natt}
  \end{prooftree}
  $

  \\[3ex]

  \multicolumn{2}{c}{
  $
  \begin{prooftree}
      \mctxtt{\mctx{C}}
             {\Gamma}{A}
             {\Delta}{\natt}
    \quad
      \Delta \vdash N : \natt
    \quad
      \pcfopm \in \{+, *\}
    \justifies
      \mctxtt{M \pcfopm \mctx{C}}
             {\Gamma}{A}
             {\Delta}{\natt}
  \end{prooftree}
  $
  }

  \\[3ex]
  \hline  


  $
  \begin{prooftree}
      \Gamma \vdash M : A
    \justifies
      \Gamma \vdash M \costleq M : A
    \using
      {(\textsf{\hypertarget{refl}{refl}})}
  \end{prooftree}
  $

  &

  $
  \begin{prooftree}
      \Gamma \vdash M \costleq N : A
    \quad
      \Gamma \vdash N \costleq P : A
    \justifies
      \Gamma \vdash M \costleq P : A
    \using
      {(\textsf{\hypertarget{trans}{trans}})}
  \end{prooftree}
  $

  \\

  $
  \begin{prooftree}
      \Gamma \vdash M : \cost
    \justifies
      \Gamma \vdash M \costleq \zeroc \plusc M : \cost
    \using
      (\hypertarget{zero}{\textsf{zero}})
  \end{prooftree}
  $

  &

  $
  \begin{prooftree}
      \Gamma \vdash M : \cost
    \quad
      \Gamma \vdash N : \cost
    \quad
      \Gamma \vdash P : \cost
    \justifies
      \Gamma \vdash M \plusc (N \plusc P) \costleq (M \plusc N) \plusc P : \cost
    \using
      (\hypertarget{assoc}{\textsf{assoc}})
  \end{prooftree}
  $

  \\

  $
  \begin{prooftree}
      \Gamma, x : A \vdash M : B
    \quad
      \Gamma \vdash N : A
    \justifies
      \Gamma \vdash N[M/x] \costleq (\lambda x.\ M)N : B
    \using
      (\hypertarget{betaarrow}{\rightarrow_\beta})
  \end{prooftree}
  $

  &
  
  $
  \begin{prooftree}
      \Gamma \vdash M_1 : A_1 
    \quad
      \Gamma \vdash M_2 : A_2
    \justifies
      \Gamma \vdash M_i \costleq \pi_i(\tuple{M_1}{M_2}) : A_i
    \using
      (\hypertarget{betaprod}{\times_\beta})
  \end{prooftree}
  $

  \\

  $
  \begin{prooftree}
      \Gamma \vdash P : A
    \quad
      \Gamma \vdash Q : A
    \justifies
      \Gamma \vdash P \costleq \pif{\num{0}}{P}{Q} : A
    \using
      (\hypertarget{iftrue}{\textsf{if}_\textsf{tt}})
  \end{prooftree}
  $

  &

  $
  \begin{prooftree}
      \Gamma \vdash P : A
    \quad
      \Gamma \vdash Q : A
    \justifies
      \Gamma \vdash Q \costleq \pif{\num{n+1}}{P}{Q} : A
    \using
      (\hypertarget{iffalse}{\textsf{if}_\textsf{ff}})
  \end{prooftree}
  $

  \\
  $
  \begin{prooftree}
      \phantom{\Gamma, x : A \vdash M : A}
    \justifies
      \Gamma \vdash 
        \num{n \pcfop m} \costleq \num{n} \pcfop \num{m} : \natt
    \using
      (\hypertarget{betanum}{\textsf{num}_\beta})
  \end{prooftree}
  $

  &

  $
  \begin{prooftree}
      \Gamma \vdash N : \natt
    \justifies
      \Gamma \vdash 
        \num{m} \bmod N \costleq N - 1 : \natt
    \using
      (\hypertarget{mod}{\textsf{mod}})
  \end{prooftree}
  $


  \\

  $
  \begin{prooftree}
      \Gamma \vdash M \costleq N : A
    \quad
      \mctxtt{\mctx{C}}{\Gamma}{A}{\Delta}{B}
    \justifies
      \Delta \vdash \mctx{C}[M] \costleq \mctx{C}[N] : B
    \using
      (\hypertarget{ctx}{\textsf{ctx}})
  \end{prooftree}
  $

  &

  $
  \begin{prooftree}
      \Gamma, x : A \vdash E : A
    \justifies
      \Gamma \vdash
        \fixn{x}{n+1}{E}
          \costleq 
        \fixn{x}{n}{E} : A
    \using
      (\textsf{\hypertarget{rat}{rat}})
  \end{prooftree}
  $

  \\
  
  \multicolumn{2}{c}{
  $
  \begin{prooftree}
      \Gamma, x : A \vdash E : A
    \quad
      \Delta, z : A \vdash \ectx{E}[z] : B
    \quad
      \forall n \geq 0.\
        \Delta \vdash M \costleq \ectx{E}[\fixn{x}{n}{E}] : B
    \justifies
      \Delta \vdash M \costleq \ectx{E}[\fix{x}{E}] : B
    \using
      (\textsf{\hypertarget{cpind}{cpind}})
  \end{prooftree}
  $
  }

\end{tabular}
}
\end{small}

%% file: cbv-extraction.tex
\begin{small}
  \begin{minipage}{0.25\textwidth}
\begin{align*}
  \cbvrec{A}               &\defeq   \cost \times \cbvpot{A} \\
  & \\
  \cbvpot{\natt}           &\defeq   \natt \\
  \cbvpot{A_1 \times A_2}  &\defeq   \cbvpot{A_1} \times \cbvpot{A_2} \\
  \cbvpot{A \rightarrow B} &\defeq   \cbvpot{A} \rightarrow \cbvrec{B} \\
\end{align*}
\end{minipage} \begin{minipage}{0.70\textwidth}
\begin{align*}
  \cbvrec{x}                 &\defeq   \tuple{\zeroc}{x} \\
  \cbvrec{\num{n}}           &\defeq   \tuple{\zeroc}{\num{n}} \\
  \cbvrec{M \{+,*\} N}       &\defeq   \tuple{\cbvrec{M}_c \plusc \cbvrec{N}_c}
                                             {\cbvrec{M}_p \{+,*\} \cbvrec{N}_p} \\
  \cbvrec{M \{-,\div\} \num{n}} &\defeq   \tuple{\cbvrec{M}_c}
                                             {\cbvrec{M}_p \{-,\div\} \num{n}} \\
  \cbvrec{M \{-,\div\} N}       &\defeq   \tuple{\cbvrec{M}_c \plusc \cbvrec{N}_c}
                                             {\cbvrec{M}_p} (\text{if } N \neq \num{n})\\
  \cbvrec{M \bmod N}       &\defeq   \tuple{\cbvrec{M}_c \plusc \cbvrec{N}_c}
                                          {\cbvrec{N}_ - \num{1}}   \\
  \cbvrec{\pif{N}{P}{Q}}     &\defeq   
    \cbvrec{N}_c \addcost \pif{\cbvrec{N}_p}{\cbvrec{P}}{\cbvrec{Q}} \\
  \cbvrec{\tuple{M}{N}}      &\defeq   \tuple{\cbvrec{M}_c \plusc \cbvrec{N}_c}
                                            {\tuple{\cbvrec{M}_p}{\cbvrec{N}_p}} \\
  \cbvrec{\pi_i(M)}          &\defeq   
    _c\tuple{\unitc \plusc \cbvrec{M}_c}{\pi_i\left(\cbvrec{M}_p\right)} \\
  \cbvrec{\lambda x.\ M}     &\defeq   \tuple{\zeroc}{\lambda x. \cbvrec{M}} \\
  \cbvrec{M\, N}             &\defeq   {\unitc \plusc \cbvrec{M}_c \plusc \cbvrec{N}_c} \addcost
                                            ({\cbvrec{M}_p \, \cbvrec{N}_p}) \\
  \cbvrec{\rec{f}{x}{M}}     &\defeq   \tuple{\zeroc}{\fix{f}{\lambda x.\ \cbvrec{M}}}
\end{align*}
\end{minipage}

\vspace{3ex}

If $\Gamma \vdash E : \cbvrec{A}$ and $\Gamma \vdash C : \cost$ we
write $ 
  \begin{cases}
    \Gamma \vdash E_c \defeq \pi_1(E) : \cost \\
    \Gamma \vdash E_p \defeq \pi_2(E) : \cbvpot{A} \\
    \Gamma \vdash C \addcost E 
      \defeq \tuple{C \plusc \pi_1(E)}{\pi_2(E)} : \cbvrec{A} \\
  \end{cases}
$

\end{small}

%% file: cbn-extraction.tex
\begin{small}
\begin{minipage}{0.25\textwidth}
\begin{align*}
  \cbnrec{\natt}           &\defeq   \cost \times \natt \\
  \cbnrec{A_1 \times A_2}  &\defeq   \cbnrec{A_1} \times \cbnrec{A_2} \\
  \cbnrec{A \rightarrow B} &\defeq   \cbnrec{A} \rightarrow \cbnrec{B} \\
  \\
\end{align*} 
\end{minipage} \begin{minipage}{0.70\textwidth}
\begin{align*}
  \cbnrec{x}                 &\defeq   x \\
  \cbnrec{\num{n}}           &\defeq   \tuple{\zeroc}{\num{n}} \\
  \cbvrec{M \pcfop N}        &\defeq   \text{(same as in CBV)}\\
  \cbnrec{\pif{N}{P}{Q}}     &\defeq   
    \cbnrec{N}_c \algp{A} \pif{\cbnrec{N}_p}{\cbnrec{P}}{\cbnrec{Q}} \\
  \cbnrec{\tuple{M}{N}}      &\defeq   \tuple{\unitc \algp{A_1} \cbnrec{M}}
                                             {\unitc \algp{A_2} \cbnrec{N}} \\
  \cbnrec{\pi_i(M)}          &\defeq   \pi_i\left(\cbnrec{M}\right) \\
  \cbnrec{\lambda x.\ M}     &\defeq   \lambda x.\ \unitc \algp{B} \cbnrec{M} \\
  \cbnrec{M\, N}             &\defeq   \cbnrec{M}\, \cbnrec{N} \\
  \cbnrec{\fix{x}{M}}        &\defeq   \fix{x}{\unitc \algp{A} \cbnrec{M}}
\end{align*}
\end{minipage}

\begin{alignat*}{3}
  c : \cost, x : \cost \times \natt
    &\vdash
      \alpha_{\natt}(c, x)
        &&\defeq 
          \tuple{c \plusc \pi_1(x)}{ \pi_2(x)} 
      &&: \cost \times \natt \\
  c : \cost, p : \cbnrec{A_1} \times \cbnrec{A_2}
    &\vdash
      \alpha_{A_1 \times A_2}(c, p)
        &&\defeq
          \tuple
            {\alpha_{A_1}(c, \pi_1(p))}
            {\alpha_{A_2}(c, \pi_2(p))}
      &&: \cbnrec{A_1} \times \cbnrec{A_2} \\
  c : \cost, f : \cbnrec{A} \rightarrow \cbnrec{B}
    &\vdash
      \alpha_{A \rightarrow B}(c, f)
        &&\defeq
          \lambda a.\ \alpha_{B}\left(c, f(a)\right)
      &&: \cbnrec{A} \rightarrow \cbnrec{B}
\end{alignat*}

\end{small}

%% file: cbpv-types.tex
\begin{small}
  \begin{align*}
  \textbf{Value Types} \qquad
    &A
      && ::= 
          &&\natt
      \mid  A_1 \times A_2
      \mid  U \ct{B}
   \\[1ex]
  \textbf{Computation Types}\qquad
    &\ct{B}
      && ::= 
            && FA
      \mid  \ct{B}_1 \with \ct{B}_2
      \mid  A \rightarrow \ct{B}
   \\[1ex]
  \textbf{Contexts}\qquad
    &\Gamma
      &&::=
          &&\cdot
      \mid  \Gamma, x : A 
  \\[1ex]
  \textbf{Numerical operations} \qquad
    &\pcfop
      && ::= 
            && +
          \mid -
          \mid \ast
          \mid \div
          \mid \bmod
  \\[1ex]
  \textbf{Terminal Computations}\qquad
    &T, S
      &&:=
             && \return V
       \mid  \tuple{M}{N}
       \mid  \lambda x.\ M
\end{align*}

\vspace{1ex}

\begin{tabular}{cc}

  $
  \begin{prooftree}
      \phantom{no premise}
    \justifies
      \Gamma, x : A, \Gamma' \vturn x : A
    \using
      {(\textsf{var})}
  \end{prooftree}
  $

  &

  $
  \begin{prooftree}
      \phantom{no premise}
    \justifies
      \Gamma \vturn \numl{n} : \natt
    \using
      {(\natt_{\numl{n}})}
  \end{prooftree}
  $

  \\[3ex]

  $
  \begin{prooftree}
      \Gamma \cturn M : \ct{B}
    \justifies
      \Gamma \cturn \charge{M} : \ct{B}
    \using
      {(\textsf{ch})}
  \end{prooftree}
  $

  &

  $
  \begin{prooftree}
      \Gamma, x : U\ct{B} \cturn M : \ct{B}
    \justifies
      \Gamma \cturn \fix{x}{M} : \ct{B}
    \using
      {(\textsf{fix})}
  \end{prooftree}
  $

  \\[3ex]

  $
  \begin{prooftree}
      \Gamma \vturn M, N : \natt
        \quad
      \Gamma, z : \natt \cturn Q : \ct{B}
    \justifies
      \Gamma \cturn \calc{z}{M \cbpvop N}{Q} : \ct{B}
    \using
      {(\natt_{\cbpvop})}
  \end{prooftree}
  $

  &

  $
  \begin{prooftree}
      \Gamma \vturn N : \natt
        \quad
      \Gamma \cturn P : \ct{B}
        \quad
      \Gamma \cturn Q : \ct{B}
    \justifies
      \Gamma \cturn \ifz{N}{P}{Q} : \ct{B}
    \using
      {(\natt_\textsf{ifz})}
  \end{prooftree}
  $

  \\[3ex]

  $
  \begin{prooftree}
      \Gamma \vturn V_1 : A_1
        \quad
      \Gamma \vturn V_2 : A_2
    \justifies
      \Gamma \vturn (V_1, V_2) : A_1 \times A_2
    \using
      {(\times\mathcal{I})}
  \end{prooftree}
  $

  &

  $
  \begin{prooftree}
      \Gamma \vturn V : A_1 \times A_2
        \quad
      \Gamma, x_1 : A_1, x_2 : A_2 \cturn N : \ct{B}
    \justifies
      \Gamma \cturn \splitprod{V}{x_1}{x_2}{N} : \ct{B}
    \using
      {(\times\mathcal{E})}
  \end{prooftree}
  $

  \\[3ex]

  $
  \begin{prooftree}
      \Gamma \cturn M : \ct{B}
    \justifies
      \Gamma \vturn \thunk{M} : U\ct{B}
    \using
      {(U\mathcal{I})}
  \end{prooftree}
  $

  &

  $
  \begin{prooftree}
      \Gamma \vturn V : U\ct{B}
    \justifies
      \Gamma \cturn \force{V} : \ct{B}
    \using
      {(U\mathcal{E})}
  \end{prooftree}
  $

  \\[3ex]


  $
  \begin{prooftree}
      \Gamma \vturn V : A
    \justifies
      \Gamma \cturn \return{V} : FA
    \using
      {(F\mathcal{I})}
  \end{prooftree}
  $

  &

  $
  \begin{prooftree}
      \Gamma \cturn M : FA
          \quad
      \Gamma, x : A \cturn N : \ct{B}
    \justifies
      \Gamma \cturn \bind{x}{M}{N} : \ct{B}
    \using
      {(F\mathcal{E})}
  \end{prooftree}
  $

  \\[3ex]

  $
  \begin{prooftree}
      \Gamma \cturn M : \ct{B}_1
        \quad
      \Gamma \cturn N : \ct{B}_2
    \justifies
      \Gamma \cturn \tuple{M}{N} : \ct{B}_1 \with \ct{B}_2
    \using
      {(\with\mathcal{I})}
  \end{prooftree}
  $

  &

  $
  \begin{prooftree}
      \Gamma \cturn M : \ct{B}_1 \with \ct{B}_2
    \justifies
      \Gamma \cturn \pi_i(M) : \ct{B}_i 
    \using
      {(\with\mathcal{E})}
  \end{prooftree}
  $

  \\[3ex]

  $
  \begin{prooftree}
      \Gamma, x : A \cturn M : \ct{B}
    \justifies
      \Gamma \cturn \lambda x.\ M : A \rightarrow \ct{B}
    \using
      {(\rightarrow\mathcal{I})}
  \end{prooftree}
  $

  &

  $
  \begin{prooftree}
      \Gamma \cturn M : A \rightarrow \ct{B}
        \quad
      \Gamma \vturn N : A
    \justifies
      \Gamma \cturn M\,N : \ct{B}
    \using
      {(\rightarrow\mathcal{E})}
  \end{prooftree}
  $

\end{tabular}
\end{small}

%% file: cbpv-bigstep.tex
\begin{small}
  \begin{tabular}{ccc}

  $
  \begin{prooftree}
      \phantom{no premise}
    \justifies
      \eval{\lambda x.\ M}{\lambda x.\ M}{0}
  \end{prooftree}
  $

  &

  $
  \begin{prooftree}
      \phantom{no premise}
    \justifies
      \eval{\return{V}}{\return{V}}{0}
  \end{prooftree}
  $

  & 

  $
  \begin{prooftree}
      \phantom{no premise}
    \justifies
      \eval{\tuple{M}{N}}{\tuple{M}{N}}{0}
  \end{prooftree}
  $

  \\\\

  $
  \begin{prooftree}
      \eval{M}{T}{m}
    \justifies
      \eval{\force{(\thunk{M})}}{T}{m}
  \end{prooftree}
  $

  &

  $
  \begin{prooftree}
      \eval{N[V_1/x_1, V_2/x_2]}{T}{m}
    \justifies
      \eval{\splitprod{(V_1, V_2)}{x_1}{x_2}{N}}{T}{m}
  \end{prooftree}
  $

  &

  $
  \begin{prooftree}
      \eval{M[\thunk{(\fix{x}{M})}/x]}{T}{n}
    \justifies
      \eval{\fix{x}{M}}{T}{n}
  \end{prooftree}
  $

  \\\\

  $
  \begin{prooftree}
      \eval{P}{T}{m}
    \justifies
      \eval{\ifz{\numl{0}}{P}{Q}}{T}{m}
  \end{prooftree}
  $

  &

  $
  \begin{prooftree}
      \eval{Q}{T}{m}
    \justifies
      \eval{\ifz{\numl{n+1}}{P}{Q}}{T}{m}
  \end{prooftree}
  $

  &

  $
  \begin{prooftree}
      \eval{P[\numl{m \cbpvop n}/z]}{T}{m}
    \justifies
      \eval{\calc{z}{\numl{m} \cbpvop \numl{n}}{P}}{T}{m}
  \end{prooftree}
  $
  
  \\\\

  $
  \begin{prooftree}
      \eval{M}{T}{n}
    \justifies
      \eval{\charge{M}}{T}{n+1}
  \end{prooftree}
  $

  &

  $
  \begin{prooftree}
      \eval{M}{\return{V}}{m}
        \qquad
      \eval{N[V/x]}{T}{n}
    \justifies
      \eval{\bind{x}{M}{N}}{T}{m + n}
  \end{prooftree}
  $

  \\\\

  $
  \begin{prooftree}
      \eval{P}{\tuple{M_1}{M_2}}{m}
        \qquad
      \eval{M_i}{T}{n}
    \justifies
      \eval{\pi_i(P)}{T}{m + n}
  \end{prooftree}
  $

  &

  $
  \begin{prooftree}
      \eval{M}{\lambda x.\ P}{m}
        \qquad
      \eval{P[V/x]}{T}{n}
    \justifies
      \eval{M V}{T}{m + n}
  \end{prooftree}
  $

\end{tabular}
\end{small}

%% file: cbpv-extraction.tex
\begin{small}

  \begin{minipage}{0.45\textwidth}
  \begin{align*}
    \potential{\natt}           &\defeq \natt     \\
    \potential{A_1 \times A_2}  &\defeq \potential{A_1} \times \potential{A_2} \\
    \potential{U\ct{B}}         &\defeq \carrier{\complexity{\ct{B}}}
  \end{align*}
\end{minipage}
\begin{minipage}{0.45\textwidth}
  \begin{align*}
    \complexity{FA}                     
      &\defeq (\cost \times \potential{A}, \alpha_{F A}) \\
    \complexity{A \rightarrow \ct{B}}     
      &\defeq (\potential{A} \rightarrow \carrier{\complexity{\ct{B}}}, 
               \alpha_{A \rightarrow \ct{B}}) \\
    \complexity{\ct{B}_1 \with \ct{B}_2}  
      &\defeq (\carrier{\complexity{\ct{B}_1}} \times \carrier{\complexity{\ct{B}_2}}, 
              \alpha_{\ct{B}_1 \with \ct{B}_2}) \\
  \end{align*}
\end{minipage}

\begin{alignat*}{3}
  c : \cost, x : \cost \times \potential{A}
    &\vdash
      \alpha_{F A}(c, x)
        &&\defeq 
          \tuple{c \plusc \pi_1(x)}{ \pi_2(x)} 
      &&: \cost \times \potential{A} \\
  c : \cost, f : \potential{A} \rightarrow \carrier{\complexity{\ct{B}}}
    &\vdash
      \alpha_{A \rightarrow\ct{B}}(c, f)
        &&\defeq
          \lambda a.\ \alpha_{\ct{B}}\left(c, f(a)\right)
      &&: \potential{A} \rightarrow \carrier{\complexity{\ct{B}}} \\
  c : \cost, p : \carrier{\complexity{\ct{B}_1}} \times \carrier{\complexity{\ct{B}_2}}
    &\vdash
      \alpha_{\ct{B}_1 \with \ct{B}_2}(c, p)
        &&\defeq
          \tuple
            {\alpha_{\ct{B}_1}(c, \pi_1(p))}
            {\alpha_{\ct{B}_2}(c, \pi_2(p))}
      &&: \carrier{\complexity{\ct{B}_1}} \times \carrier{\complexity{\ct{B}_2}}
\end{alignat*}

\textbf{Note}: whenever $\Gamma \vdash E : \complexity{FA}$, we
let $ 
  \begin{cases}
    \Gamma \vdash E_c \defeq \pi_1(E) : \cost \\
    \Gamma \vdash E_p \defeq \pi_2(E) : \potential{A}
  \end{cases}
$

\begin{tabular}{ccc}

  $
  \begin{prooftree}
      \phantom{no premise}
    \justifies
      \cbpvtv{\Gamma, x : A, \Gamma'}{x}{x}{A}
    \using
      {(\textsf{var})}
  \end{prooftree}
  $

  &

  $
  \begin{prooftree}
      \phantom{no premise}
    \justifies
      \cbpvtv{\Gamma}{\numl{n}}{\num{n}}{\natt}
  \end{prooftree}
  $

  &

  $
  \begin{prooftree}
      \cbpvtc{\Gamma, x : U\ct{B}}{M}{P}{\ct{B}}
    \justifies
      \cbpvtc{\Gamma}{\fix{x}{M}}{\fix{x}{P}}{\ct{B}}
  \end{prooftree}
  $

  \\\\

  $
  \begin{prooftree}
      \cbpvtv{\Gamma}{V_1}{W_1}{A_1}
        \quad
      \cbpvtv{\Gamma}{V_2}{W_2}{A_2}
    \justifies
      \cbpvtv{\Gamma}{(V_1, V_2)}{\tuple{W_1}{W_2}}{A_1 \times A_2}
  \end{prooftree}
  $

  & 

  \multicolumn{2}{c}{
  $
  \begin{prooftree}
      \cbpvtv{\Gamma}{V}{P}{A_1 \times A_2}
        \quad
      \cbpvtc{\Gamma, x_1 : A_1, x_2 : A_2}{N}{Q}{\ct{B}}
    \justifies
      \cbpvtc{\Gamma}
             {\splitprod{V}{x_1}{x_2}{N}}
             {Q[\pi_1(P)/x_1, \pi_2(P)/x_2]}{\ct{B}}
  \end{prooftree}
  $
  }

  \\\\

  $
  \begin{prooftree}
      \cbpvtv{\Gamma}{V}{P}{A}
    \justifies
      \cbpvtc{\Gamma}{\return{V}}{\tuple{\zeroc}{P}}{FA}
  \end{prooftree}
  $

  &

  \multicolumn{2}{c}{
  $
  \begin{prooftree}
      \cbpvtc{\Gamma}{M}{Q}{FA}
          \quad
      \cbpvtc{\Gamma, x : A}{N}{R}{\ct{B}}
    \justifies
      \cbpvtc{\Gamma}
             {\bind{x}{M}{N}}
             {\costof{Q} +_{\ct{B}} R[\potentialof{Q}/x]}
             {\ct{B}}
  \end{prooftree}
  $
  }

  \\\\

  $
  \begin{prooftree}
      \cbpvtc{\Gamma, x : A}{M}{P}{\ct{B}}
    \justifies
      \cbpvtc{\Gamma}
             {\lambda x.\ M}
             {\lambda x.\ P}
             {A \rightarrow \ct{B}}
  \end{prooftree}
  $

  &

  \multicolumn{2}{c}{
  $
  \begin{prooftree}
      \cbpvtc{\Gamma}{M}{P}{A \rightarrow \ct{B}}
        \quad
      \cbpvtv{\Gamma}{N}{Q}{A}
    \justifies
      \cbpvtc{\Gamma}{M\,N}{P\,Q}{\ct{B}}
  \end{prooftree}
  $
  }

  \\\\
  
  $
  \begin{prooftree}
      \cbpvtc{\Gamma}{M}{P}{\ct{B}_1}
        \quad
      \cbpvtc{\Gamma}{N}{Q}{\ct{B}_2}
    \justifies
      \cbpvtc{\Gamma}{\tuple{M}{N}}
                     {\tuple{P}
                            {Q}
                     }
                     {\ct{B}_1 \with \ct{B}_2}
  \end{prooftree}
  $

  &

  \multicolumn{2}{c}{
  $
  \begin{prooftree}
      \cbpvtc{\Gamma}{M}{P}{\ct{B}_1 \with \ct{B}_2}
    \justifies
      \cbpvtc{\Gamma}{\pi_i(M)}{\pi_i(P)}{\ct{B}_i}
  \end{prooftree}
  $
  }

  \\\\

  $
  \begin{prooftree}
      \cbpvtc{\Gamma}{M}{P}{\ct{B}}
    \justifies
      \cbpvtv{\Gamma}{\thunk{M}}{P}{U\ct{B}}
  \end{prooftree}
  $

  &

  \multicolumn{2}{c}{
  $
  \begin{prooftree}
      \cbpvtv{\Gamma}{V}{P}{U\ct{B}}
    \justifies
      \cbpvtc{\Gamma}{\force{V}}{P}{\ct{B}}
  \end{prooftree}
  $
  }

  \\\\

  $
  \begin{prooftree}
      \cbpvtc{\Gamma}{M}{P}{\ct{B}}
    \justifies
      \cbpvtc{\Gamma}{\charge{M}}{\unitc +_{\ct{B}} P}{\ct{B}}
  \end{prooftree}
  $

  &

  \multicolumn{2}{c}{
  $
  \begin{prooftree}
      \cbpvtv{\Gamma}{V}{W}{\natt}
    \quad
      \cbpvtc{\Gamma}{M}{P}{\ct{B}}
    \quad
      \cbpvtc{\Gamma}{N}{Q}{\ct{B}}
    \justifies
      \cbpvtv{\Gamma}{\ifz{V}{M}{N}}
                     {\pif{W}{P}{Q}}
                     {\ct{B}}
  \end{prooftree}
  $
  }

  \\\\

  \multicolumn{3}{c}{
  $
  \begin{prooftree}
      \cbpvtv{\Gamma}{V}{M}{\natt}
    \quad
      \cbpvtv{\Gamma}{W}{N}{\natt}
    \quad
      \cbpvtc{\Gamma, z : \natt}{P}{Q}{\ct{B}}
    \quad
      \cbpvopm \in \{ +, * \}
    \justifies
      \cbpvtc{\Gamma}{\calc{z}{V \cbpvopm W}{P}}
                     {Q[M \pcfopm N/z]}
                     {\ct{B}}
  \end{prooftree}
  $
  }

  \\\\

  \multicolumn{3}{c}{
  $
  \begin{prooftree}
      \cbpvtv{\Gamma}{V}{M}{\natt}
    \quad
      \cbpvtc{\Gamma, z : \natt}{P}{Q}{\ct{B}}
    \quad
      \cbpvopa \in \{ -, \div \}
    \justifies
      \cbpvtc{\Gamma}{\calc{z}{V \cbpvopa \numl{n}}{P}}
                     {Q[M \cbpvopa \numl{n}/z]}
                     {\ct{B}}
  \end{prooftree}
  $
  }

  \\[3ex]

  \multicolumn{3}{c}{
  $
  \begin{prooftree}
      \cbpvtv{\Gamma}{V}{M}{\natt}
    \quad
      \cbpvtv{\Gamma}{W}{N}{\natt}
    \quad
      \cbpvtc{\Gamma, z : \natt}{P}{Q}{\ct{B}}
    \quad
      \cbpvopa \in \{ -, \div \}
    \quad
      N \neq \numl{n}
    \justifies
      \cbpvtc{\Gamma}{\calc{z}{V \cbpvopa W}{P}}
                     {Q[M/z]}
                     {\ct{B}}
  \end{prooftree}
  $
  }

  \\[3ex]

  \multicolumn{3}{c}{
  $
  \begin{prooftree}
      \cbpvtv{\Gamma}{V}{M}{\natt}
    \quad
      \cbpvtv{\Gamma}{W}{N}{\natt}
    \quad
      \cbpvtc{\Gamma, z : \natt}{P}{Q}{\ct{B}}
    \quad
    \justifies
      \cbpvtc{\Gamma}{\calc{z}{V \bmod W}{P}}
                     {Q[N - \num{1}/z]}
                     {\ct{B}}
  \end{prooftree}
  $
  }

\end{tabular}
\end{small}

%% file: cbpv-logrel.tex
\renewcommand{\arraystretch}{3}

\[
\begin{aligned}[t]
  \vrel{\numl{n}}{E}{\natt}        
    &\defiff \num{n} \costleq E \\
  \vrel{(V_1, V_2)}{E}{A_1 \times A_2}  
    &\defiff 
      \begin{cases}
        \vrel{V_1}{\pi_1(E)}{A_1} \\
        \vrel{V_2}{\pi_2(E)}{A_2} \\
      \end{cases} \\
  \vrel{\thunk{M}}{E}{U\ct{B}}
    &\defiff \crel{M}{E}{\ct{B}} \\
\end{aligned}
\quad
\begin{aligned}[t]
  \crel{M}{E}{FA} 
    &\defiff  
    \bounded{E_c}\ \Longrightarrow\
        \exists n, V.\ 
          \begin{cases}
            \eval{M}{\return{V}}{n} \\
            \numc{n} \costleq E_c \\
            \vrel{V}{E_p}{A}
          \end{cases} \\
  \crel{M}{E}{A \rightarrow \ct{B}}
    &\defiff
    \forall(\vrel{N}{X}{A}).\ \crel{M\,N}{E\,X}{\ct{B}} \\
  \crel{M}{E}{\ct{B}_1 \with \ct{B}_2}
    &\defiff
     \begin{cases}
          \crel{\pi_1(M)}{\pi_1(E)}{\ct{B}_1} \\
          \crel{\pi_2(M)}{\pi_2(E)}{\ct{B}_2}
        \end{cases}
\end{aligned}
\]

%% file: cbn-embedding.tex
\begin{small}
  \begin{tabular}{|c|c|}
  \hline

  $A$ (PCF type, CBN)
    &
  $\cbnt{A}$ (CBPV+ computation type) \\ 

  \hline

  $\natt$                & $F(\natt)$ \\
  $A_1 \times A_2$       & $\cbnt{A} \with \cbnt{B}$ \\
  $A \rightarrow B$      & $U(\cbnt{A}) \rightarrow \cbnt{B}$ \\

  \hline
\end{tabular}

\vspace{2ex}

\makebox[0pt]{

\begin{tabular}{|c|c|}
  \hline

  $x_1 : A_1, \dots, x_n : A_n \vdash M : A$ 
    &
  $x_1 : U(\cbnt{A_1}), \dots, x_n : U(\cbnt{A_n}) 
          \cturn \cbnt{M} : \cbnt{A}$ \\

  \hline

  $x$               & $\force{x}$ \\
  $\num{n}$         & $\return{\numl{n}}$ \\
  $M \pcfop N$  &
    $\bind{m}{\cbnt{M}}
        \bind{n}{\cbnt{N}}
        {\calc{v}{m \cbpvop n}
          {\return{v}}}$ \\
  $\pif{N}{P}{Q}$   &
    $\bind{n}{\cbnt{N}}{\ifz{n}{\cbnt{P}}{\cbnt{Q}}} $ \\
  $\tuple{M}{N}$    & $\tuple{\charge{\cbnt{M}}}{\charge{\cbnt{N}}}$ \\
  $\pi_i(M)$        & $\pi_i\left(\cbnt{M}\right)$ \\
  $\lambda x.\ M$   & $\lambda x.\ \left(\charge{\cbnt{M}}\right)$ \\
  $M N$             & $\cbnt{M} (\thunk{\cbnt{N}})$ \\
  $\fix{x}{M}$      & $\fix{x}{(\charge{\cbnt{M}})}$ \\
  
  \hline
\end{tabular}
}
\end{small}

%% file: cbv-embedding.tex
\begin{small}

\begin{tabular}{|c|c|}
  \hline

  $A$ (PCF type, CBV)
    &
  $\cbvt{A}$ (CBPV+ value type) \\ 

  \hline

  $\natt$                & $\natt$ \\
  $A_1 \times A_2$       & $\cbvt{A} \times \cbvt{B}$ \\
  $A_1 \rightarrow A_2$  & $U\left(\cbvt{A} \rightarrow F(\cbvt{B})\right)$ \\

  \hline
\end{tabular}

\vspace{2ex}

\makebox[0pt]{

\begin{tabular}{|c|c|}
  \hline

  $x_1 : A_1, \dots, x_n : A_n \vdash M : A$ 
    &
  $x_1 : \cbvt{A_1}, \dots, x_n : \cbvt{A_n} \cturn \cbvt{M} : F(\cbvt{A})$ \\

  \hline

  $x$               & $\return{x}$ \\
  $\num{n}$         & $\return{\numl{n}}$ \\
  $M \pcfop N$  &
    $\bind{m}{\cbvt{M}}
      \bind{n}{\cbvt{N}}
        {\calc{v}{m \cbpvop n}{\return{v}}}$ \\
  $\pif{N}{P}{Q}$   &
    $\bind{n}{\cbvt{N}}
      {\ifz{n}{\cbvt{P}}{\cbvt{Q}}}$ \\
  $\tuple{M}{N}$    & 
    $\bind{x}{\cbvt{M}}
      {\bind{y}{\cbvt{N}}
        {\return (x, y)}}$ \\
  $\pi_i(M)$        &
    $\bind{p}{\cbvt{M}}
      {\charge{\splitprod{p}{x_1}{x_2}{\return{x_i}}}}$ \\
  $\lambda x.\ M$   & $\return \thunk \lambda x.\ \cbvt{M}$ \\
  $M N$             & 
    $\bind{f}{\cbvt{M}}{\bind{x}{\cbvt{N}}{\charge{(\force f) x}}}$ \\
  $\rec{f}{x}{M}$   & 
    $\return \thunk \left(\fix{f} \lambda x. \; \cbvt{M}\right)$ \\
  
  \hline
\end{tabular}
}

\vspace{2ex}


\makebox[0pt]{

\begin{tabular}{|c|c|}
  \hline

  $x_1 : A_1, \dots, x_n : A_n \vdash V : A$ 
    &
  $x_1 : \cbvt{A_1}, \dots, x_n : \cbvt{A_n} \vturn \val{V} : \cbvt{A}$ \\

  \hline

  $x$               & $x$ \\
  $\num{n}$         & $\numl{n}$ \\
  $\lambda x.\ M$   & $\thunk \lambda x.\ \cbvt{M}$ \\
  $\rec{f}{x}{M}$   & 
    $\thunk \left(\fix{f} \lambda x. \; \cbvt{M}\right)$ \\
  
  \hline
\end{tabular}
}

\end{small}

%% file: exp.tex
\newcommand{\reccall}{\prog{exp} (n / \num{2})}
\newcommand{\multiplier}{\text{if} \, \num{1} \, \text{then} \, \num{1} \, \text{else} \, \num{2}}
\newcommand{\constzero}{\text{if} \, \num{1} \, \text{then} \, \num{0} \, \text{else} \, \num{0}}
\[
\begin{array}{rcl}
\textsf{rec}\ \prog{exp}(n) &=&
  \pif n 1 {
    \pcflet {\valbind z {\prog{exp}(n/2)}, \valbind y {
      \pif {n\bmod 2} 1 2
    }} {
      z*z*y
    }
  }
\\ \\
\complexity{\prog{exp}} &=&
\big\langle
0,
\fix
  {\prog{exp}}{\lambda n.
    \pif*{n}{\langle{\zeroc}, {\num{1}}\rangle}
                {\langle
                 \begin{aligned}[t]
                 &{\numc{3} \plusc \costof{(\reccall)}}, \\
                 &{\potentialof{(\reccall)} * \potentialof{(\reccall)} *
                 (\multiplier)}\rangle\big\rangle
                 \end{aligned}
                }
  }
\end{array}
\]

%% file: ind-types.tex
\begin{small}

\textbf{CBV \pcf}


\[
  \textbf{Types} \quad
    A\ ::=\ \dots \mid  \pcflist A
  \qquad
  \textbf{Canonical forms (CBV)} \quad
    V, W\ ::=\ \dots \mid  \pcfnil \mid  \pcfcons(V, W)
\]

\vspace{1ex}


\begin{tabular}{c}

  \begin{prooftree}
      \phantom{no premise} 
    \justifies
      \Gamma \vdash \pcfnil : \pcflist A
  \end{prooftree}

  \quad

  \begin{prooftree}
      \Gamma \vdash M : A 
    \quad 
      \Gamma \vdash N : \pcflist A 
    \justifies
      \Gamma \vdash \pcfcons(M, N) : \pcflist A
  \end{prooftree}

  \quad

  \begin{prooftree}
      \Gamma \vdash N : \pcflist A
    \quad
      \Gamma \vdash M_{\pcfnil} : B
    \quad
      \Gamma, x : A, xs : \pcflist A \vdash M_\pcfcons : B
    \justifies
      \Gamma \vdash \pcflcase n {M_\pcfnil} x {xs} {M_\pcfcons} : B
  \end{prooftree}

  \\\\


  \begin{tabular}{c}

  \begin{prooftree}
      \phantom{no premise}
    \justifies
      \evalpcf{\pcfnil}{\pcfnil}{0}
  \end{prooftree}

  \\ \\

  \begin{prooftree}
      \evalpcf{M}{V}{m}
    \quad
      \evalpcf{N}{W}{n}
    \justifies
      \evalpcf{\pcfcons(M, N)}{\pcfcons(V, W)}{m+n}
  \end{prooftree}

  \end{tabular}

  \begin{tabular}{c}

  \begin{prooftree}
      \evalpcf{N}{\pcfnil}{n}
    \quad
      \evalpcf{M_\pcfnil}{Z}{m}
    \justifies
      \evalpcf{\pcflcase N {M_\pcfnil} x {xs} {M_\pcfcons}}{Z}{n+m}
  \end{prooftree}

  \\ \\

  \begin{prooftree}
      \evalpcf{N}{\pcfcons(V, W)}{n}
    \quad
      \evalpcf{M_\pcfcons[V/x, W/xs]}{Z}{m}
    \justifies
      \evalpcf{\pcflcase N {M_\pcfnil} x {xs} {M_\pcfcons}}{Z}{n+m}
  \end{prooftree}

  \end{tabular}

\end{tabular}

\vspace{3ex}

\textbf{\cbpv}


\begin{align*}
  \textbf{Value types} \quad
    A ::= \dots \mid  \cbpvlist A
\end{align*}

\vspace{2ex}


\begin{tabular}{c}

  \begin{prooftree}
      \phantom{no premise} 
    \justifies
      \Gamma \vturn \cbpvnil : \cbpvlist A
  \end{prooftree}

  \quad

  \begin{prooftree}
      \Gamma \vturn M : A 
    \quad 
      \Gamma \vturn N : \cbpvlist A 
    \justifies
      \Gamma \vturn \cbpvcons(M, N) : \cbpvlist A
  \end{prooftree}

  \\ \\

  \begin{prooftree}
      \Gamma \vturn N : \cbpvlist A
    \quad
      \Gamma \cturn M_{\cbpvnil} : \ct{B}
    \quad
      \Gamma, x : A, xs : \cbpvlist A \cturn M_\cbpvcons : \ct{B}
    \justifies
      \Gamma \cturn \cbpvlcase N {M_\cbpvnil} x {xs} {M_\cbpvcons} : \ct{B}
  \end{prooftree}

  \\\\

  \quad


  \begin{prooftree}
      \eval{M_\pcfnil}{T}{n}
    \justifies
      \eval{\cbpvlcase \cbpvnil {M_\cbpvnil} x {xs} {M_\cbpvcons}}{T}{n}
  \end{prooftree}

  \quad

  \begin{prooftree}
      \eval{M_\cbpvcons[V/x, W/xs]}{T}{n}
    \justifies
      \eval{\cbpvlcase {\cbpvcons(V, W)} {M_\cbpvnil} x {xs} {M_\cbpvcons}}{T}{n}
  \end{prooftree}

  \\

\end{tabular}

\vspace{2ex}
\textbf{CBV \pcf\ translation to \cbpv}
\vspace{2ex}

\begin{tabular}{|c|c|}
  \hline

  $A$ (PCF type, CBV)
    &
  $\cbvt{A}$ (CBPV+ value type) \\ 

  \hline

  $\pcflist{A}$           & $\cbpvlist{\cbvt{A}}$ \\

  \hline
\end{tabular}

\begin{tabular}{|c|c|}
  \hline

  $x_1 : A_1, \dots, x_n : A_n \vdash M : A$ 
    &
  $x_1 : \cbvt{A_1}, \dots, x_n : \cbvt{A_n} \cturn \cbvt{M} : F(\cbvt{A})$ \\

  \hline

  $\pcfnil$     & $\return{\cbpvnil}$ \\
  $\pcfcons(M, N)$  &
    $\bind{m}{\cbvt{M}}
      \bind{n}{\cbvt{N}}
        {\return{\cbpvcons(m, n)}}$ \\
  $\pcflcase M {M_\pcfnil} x {xs} {M_\pcfcons}$   &
    $\bind{m}{\cbvt{M}}
      {\cbpvlcase m {\cbvt{M_\pcfnil}} x {xs} {\cbvt{M_\pcfcons}}}$ \\
  \hline
\end{tabular}

\vspace{2ex}
\textbf{\cbpv\ recurrence extraction}
\vspace{2ex}

\begin{tabular}{c}

  $\potential{\cbpvlist A} \defeq \natt$

  \qquad 

  \begin{prooftree}
      \phantom{no premise}
    \justifies
      \cbpvtv{\Gamma}{\cbpvnil}{\num 0}{\cbpvlist A}
  \end{prooftree}

  \qquad

  \begin{prooftree}
      \cbpvtv{\Gamma}{W}{P}{\cbpvlist A}
    \justifies
      \cbpvtv{\Gamma}{\cbpvcons(V, W)}{P+\num 1}{\cbpvlist A}
  \end{prooftree}

  \\\\

  \begin{prooftree}
      \cbpvtv{\Gamma}{V}{W}{\cbpvlist A}
    \quad
      \cbpvtc{\Gamma}
             {M_\cbpvnil}
             {P}
             {\ct{B}}
    \quad
      \cbpvtc{\Gamma, x : A, xs : \cbpvlist A}
              {M_\cbpvcons}
              {Q}
              {\ct B}
    \justifies
      \cbpvtc{\Gamma}
             {\cbpvlcase V {M_{\cbpvnil}} x {xs} {M_{\cbpvcons}}}
             {\pif{W}{P}{Q[\fix x x/x, W - \num{1}/xs]}}
             {\ct{B}}
  \end{prooftree}

\end{tabular}

\[
  \vrel{\cbpvnil}{E}{\cbpvlist A}
    \defiff
      \num 0 \costleq E
  \qquad\qquad
  \vrel{\cbpvcons(V, W)}{E}{\cbpvlist A}
    \defiff
       \num 1 \costleq E \land \vrel{W}{E - \num{1}}{\cbpvlist A}
\]

\end{small}

%% file: merge-sort.tex
\[
\begin{array}{rcl}
\textsf{rec}\ \prog{sort}(xs) &=& \\
\multicolumn{3}{l}{
\pcflcase* 
  {xs} 
  {\pcfnil} 
  {y} 
  {ys} 
  {\pcflcase*
     {ys}
     {\pcfcons(y, \pcfnil)}
     {z}
     {zs}
     {
     \begin{aligned}[t]
     &\textsf{let}\ q = \prog{divide}(\pcfcons(y, ys))\ \textsf{in} \\
     &\prog{merge}(\prog{sort}(\pi_1\,q), \prog{sort}(\pi_2\, q))
     \end{aligned}
     }
  }
}
\\[2\baselineskip]
\complexity{\prog{sort}} &=&
\langle
0,
{
  \fix
    {\prog{sort}}{\lambda xs.
    {\pif*{xs}
          {\tuple 0 0}
          {\pif{xs-1}
               {\tuple 0 1 }
               {F(xs-1)\rangle}
          }
    }}
} \\
F(xs') &=&
  (\numc 2 \plusc (\prog{divide}(xs'+1))_c) +_c E((\prog{divide}(xs'+1))_p) \\
E(q) &=&
  (\numc 5 \plusc (\prog{sort}(\pi_1\,q))_c  \plusc  (\prog{sort}(\pi_2\,q))_c) +_c
     \prog{merge}\tuple{(\prog{sort}(\pi_1\,q))_p}
                {(\prog{sort}(\pi_2\,q))_p}
\end{array}
\]

%% file: acks.tex
This material is based upon work supported by the
\grantsponsor{afosr}{Air Force Office of Scientific Research}{}
under award number \grantnum{afosr}{FA9550-16-1-0292}. Any
opinions, finding, and conclusions or recommendations expressed in
this material are those of the author(s) and do not necessarily
reflect the views of the United States Air Force. 
Additionally, this material is based upon work supported by the 
\grantsponsor{nsf}
            {National Science Foundation}
            {https://nsf.gov}
under Grant Number~\grantnum{nsf}{1618203}.

%% file: popl2020.bbl

%% file: supp.tex
\section{An Assortment of Lemmas}

The following results are needed in the proof of the bounding
theorem.

\begin{theorem}[Determinacy, Theorem~\ref{theorem:PCF-determinacy}]
  There is a most one pair $n$, $V$ such that $\evalpcf{M}{V}{n}$.
\end{theorem} 

\begin{lemma}[Bounded terms are a lower set, Lemma~\ref{lemma:bounded}]
  $
    \bounded{M}\ 
      \land\ 
    N \costleq M : A\ 
      \Longrightarrow\ 
    \bounded{N}
  $
\end{lemma}

\begin{proof}
  By induction on $N \costleq M : A$.
\end{proof}

\begin{lemma}[Strictness of the Free Algebra]
  \label{lemma:strict-free}
  For closed $L : \cost$ and $E : \complexity{FA}$,
  $\bounded{\left(L \algp{FA} E\right)_c}$ implies that
  $\bounded{L}$ and $\bounded{E_c}$.
\end{lemma}
\begin{proof}
  We have that $
    L \plusc E_c
      \costleq \pi_1\left(\tuple{L \plusc E_c}{E_p}\right)
      \aequiv  \left(L \algp{FA} E\right)_c
  $. Hence, $\bounded{L \plusc E_c}$ by Lemma \ref{lemma:bounded}.
  It follows that $\bounded{L}$ and $\bounded{E_c}$, as $\plusc$
  is strict.
\end{proof}

\begin{lemma}[Algebra Monotonicity, Lemma~\ref{lemma:algebra-monotone}]
  If $M : \complexity{\ct{B}}$, \[
    E \costleq E' : \cost
      \quad\Longrightarrow\quad
    E \algp{\ct{B}} M \costleq E' \algp{\ct{B}} M
  \]
\end{lemma}
\begin{proof}
  By induction on $\ct{B}$. Use the rule
  $(\textsf{\hyperlink{ctx}{\textsf{ctx}}})$, after noticing that
  all the contexts involved are monotone.
\end{proof}

\begin{theorem}[Compactness, Theorem~\ref{theorem:compactness}]
  Suppose $x : A \vdash M : A$ and $z : A \vdash \ectx{C}[z] :
  \cost$ for an eliminative context $\ectx{E}$. Then \[
    \bounded{\ectx{E}[\fix{x}{M}]}\
      \Longrightarrow\
    \exists n \geq 0.\ \bounded{\ectx{E}[\fixn{x}{n}{M}]}
  \]
\end{theorem}

\section{The bounding relation}
\label{sec:bounding-relation-proofs}


Figure \ref{fig:cbpv-logrel} presents a logical relation that
relates terms of the intermediate language (\cbpv{}) to terms of
the recurrence language (\pcfc{}). The relation \[
  \crel{M}{E}{\ct{B}}
\] intuitively means that the closed recurrence $E$ is an upper
bound on the cost incurred when evaluating the closed program $M$.

\begin{figure}
  \caption{Costless weak head reduction for CBPV}
  \centering
  \label{fig:wh}
  \input{wh}
\end{figure}

In order to prove the bounding theorem, we will need an auxiliary
notion of small-step weak head reduction, the steps of which do
not incur any cost. We denote it by $\wh{0}$.  This
\emph{costless} weak head reduction is defined by the rules of
Figure \ref{fig:wh}. Most of the rules are standard. The only
unusual ones are the congruence rule, which allows weak head
reduction within any eliminative context, and the two rules that
permute charges with application and projection. It is easy to
show that

\begin{lemma} \hfill
  \label{lemma:wheval}
  \begin{enumerate}
    \item
      $
        \eval{M'}{T}{n}\
          \land\
        M \wh{0} M'
          \Longrightarrow\
        \eval{M}{T}{n}
      $
    \item
      $
        \eval{M}{T}{n}\
          \land\
        M \wh{0} M'
          \Longrightarrow\
        \eval{M'}{T}{n}
      $
  \end{enumerate}
\end{lemma}
\begin{proof}
  By induction on the definition of $\wh{0}$, as given in Figure
  \ref{fig:wh}. For the case of congruence rules and the commuting
  conversion that permutes the unit charge, perform a subsidiary
  induction on eliminative contexts for CBPV.
\end{proof}

\begin{lemma}[Bound weakening, Lemma~\ref{lemma:bound}] \hfill
  \begin{enumerate}
    \item
      $
      \vrel{V}{E}{A}\ \land\ E \costleq E' : \potential{A}\
        \Longrightarrow\
      \vrel{V}{E'}{A}
      $
    \item
      $
      \crel{M}{E}{\ct{B}}\ \land\ E \costleq E' : \complexity{\ct{B}}\
        \Longrightarrow\
      \crel{M}{E'}{\ct{B}}
      $
  \end{enumerate}
\end{lemma}

\begin{proof}
  By induction on types.
  \begin{indproof}
    \indcase{$\natt$}
      
      $\num{n} \costleq E :
      \natt$, so---by transitivity of $\costleq$---we obtain
      $\num{n} \costleq E' : \natt$, and hence
      $\vrel{V}{E'}{\natt}$.

    \indcase{$A_1 \times A_2$}

      Write $V \aequiv (V_1, V_2)$. Then
      $\vrel{V_i}{\pi_i(E)}{A_i}$ for $i \in \{1, 2\}$. As $E
      \costleq E'$ implies $\pi_i(E) \costleq \pi_i(E')$ by
      $(\textsf{\hyperlink{ctx}{\textsf{ctx}}})$, the IH yields
      that $\vrel{V_i}{\pi_i(E')}{A_i}$ for $i \in \{1, 2\}$, and
      hence the result.

    \indcase{$U\ct{B}$}

      Write $V \aequiv \thunk M$. Then the assumption implies that
      $\crel{M}{E}{\ct{B}}$. From that, the IH yields that
      $\crel{M}{E'}{\ct{B}}$, and hence the result.

    \indcase{$FA$}

      As $E \costleq E' : \cost \times \potential{A}$, we deduce
      by $(\textsf{\hyperlink{ctx}{\textsf{ctx}}})$ that that
      $\pi_i(E) \costleq \pi_i(E')$ for $i \in \{ 1, 2 \}$.

      If $\pi_1(E') \aequiv \bounded{E'_c}$, we have
      by Lemma \ref{lemma:bounded} that $\pi_1(E) \aequiv
      \bounded{E_c}$ as well. Thus, we have $m$ and $V$ such that
      $\eval{M}{\return{V}}{m}$, and \begin{align*}
        &\numc{m} \costleq E_c \aequiv \pi_1(E)
        &\vrel{V}{E_p}{A} \aequiv \pi_2(E)
      \end{align*} As $\pi_i(E) \costleq \pi_i(E')$, we can use
      the transitivity of $\costleq$ and the IH for $A$ to obtain
      \begin{align*}
        &\numc{m} \costleq \pi_1(E') \aequiv E'_c
        &\vrel{V}{\pi_2(E')}{A}     \aequiv E'_p
      \end{align*} and hence we have the result.

    \indcase{$A \rightarrow \ct{B}$}

      Suppose $\vrel{V}{X}{A}$. By the fact $\crel{M}{E}{A
      \rightarrow \ct{B}}$, we know that $\crel{M V}{E
      X}{\ct{B}}$. As $E \costleq E'$, we deduce that $E X
      \costleq E' X$ by
      $(\textsf{\hyperlink{ctx}{\textsf{ctx}}})$.  By the IH, it
      follows that $\crel{M V}{E' X}{\ct{B}}$, whence the result.

    \indcase{$\ct{B}_1 \with \ct{B}_2$}

      We have that $\crel{\pi_i(M)}{\pi_i(E)}{\ct{B}_i}$ for $i
      \in \{1, 2\}$. As $E \leq E'$, it is by
      $(\textsf{\hyperlink{ctx}{\textsf{ctx}}})$ the case that
      $\pi_i(E) \leq \pi_i(E')$ for $i \in \{1, 2\}$. Hence, by
      the IH we have that $\crel{\pi_i(M)}{\pi_i(E')}{\ct{B}_i}$
      for $i \in \{1, 2\}$, and hence the result.

  \end{indproof}
\end{proof}

\begin{lemma}[Head expansion, Lemma~\ref{lemma:headexp}(i)]
  $
    \crel{M'}{E}{\ct{B}}\
      \land\
    M \wh{0} M'
      \Longrightarrow
    \crel{M}{E}{\ct{B}}
  $
\end{lemma}
\begin{proof}
  By induction on $\ct{B}$.

  \begin{indproof}

    \indcase{$FA$}

      Suppose $\bounded{E_c}$. Then, as $\crel{M'}{E}{FA}$, there
      exist $n$ and $V$ such that \begin{align}
        \label{eq:wheval}
          \eval{M'}{\return{V}}{n} \\
        \label{eq:whcost}
          \numc{n} \costleq E_c \\
        \label{eq:whvalue}
          \vrel{V}{E_p}{A}
      \end{align} By \eqref{eq:wheval}, $M \wh{0} M'$, and Lemma
      \ref{lemma:wheval}, \begin{equation}
        \label{eq:wheval2}
          \eval{M}{\return{V}}{n}
      \end{equation} Then, by \eqref{eq:wheval2},
      \eqref{eq:whcost} and \eqref{eq:whvalue}, we obtain
      $\crel{M}{E}{FA}$.

    \indcase{$A \rightarrow \ct{B}$}

      Suppose $\vrel{V}{X}{A}$. As $\crel{M'}{E}{A
      \rightarrow \ct{B}}$ we know that $\crel{M' V}{E
      X}{\ct{B}}$. But as $M \wh{0} M'$ we have that $MV
      \wh{0} M'V$. By the IH, we deduce that $\crel{M V}{E
      X}{\ct{B}}$, whence the result.

    \indcase{$\ct{B}_1 \with \ct{B}_2$}

      By the assumption we have that
      $\crel{\pi_i(M')}{\pi_i(E)}{\ct{B}_i}$ for $i \in \{1, 2\}$.
      As $M \wh{0} M'$, it follows that $\pi_i(M) \wh{0}
      \pi_i(M')$. Hence, by the IH for $\ct{B}_i$,
      $\crel{\pi_i(M)}{\pi_i(E)}{\ct{B}_i}$ for $i \in \{1, 2\}$,
      whence the result.

  \end{indproof}
\end{proof}

\begin{lemma}[Head reduction, Lemma~\ref{lemma:headexp}(ii)]
  $
    \crel{M}{E}{\ct{B}}\
      \land\
    M \wh{0} M'
      \Longrightarrow
    \crel{M'}{E}{\ct{B}}
  $
\end{lemma}
\begin{proof}
  Similar.
\end{proof}

\begin{lemma}[Unit charge, Lemma~\ref{lemma:unit}]
  $
    \crel{M}{E}{\ct{B}}\
      \Longrightarrow
    \crel{\charge{M}}{\unitc +_{\ct{B}} E}{\ct{B}}
  $
\end{lemma}
\begin{proof}

  By induction on $\ct{B}$.

  \begin{indproof}

    \indcase{$FA$}

    Suppose that $\bounded{(\unitc \algp{FA} E)_c}$. Then, by
    Lemma \ref{lemma:strict-free}, we know that $\bounded{E_c}$.
    It follows that there exist $V$ and $n$ such that
    \begin{align}
        \label{eq:chg-eval}
          \eval{M}{\return{V}}{n} \\
        \label{eq:chg-cost}
          \numc{n} \costleq E_c \\
        \label{eq:chg-value}
          \vrel{V}{E_p}{A}
      \end{align} By \eqref{eq:chg-eval} and the definition of
      the big-step semantics of CBPV, we have
      \begin{equation}
        \label{eq:chg-eval'}
          \eval{\charge{M}}{\return{V}}{n+1}
      \end{equation} By \eqref{eq:chg-cost},
      $(\textsf{\hyperlink{ctx}{\textsf{ctx}}})$, and
      $(\hyperlink{betaprod}{\times_\beta})$,
      we get that
      \begin{equation}
        \label{eq:chg-cost'}
        \numc{n+1}
          \aequiv \unitc \plusc \numc{n}
          \costleq \unitc \plusc E_c
          \costleq \pi_1\left(\tuple{\unitc \plusc E_c}{E_p}\right)
          \aequiv (\unitc \algp{FA} E)_c
      \end{equation} By
      $(\hyperlink{betaprod}{\times_\beta})$, we have that
      \begin{equation}
        E_p
          \costleq \pi_2\left(\tuple{\unitc \plusc E_c}{E_p}\right)
          \aequiv (\unitc \algp{FA} E)_p
      \end{equation} 
      and so by \eqref{eq:chg-value} and bound weakening,
      \begin{equation}
      \vrel{V}{(\unitc \algp{FA} E)_p}{A}.
      \label{eq:chg-val-bound}
      \end{equation}
      Hence, by \eqref{eq:chg-eval'},
      \eqref{eq:chg-cost'}, and \eqref{eq:chg-val-bound},
      $\crel{\charge{M}}{\unitc \algp{FA} E}{FA}$.

    \indcase{$A \rightarrow \ct{B}$}

    Suppose $\vrel{N}{X}{A}$. Then, we have by assumption that
    $\crel{MN}{EX}{\ct{B}}$. By the IH, we may infer that $
      \crel{\charge{(MN)}}
           {\unitc \algp{\ct{B}} EX}
           {\ct{B}}
    $, and hence $
      \crel{(\charge{M})N}
           {\unitc \algp{\ct{B}} EX}
           {\ct{B}}
    $ by head expansion. But \[
      \unitc \algp{\ct{B}} EX
        \costleq \left(\lambda a.\ \unitc \algp{\ct{B}} E(a)\right)X
        \aequiv (\unitc \algp{A \rightarrow \ct{B}} E)X
    \] so we obtain the result by bound weakening.

    \indcase{$\ct{B}_1 \with \ct{B}_2$}

    By the IH, we have $
      \crel{\charge{\pi_i(M)}}
           {\unitc \algp{\ct{B}_i} \pi_i(E)}
           {\ct{B}_i}
    $ for $i \in \{1, 2\}$. By head expansion, $
      \crel{\pi_i\left(\charge{M}\right)}
           {\unitc \algp{\ct{B}_i} \pi_i(E)}
           {\ct{B}_i}
    $. But \[
      \unitc \algp{\ct{B}_i} \pi_i(E)
        \costleq \pi_i\left(\tuple{\unitc \algp{\ct{B}_1} \pi_1(E)}
                                  {\unitc \algp{\ct{B}_2} \pi_2(E)}\right)
        \aequiv  \pi_i(\unitc \algp{\ct{B}_1 \with \ct{B}_2} E)
    \] by $(\hyperlink{betaprod}{\times_\beta})$. Hence, by bound
    weakening, \[
      \crel{\pi_i\left(\charge{M}\right)}
           {\pi_i(\unitc \algp{\ct{B}_1 \with \ct{B}_2} E)}
           {\ct{B}_1 \with \ct{B}_2}
    \] for $i \in \{1, 2\}$, whence the result.

  \end{indproof}
\end{proof}

\begin{lemma}[Bind expansion]
  \label{lemma:bind-expansion}
  \[
    \crel{N[V/x]}{E}{\ct{B}}\
      \land\
    \eval{M}{\return{V}}{n}\
      \Longrightarrow
    \crel{\bind{x}{M}{N}}{\numc{n} +_{\ct{B}} E}{\ct{B}}
  \]
\end{lemma}
\begin{proof}
  Recall the definition of eliminative contexts for CBPV, as
  presented in Figure \ref{fig:wh}. We first show that for any
  eliminative context $\ectx{E}$, \[
    \eval{M}{\return{V}}{m}\
      \land\
    \eval{\ectx{E}[N[V/x]]}{T}{n}\
      \Longrightarrow\
    \eval{\ectx{E}[\bind{x}{M}{N}]}{T}{m+n}
  \] This can be shown straightforwardly by induction on
  $\ectx{E}$. We then strengthen the IH: we prove that for any
  eliminative context $\ectx{E}$ it is the case that \[
    \crel{\ectx{E}\left[N[V/x]\right]}{E}{\ct{B}}\
      \land\
    \eval{M}{\return{V}}{n}\
      \Longrightarrow\
    \crel{\ectx{E}\left[\bind{x}{M}{N}\right]}{\numc{n} +_{\ct{B}} E}{\ct{B}}
  \] The result follows by taking $\ectx{E} \defeq []$. We proceed
  by induction on $\ct{B}$.

  \begin{indproof}
    \indcase{$FA$}

      If $\bounded{(\numc{n} +_{FA} E)_c}$, then by Lemma
      \ref{lemma:strict-free} it follows that $\bounded{E_c}$.
      Thus, there exist $W$ and $m$ such that
      $\eval{\ectx{E}[N[V/x]]}{\return{W}}{m}$, and \begin{align*}
        &\numc{m} \costleq E_c \\
        &\vrel{W}{E_p}{A}
      \end{align*} As $\eval{M}{\return{V}}{n}$, the
      aforementioned lemma shows that
      $\eval{\ectx{E}[\bind{x}{M}{N}]}{\return{W}}{n + m}$.
      Moreover, we have \[
        \numc{n + m}
          \costleq \numc{n} \plusc E_c
          \costleq (\numc{n} +_{FA} E)_c
      \] (The first inequality can be shown by induction on $n$,
      and requires the rules $(\hyperlink{zero}{\textsf{zero}})$,
      $(\hyperlink{assoc}{\textsf{assoc}})$ and
      $(\hyperlink{ctx}{\textsf{ctx}})$.) Moreover, as $E_p
      \costleq (\numc{n} +_{FA} E)_p$, we have by bound weakening
      at $A$ that \[
        \vrel{W}{(\numc{n} +_{FA} E)_p}{A} 
      \] and hence that $\crel{\ectx{E}[\bind{x}{M}{N}]}
      {\numc{n} +_{FA} E}{FA}$.

    \indcase{$A \rightarrow \ct{B}$}

      Suppose $\vrel{U}{X}{A}$. Then, it follows by assumption
      that \[
        \crel{\ectx{E}[N[V/x]]U}{EX}{\ct{B}}
      \] The LHS is $\alpha$-equivalent to $(\ectx{E}U)[N[V/x]]$,
      so by the IH we have \[
        \crel{(\ectx{E}U)[\bind{x}{M}{N}]}
             {\numc{n} +_{\ct{B}} EX}{\ct{B}}
      \] But this LHS is $\alpha$-equivalent to
      $(\ectx{E}[\bind{x}{M}{N}])U$, and \[
        \numc{n} +_{\ct{B}} EX
          \costleq (\numc{n} +_{A \rightarrow \ct{B}} E)X
      \] so the result follows by bound weakening.

    \indcase{$\ct{B}_1 \with \ct{B}_2$}

    We have that \[
      \crel{\pi_i(\ectx{E}[N[V/x]])}
           {\pi_i(E)}
           {\ct{B}_i}
    \] for $i \in \{1, 2\}$. But the LHS is $\alpha$-equivalent to
    $(\pi_i(\ectx{E}))[N[V/x]]$, so by the IH we have that \[
      \crel{\pi_i(\ectx{E})[\bind{x}{M}{N}]}
           {\numc{n} +_{\ct{B}_i} \pi_i(E)}
           {\ct{B}_i}
    \] for $i \in \{1, 2\}$.  But this LHS is $\alpha$-equivalent
    to $\pi_i(\ectx{E}[\bind{x}{M}{N}])$, and \[
      {\numc{n} +_{\ct{B}_i} \pi_i(E)}
        \costleq \pi_i(\numc{n} +_{\ct{B}_1 \with \ct{B}_2} E)
    \] so we obtain the result.
  \end{indproof}
\end{proof}

\begin{lemma}[Bounds for Recursion, Lemma~\ref{lemma:recbounds}] \hfill
  \begin{enumerate}[label=(\roman*)]
    \item (Infinity)
      If $M : \ct{B}$, $E : \complexity{\ct{B}}$, and
      $\unbounded{E}$ then $\crel{M}{E}{\ct{B}}$.
    \item (Infinity-Algebra)
      If $M : \ct{B}$, $E : \complexity{\ct{B}}$, $L : \cost$, and
      $\unbounded{L}$, then $\crel{M}{L \algp{\ct{B}}
      E}{\ct{B}}$.
    \item (Fixed Point Induction)
      If $\cturn M : \ct{B}$ then it is the case that \[
          \left(\forall n \geq 0.\
            \crel{M}{\fixn{x}{n}{E}}{\ct{B}}\right)
        \quad\Longrightarrow\quad
          \crel{M}{\fix{x}{E}}{\ct{B}}
      \]
  \end{enumerate}
\end{lemma}

\begin{proof} \hfill
  \begin{enumerate}[label=(\roman*)]
    \item 
      By induction on $\ct{B}$.
      
      \begin{indproof}

        \indcase{$FA$}
          Trivial.

        \indcase{$A \rightarrow \ct{B}$}

        Then, we need to show that for any $\vrel{N}{X}{A}$ we
        have $
          \crel{M\,N}
               {E\,X}
               {\ct{B}}
        $. But if $\unbounded{E}$ then $\unbounded{E\,X}$ as well,
        so we may invoke the IH.

        \indcase{$\ct{B}_1 \with \ct{B}_2$}

        Similar.

      \end{indproof}

    \item 
      By induction on $\ct{B}$.

      \begin{indproof}

        \indcase{$FA$}

        If $\bounded{(L +_{FA} N)_c}$, we have by Lemma
        \ref{lemma:strict-free} that $\bounded{L}$, which is
        impossible. Therefore $\unbounded{(L +_{FA} N)_c}$, and
        there is nothing to prove.
        
        \indcase{$A \rightarrow \ct{B}$}

        Suppose $\vrel{W}{X}{A}$. Then, by the IH, $
          \crel{M W}{L +_{\ct{B}} N X}{\ct{B}}
        $. But $
          L +_{\ct{B}} N X
            \costleq (L +_{A \rightarrow \ct{B}} N)X
        $ so we obtain $\crel{M W}{(L +_{A \rightarrow \ct{B}}
        N)X}{\ct{B}}$ by bound weakening (Lemma
        \ref{lemma:bound}).

        \indcase{$\ct{B}_1 \with \ct{B}_2$}
        
        By the IH, we have that $
          \crel{\pi_i(M)}
               {L +_{\ct{B}_i} \pi_i(N)}
               {\ct{B}_i}
        $ for $i \in \{1, 2\}$. However, $
            L +_{\ct{B}_i} \pi_i(N)
             \costleq \pi_i(L +_{\ct{B}_1 \with \ct{B}_2} N)
        $ Hence, we obtain $
          \crel{\pi_i(M)}
               {\pi_i(L +_{\ct{B}_1 \with \ct{B}_2} N)}
               {\ct{B}_i}
        $ for $i \in \{1, 2\}$ by bound weakening (Lemma
        \ref{lemma:bound}).

      \end{indproof}

    \item
      First, we strengthen the IH: we show instead that
      \begin{alignat*}{3}
          &\left(\forall n \geq 0.\
            \crel{M}{\ectx{E}[\fixn{x}{n}{E}]}{\ct{B}}
          \right) 
        &&\Longrightarrow\
          &&\crel{M}{\ectx{E}[\fix{x}{E}]}{\ct{B}}
        \\
          &\left(\forall n \geq 0.\
            \vrel{V}{\ectx{E}[\fixn{x}{n}{E}]}{A}
          \right) 
        &&\Longrightarrow\
          &&\vrel{V}{\ectx{E}[\fix{x}{E}]}{A}
      \end{alignat*} for any PCF eliminative context $\ectx{E}$
      (cf. Fig. \ref{fig:pcf-inequality}). We obtain the result by
      setting $\ectx{E} \defeq []$.  To show the stronger IH, we
      proceed by induction on types.

      \begin{indproof}
        \indcase{$\natt$}

        Let $V \aequiv \num{n}$. Then, for any $m \geq 0$ we
        have that $\num{n} \costleq \ectx{E}[\fixn{x}{m}{E}]$. By
        the rule $(\textsf{\hyperlink{cpind}{cpind}})$, it follows
        that $ \num{n} \costleq \ectx{E}[\fix{x}{E}]$, whence the
        result.

        \indcase{$A_1 \times A_2$}

        Given any $m \geq 0$, the assumption implies that \[
          \vrel{V_i}
               {\pi_i(\ectx{E}[\fixn{x}{m}{E}])}
               {A_i}
            \aequiv
               {\pi_i(\ectx{E})[\fixn{x}{m}{E}]}
        \] for $i \in \{1, 2\}$. By the IH for $A_i$, this
        implies that \[
          \vrel{V_i}
               {\pi_i(\ectx{E})[\fix{x}{E}]}
               {A_i}
            \aequiv
               {\pi_i(\ectx{E}[\fix{x}{E}])}
        \] for $i \in \{1, 2\}$, and hence the result.

        \indcase{$U\ct{B}$}

        Let $V \aequiv \thunk{M}$. The assumption implies that
        $\crel{M}{\ectx{E}[\fixn{x}{m}{E}]}{\ct{B}}$ for any $m
        \geq 0$. By the IH for $\ct{B}$, we obtain
        $\crel{M}{\ectx{E}[\fix{x}{E}]}{\ct{B}}$ and hence the
        result.

        \indcase{$FA$}

        First, notice that all PCF eliminative contexts are, in
        fact, monotone. Hence, by the rules
        $(\textsf{\hyperlink{rat}{rat}})$ and
        $(\textsf{\hyperlink{ctx}{ctx}})$, \begin{equation}
          \label{eq:ratchain}
          \dots
            \costleq
          \ectx{E}[\fixn{x}{2}{E}]
            \costleq
          \ectx{E}[\fixn{x}{1}{E}]
            \costleq
          \ectx{E}[\fixn{x}{0}{E}]
        \end{equation} Suppose
        $\bounded{\ectx{E}[\fix{x}{E}]_c}{}{}$. Then---by
        compactness (Theorem \ref{theorem:compactness})---there
        exists some $j \geq 0$ such that
        $\bounded{\ectx{E}[\fixn{x}{j}{E}]_c}$. Hence, by
        \eqref{eq:ratchain}, the rule 
        $(\textsf{\hyperlink{rat}{rat}})$, and Lemma
        \ref{lemma:bounded}, \begin{equation}
          \label{eq:compact-chain}
          \forall k \geq j.\
            \bounded{\ectx{E}[\fixn{x}{k}{E}_c]}
        \end{equation} By the definition of $\crel{}{}{FA}$ and
        the determinacy of evaluation for PCF (Theorem
        \ref{theorem:PCF-determinacy}), this implies that there
        exist unique $n$ and $V$ such that for all $k \geq j$
        \begin{align}
          \label{eq:cpt-eval}
            \eval{M}{\return{V}}{n} \\
          \label{eq:cpt-cost}
            \numc{n} \costleq \ectx{E}[\fixn{x}{k}{E}]_c \\
          \label{eq:cpt-value}
            \vrel{V}{\ectx{E}[\fixn{x}{k}{E}]_p}{A}
        \end{align}

        Moreover, by \eqref{eq:cpt-cost}, \eqref{eq:ratchain}, and
        $(\textsf{\hyperlink{trans}{trans}})$, we can in fact
        establish that $\numc{n} \costleq
        \ectx{E}[\fixn{x}{k}{E}]_c \aequiv
        \pi_1(\ectx{E})[\fixn{x}{k}{E}]$ for all $k \geq 0$.
        Hence, \begin{equation}
          \label{eq:cpt-cost-final}
              \numc{n} \costleq \ectx{E}[\fix{x}{E}]_c
        \end{equation} by $(\textsf{\hyperlink{cpind}{cpind}})$.

        Similarly, by \eqref{eq:cpt-value}, \eqref{eq:ratchain}
        and bound weakening (Lemma \ref{lemma:bound}), we get
        that $\vrel{V}{\ectx{E}[\fixn{x}{k}{E}]_p}{A} \aequiv
        \pi_2(\ectx{E})[\fixn{x}{k}{E}]$ for all $k \geq 0$.
        Hence, by the IH \begin{equation}
          \label{eq:cpt-value-final}
            \vrel{V}{\ectx{E}[\fix{x}{E}]_p}{A}
        \end{equation}

        \noindent In conclusion, by \eqref{eq:cpt-eval},
        \eqref{eq:cpt-cost-final} and \eqref{eq:cpt-value-final},
        we obtain $\crel{M}{\ectx{E}[\fix{x}{E}]}{F A}$.

        \indcase{$A \rightarrow \ct{B}$}

        Let $\vrel{N}{X}{A}$. Given any $m \geq 0$, it is the case
        by the assumption that
        $\crel{M}{\ectx{E}[\fixn{x}{m}{E}]}{A \rightarrow
        \ct{B}}$. By the definition of $\crel{}{}{A \rightarrow
        \ct{B}}$ it follows that \[
          \crel{M\,N}
               {\ectx{E}[\fixn{x}{m}{E}]\,X}
               {\ct{B}}
            \aequiv
               {(\ectx{E}\,X)[\fixn{x}{m}{E}]}
        \] Hence, by the IH for $\ct{B}$, \[
          \crel{M\,N}
               {(\ectx{E}\,X)[\fix{x}{E}]}
               {\ct{B}}
            \aequiv
               {\ectx{E}[\fix{x}{E}]\,X}
        \]

        \indcase{$\ct{B}_1 \with \ct{B}_2$}

        Given any $m \geq 0$, it is the case by the assumption and
        the definition of $\crel{}{}{\ct{B}_1 \with \ct{B}_2}$
        that, for $i \in \{1, 2\}$, \[
          \crel{\pi_i(M)}
               {\pi_i(\ectx{E}[\fixn{x}{m}{E}])}
               {\ct{B}_i}
            \aequiv
               {\pi_i(\ectx{E})[\fixn{x}{m}{E}]}
        \] Therefore, by the IH for $\ct{B}_i$, we get \[
          \crel{\pi_i(M)}
               {\pi_i(\ectx{E})[\fix{x}{E}]}
               {\ct{B}_i}
            \aequiv
               {\pi_i(\ectx{E}[\fix{x}{E}])}
        \] for $i \in \{1, 2\}$, and hence $
          \crel{M}
               {\ectx{E}[\fix{x}{E}]}
               {\ct{B}_1 \with \ct{B}_2}
        $.
      \end{indproof}
  \end{enumerate}
\end{proof}

\section{The bounding theorem}
\label{sec:bounding-theorem-proof}

For a substitution $\gamma$, write $\gamma : \Gamma$ to mean \[
  \forall x \in \dom(\Gamma).\ \vturn \gamma(x) : \Gamma(x)
\] Write $\subst{M}{\gamma}$ for the application of this
substitution to a term. Moreover, let \[
  \vrel{\gamma}{\delta}{} : \Gamma\
    \defiff
  \forall x \in \dom(\gamma).\ 
    x \in \dom(\delta) \land \vrel{\gamma(x)}{\delta(x)}{\Gamma(x)}
\]

\begin{theorem}[Bounding theorem]\label{thm:bounding-theorem}
  If $\vrel{\gamma}{\delta}{} : \Gamma$ then for all $\Gamma
  \vturn V : A$ and $\Gamma \cturn M : \ct{B}$ we have
  \begin{align*}
    &\vrel{\subst{V}{\gamma}}{\subst{\potential{V}}{\delta}}{A} \\
    &\crel{\subst{M}{\gamma}}{\subst{\complexity{M}}{\delta}}{\ct{B}}
  \end{align*}
\end{theorem}

\begin{proof}
  By induction on the derivations of $\Gamma \vturn V : A$ and
  $\Gamma \cturn M : \ct{B}$.

  \begin{indproof}

  \indcase{$\Gamma, x : A, \Gamma' \vturn x : A$}
    By the assumption that $\vrel{\gamma}{\delta}{} : (\Gamma, x :
    A, \Gamma')$.

  \indcase{$\Gamma \vturn (V_1, V_2) : A_1 \times A_2$}
    By the IH we have that \begin{align*}
      &\vrel{\subst{V_1}{\gamma}}{\subst{\potential{V_1}}{\delta}}{A_1} \\
      &\vrel{\subst{V_2}{\gamma}}{\subst{\potential{V_2}}{\delta}}{A_2}
    \end{align*} By the rule
    $(\hyperlink{betaprod}{\times_\beta})$ and bound weakening, we
    obtain that \begin{align*}
      &\vrel{\subst{V_1}{\gamma}}
            {\pi_1(\tuple{\subst{\potential{V_1}}{\delta}}
                         {\subst{\potential{V_2}}{\delta}}
                  )
            }
            {A_1} \\
      &\vrel{\subst{V_2}{\gamma}}
            {\pi_2(\tuple{\subst{\potential{V_1}}{\delta}}
                         {\subst{\potential{V_2}}{\delta}}
                  )
            }
            {A_1}
    \end{align*} But \begin{align*}
      \tuple{\subst{V_1}{\gamma}}
            {\subst{V_2}{\gamma}}
        &\aequiv
      \subst{\tuple{V_1}{V_2}}{\gamma} \\
      \tuple{\subst{\potential{V_1}}{\delta}}
            {\subst{\potential{V_2}}{\delta}}
        &\aequiv
      \subst{\tuple{\potential{V_1}}{\potential{V_2}}}{\delta}
    \end{align*} so we immediately obtain the result.

  \indcase{$\Gamma \vturn \thunk{M} : U\ct{B}$}
    By the IH we have that
    $\crel{\subst{M}{\gamma}}{\subst{\complexity{M}}{\delta}}{\ct{B}}$,
    which immediately yields the result, as $\potential{\thunk{M}}
    \defeq \complexity{M}$.

  \indcase{$\Gamma \cturn \tuple{M}{N} : \ct{B}_1 \with \ct{B}_2$}
    We have by the IH that \begin{align*}
      &\crel{\subst{M}{\gamma}}
            {\subst{\complexity{M}}{\delta}}{\ct{B}_1} \\
      &\crel{\subst{N}{\gamma}}
            {\subst{\complexity{N}}{\delta}}{\ct{B}_2}
    \end{align*}
    Hence by head expansion we have that \begin{align*}
      &\crel{\pi_1(\tuple{\subst{M}{\gamma}}{\subst{N}{\gamma}})}
        {\subst{\complexity{M}}{\delta}}{\ct{B}_1} \\
      &\crel{\pi_2(\tuple{\subst{M}{\gamma}}{\subst{N}{\gamma}})}
        {\subst{\complexity{N}}{\delta}}{\ct{B}_2}
    \end{align*} By the rule
    $(\hyperlink{betaprod}{\times_\beta})$ and bound weakening, we
    see that \begin{align*}
      &\crel{\pi_1(\tuple{\subst{M}{\gamma}}{\subst{N}{\gamma}})}
        {\pi_1
          \left(
            \tuple{\subst{\complexity{M}}{\delta}}
                  {\subst{\complexity{N}}{\delta}}
          \right)
        }
        {\ct{B}_1} \\
      &\crel{\pi_2(\tuple{\subst{M}{\gamma}}{\subst{N}{\gamma}})}
        {\pi_2
          \left(
            \tuple{\subst{\complexity{M}}{\delta}}
                  {\subst{\complexity{N}}{\delta}}
          \right)
        }
        {\ct{B}_2}
    \end{align*} 
    Hence, as \begin{align*}
      \subst{\tuple{M}{N}}{\gamma}
        &\aequiv
      \tuple{\subst{M}{\gamma}}{\subst{N}{\gamma}} \\
      \subst{\complexity{\tuple{M}{N}}}{\delta}
        &\aequiv
      \tuple{\subst{\complexity{M}}{\delta}}
            {\subst{\complexity{N}}{\delta}}
    \end{align*} the result follows by the definition of
    $\crel{}{}{\ct{B}_1 \with \ct{B}_2}$.

    \indcase{$\Gamma \cturn \lambda x.\ M : A \rightarrow \ct{B}$}

    We need to show that \[
      \subst{(\lambda x.\ M)}{\gamma}
        \aequiv
      \crel{\lambda x.\ \subst{M}{\gamma}}
           {\lambda x.\ \subst{\complexity{M}}{\delta}}
           {A \rightarrow \ct{B}}
        \aequiv
      \subst{\complexity{\lambda x.\ M} }{\delta}
    \] Hence, we need to show that for all $\vrel{N}{X}{A}$ it is
    the case that \[
      \crel{(\lambda x.\ \subst{M}{\gamma})N}
           {(\lambda x.\ \subst{\complexity{M}}{\delta})X}
           {\ct{B}}
    \] As $\vrel{N}{X}{A}$, we have that $\vrel{(\gamma,
    N/x)}{(\delta, X/x)}{} : (\Gamma, x : A)$, so by the IH we
    have that \[
      \crel{\subst{M}{\gamma, N/x}}
           {\subst{\complexity{M}}{\delta, X/x}}
           {\ct{B}}
    \] By head expansion, we get \[
      \crel{(\lambda x.\ \subst{M}{\gamma})N}
           {\subst{\complexity{M}}{\delta, X/x}}
           {\ct{B}}
    \] and then the rule
    $(\hyperlink{betaarrow}{\rightarrow_\beta})$ along with bound
    weakening yields \[
      \crel{(\lambda x.\ \subst{M}{\gamma})N}
           {(\lambda x.\ \subst{\complexity{M}}{\delta})X}
           {\ct{B}}
    \] which is the desired result.

  \indcase{$\Gamma \cturn \return{V} : FA$}

  We have to show that \[
    \crel{\return{\subst{V}{\gamma}}}
         {\tuple{\zeroc}{\subst{\potential{V}}{\delta}}}
         {FA}
  \] As
  $\bounded{\tuple{\zeroc}{\subst{\potential{V}}{\delta}}_c}$,
  it suffices to show that $\subst{\return{V}}{\gamma}$ converges,
  and that $\tuple{\zeroc}{\subst{\potential{V}}{\delta}}$ is an
  appropriate bound for it. But
  $\eval{\subst{\return{V}}{\gamma}}{\subst{\return{V}}{\gamma}}{0}$,
  and \[
    \zeroc 
      \costleq
        \pi_1\left(\tuple{\zeroc}{\subst{\potential{V}}{\delta}}\right)
      \aequiv
        \left(\tuple{\zeroc}{\subst{\potential{V}}{\delta}}\right)_c
  \] by rule $(\hyperlink{betaprod}{\times_\beta})$. Moreover,
  $\vrel{\subst{V}{\gamma}}{\subst{\potential{V}}{\delta}}{A}$ by
  the IH, and \[
    \subst{\potential{V}}{\delta}
      \costleq
        \pi_2\left(\tuple{\zeroc}{\subst{\potential{V}}{\delta}}\right)
      \aequiv
        \left(\tuple{\zeroc}{\subst{\potential{V}}{\delta}}\right)_p
  \] by rule $(\hyperlink{betaprod}{\times_\beta})$, and hence
  $\vrel{\subst{V}{\gamma}}
        {\tuple{\zeroc}
               {\subst{\potential{V}}
                      {\delta}}_p}
        {A}$ by bound weakening.

  \indcase{$\Gamma \vturn \numl{n} : \natt$}

  Then the result folows by $(\textsf{refl})$.

  \indcase{$\Gamma \cturn \force{V} : \ct{B}$}

  By the IH we have that \[
    \vrel{\subst{V}{\gamma}}{\subst{\potential{V}}{\delta}}{U\ct{B}}
  \] Hence, it must be that $\subst{V}{\gamma} \aequiv
  \thunk{M}$ for some $M$ with \[
    \crel{M}{\subst{\potential{V}}{\delta}}{\ct{B}}
  \] It then follows by head expansion that \[
    \subst{(\force{V})}{\gamma}
      \aequiv
    \crel{\force{(\thunk{M})}}{\subst{\potential{V}}{\delta}}{\ct{B}}
      \aequiv
      \subst{\complexity{\force{V}}}{\delta}
  \]

  \indcase{$\Gamma \cturn \pi_i(M) : \ct{B}_i$}

  By the IH we get that \[
    \crel{\subst{M}{\gamma}}
         {\subst{\complexity{M}}{\delta}}
         {\ct{B}_1 \with \ct{B}_2}
  \] Expanding the definition of $\crel{}{}{\ct{B}_1 \with
  \ct{B}_2}$ yields \[
    \subst{\pi_i(M)}{\gamma}
      \aequiv
    \crel{\pi_i(\subst{M}{\gamma})}
         {\pi_i(\subst{\complexity{M}}{\delta})}
         {\ct{B}_i}
      \aequiv
    \subst{\complexity{\pi_i(M)}}{\delta}
  \]

  \indcase{$\Gamma \cturn M V : \ct{B}$}

  By the IH we have \begin{align*}
    &\crel{\subst{M}{\gamma}}
          {\subst{\complexity{M}}{\delta}}
          {A \rightarrow \ct{B}} \\
    &\vrel{\subst{V}{\gamma}}
          {\subst{\potential{V}}{\delta}}
          {A}
  \end{align*}

  Hence, by the definition of $\crel{}{}{A \rightarrow \ct{B}}$,
  we get that \[
    \subst{(M\,V)}{\gamma}
      \aequiv
    \crel{\subst{M}{\gamma}\subst{V}{\gamma}}
         {\subst{\complexity{M}}{\delta}\subst{\potential{V}}{\delta}}
         {\ct{B}}
      \aequiv
    \subst{\complexity{M\,V}}{\delta}
  \]

  \indcase{$\Gamma \cturn \bind{x}{M}{N} : \ct{B}$}

  Recall that \begin{equation}
    \label{eq:bindRHS}
    \subst{\complexity{\bind{x}{M}{N}}}{\delta}
      \aequiv
    \subst{\complexity{M}}{\delta}_c
      +_{\ct{B}} 
    \subst{\complexity{N}}
          {\delta, \subst{\complexity{M}}{\delta}_p/x)}
  \end{equation} By the IH we have \[
    \crel{\subst{M}{\gamma}}{\subst{\complexity{M}}{\delta}}{FA}
  \] If $\unbounded{\subst{\complexity{M}}{\delta}_c}$, then by
  Lemma \ref{lemma:recbounds}(ii) we have that \[
    \crel{\subst{(\bind{x}{M}{N})}{\gamma}}
         {(\subst{\complexity{M}}{\delta})_c
            +_{\ct{B}} 
          \subst{\complexity{N}}
                {\delta, \subst{\complexity{M}}{\delta}_p/x)}
         }
         {\ct{B}}
  \] Recalling \eqref{eq:bindRHS}, we obtain the result.
  
  Otherwise $\bounded{\subst{\complexity{M}}{\delta}}$, and the
  definition of $\crel{}{}{FA}$ implies that there exist $n$ and
  $V$ such that \begin{align}
    \label{eq:bindeval}
      \eval{\subst{M}{\gamma}}{\return{V}}{n} \\
    \label{eq:bindbound}
      \numc{n} \costleq
        \subst{\complexity{M}}{\delta}_c \\
    \label{eq:bindvalue}
      \vrel{V}{\subst{\complexity{M}}{\delta}_p}{A} 
  \end{align} Hence, it follows by \eqref{eq:bindvalue} that \[
    \vrel{(\gamma, V/x)}
         {(\delta, \subst{\complexity{M}}{\delta}_p/x)}
         {}
         : (\Gamma, x : A)
  \] So, by the IH for $N$ we obtain that \[
    \crel{\subst{N}{\gamma, V/x}}
         {\subst{\complexity{N}}
                {\delta, \subst{\complexity{M}}{\delta}_p/x)}
         }
         {\ct{B}}
  \] Hence, by bind expansion (Lemma \ref{lemma:bind-expansion})
  and \eqref{eq:bindeval}, we get \[
    \crel{\bind{x}{\subst{M}{\gamma}}{\subst{N}{\gamma}}}
         {
          \numc{n}
            +_{\ct{B}} 
          \subst{\complexity{N}}
                {\delta, \subst{\complexity{M}}{\delta}_p/x)}
         }
         {\ct{B}}
  \] Hence, by \eqref{eq:bindbound}, algebra monotonicity (Lemma
  \ref{lemma:algebra-monotone}), and bound weakening, we obtain \[
    \crel{\bind{x}{\subst{M}{\gamma}}{\subst{N}{\gamma}}}
         {
          \subst{\complexity{M}}{\delta}_c
            +_{\ct{B}} 
          \subst{\complexity{N}}
                {\delta, \subst{\complexity{M}}{\delta}_p/x)}
         }
         {\ct{B}}
  \] But the LHS is $\alpha$-equivalent to
  $(\subst{\bind{x}{M}{N})}{\gamma}$, so---recalling
  \eqref{eq:bindRHS}---we obtain the result.

  \indcase{$\Gamma \cturn \splitprod{V}{x_1}{y_2}{N} : \ct{B}$}

  By the IH we have that \[
    \vrel{\subst{V}{\gamma}}
         {\subst{\potential{V}}{\delta}}
         {A_1 \times A_2}
  \] Then $\subst{V}{\gamma} \aequiv (V_1, V_2)$, and
  \begin{align*}
    \vrel{V_1}{\pi_1(\subst{\potential{V}}{\delta})}{A_1} \\
    \vrel{V_2}{\pi_2(\subst{\potential{V}}{\delta})}{A_2}
  \end{align*} Consequently, \begin{align*}
    \vrel{(\gamma, V_1/x_1, V_2/x_2)}
         {(\delta, \pi_1(\subst{\potential{V}}{\delta})/x_1,
                   \pi_2(\subst{\potential{V}}{\delta})/x_2)}
         {} : (\Gamma, x_1 : A_1, x_2 : A_2)
  \end{align*} By the IH for $N$, we have \[
    \crel{\subst{N}{\gamma, V_1/x_1, V_2/x_2}}
         {\subst{\complexity{N}}
                {\delta, \pi_1(\subst{\potential{V}}{\delta})/x_1,
                         \pi_2(\subst{\potential{V}}{\delta})/x_2}
         }
         {\ct{B}}
  \] Furthermore, by head expansion we get \[
    \crel{\splitprod{\subst{V}{\gamma}}{x}{y}{\subst{N}{\gamma}}}
         {\subst{\complexity{N}}
                {(\delta, \pi_1(\subst{\potential{V}}{\delta})/x_1,
                          \pi_2(\subst{\potential{V}}{\delta})/x_2)}
         }
         {\ct{B}}
  \] where the RHS is $\alpha$-equivalent to
  $\subst{\complexity{\splitprod{V}{x_1}{x_2}{N}}}{\delta}$.

  \indcase{$\Gamma \cturn \ifz{V}{P}{Q} : \ct{B}$}

  By the IH we have that \[
    \vrel{\subst{V}{\gamma}}
         {\subst{\potential{V}}{\delta}}
         {\natt}
  \] Hence, $\subst{V}{\gamma} \aequiv \numl{n}$. There are two
  cases.
  \begin{indproof}

    \indcase{$n = 0$}
      Then $\subst{V}{\gamma} \aequiv \numl{0}$. As
      $\vrel{\subst{V}{\gamma}} {\subst{\potential{V}}{\delta}}
      {\natt}$, we have that $\num{n} \costleq
      \subst{\potential{V}}{\delta}$.  By the IH it also follows
      that \[
        \crel{\subst{P}{\gamma}}
             {\subst{\complexity{P}}{\delta}}
             {\ct{B}}
      \] By head expansion, \[
        \ifz{\subst{V}{\gamma}}{\subst{P}{\gamma}}{\subst{Q}{\gamma}}
          \aequiv
        \crel{\ifz{\numl{0}}{\subst{P}{\gamma}}{\subst{Q}{\gamma}}}
             {\subst{\complexity{P}}{\delta}}
             {\ct{B}}
      \] By $(\hyperlink{iftrue}{\textsf{if}_\textsf{tt}})$ along
      with bound expansion, it follows that \[
        \crel{\ifz{\subst{V}{\gamma}}{\subst{P}{\gamma}}{\subst{Q}{\gamma}}}
             {\pif{\num{0}}
                  {\subst{\complexity{P}}{\delta}}
                  {\subst{\complexity{Q}}{\delta}}
             }
             {\ct{B}}
      \] By $(\hyperlink{ctx}{\textsf{ctx}})$, the above
      inequality, and bound expansion, we get \[
        \crel{\ifz{\subst{V}{\gamma}}{\subst{P}{\gamma}}{\subst{Q}{\gamma}}}
             {\pif{\subst{\potential{V}}{\delta}}
                  {\subst{\complexity{P}}{\delta}}
                  {\subst{\complexity{Q}}{\delta}}
             }
             {\ct{B}}
      \] But the RHS is $\alpha$-equivalent to
      $\subst{\complexity{\pif{V}{P}{Q}}}{\delta}$.

    \indcase{$n > 0$} Similar.

  \end{indproof}

  \indcase{$\Gamma \cturn \calc{v}{V \cbpvop W}{N} : \ct{B}$}

  By the IH we have that \begin{align*}
    &\vrel{\subst{V}{\gamma}}
          {\subst{\potential{V}}{\delta}}
          {\natt} \\
    &\vrel{\subst{W}{\gamma}}
          {\subst{\potential{W}}{\delta}}
          {\natt}
  \end{align*} This implies that there exist $n$ and $m$ such that
  $\subst{V}{\gamma} \aequiv \numl{n}$, and $\subst{V}{\gamma}
  \aequiv \numl{m}$, so that \begin{align*}
    \num{n} &\costleq \subst{\potential{V}}{\delta} \\
    \num{m} &\costleq \subst{\potential{W}}{\delta}
  \end{align*} Suppose without loss of generality that $\cbpvop
  \in \{+, *\}$. Then by the rules $(\hyperlink{ctx}{\textsf{ctx}})$ 
  and $(\hyperlink{betanum}{\textsf{num}_\beta})$ we have \[
    \num{n \pcfop m}
      \costleq
    \num{n} \pcfop \num{m}
      \costleq 
    \subst{\potential{V}}{\delta} \pcfop \subst{\potential{V}}{\delta}
  \] Hence \[
      \vrel{\numl{n \cbpvop m}}
           {\subst{\potential{V}}{\delta} \pcfop \subst{\potential{V}}{\delta}}
           {\natt}
  \] whence \[
    \vrel{(\gamma, \numl{n \cbpvop m}/v)}
         {(\delta, \subst{\potential{V}}{\delta} 
                     \pcfop
                   \subst{\potential{W}}{\delta}/v)}
         {} : \left(\Gamma, v : \natt\right)
  \] Therefore, by the IH for $N$ we obtain \[
    \crel{\subst{N}{\gamma, \numl{n \cbpvop m}/v}}
         {\subst{\complexity{N}}
                {\delta, \subst{\potential{V}}{\delta} 
                           \pcfop
                         \subst{\potential{W}}{\delta}}
         }
         {\ct{B}}
  \] The RHS is $\alpha$-equivalent to
  $\subst{\complexity{\calc{v}{V \cbpvop \numl{m}}{N}}}{\delta}$,
  and by head expansion we get \[
    \crel{\calc{v}
               {\subst{V}{\gamma} \cbpvop \subst{W}{\gamma}}
               {\subst{N}{\gamma}}
         }
         {\subst{\complexity{\calc{v}{V \cbpvop W}{N}}}{\delta}}
         {\ct{B}}
  \] The cases for $\cbpvopa$ and $\bmod$ are similar.

  \indcase{$\Gamma \cturn \charge{M} : \ct{B}$}

  We have by the IH that \[
    \crel{\subst{M}{\gamma}}
         {\subst{\complexity{M}}{\delta}}
         {\ct{B}}
  \] By applying a unit charge (Lemma \ref{lemma:unit}) we
  obtain \[
    \subst{(\charge{M})}{\gamma}
      \aequiv
    \crel{\charge{\subst{M}{\gamma}}}
         {\unitc \algp{\ct{B}} \subst{\complexity{M}}{\delta}}
         {\ct{B}}
      \aequiv
    \subst{\complexity{\charge{M}}}{\delta}
  \]

  \indcase{$\Gamma \cturn \fix{x}{M} : \ct{B}$}

  Using the IH, we prove by induction on $n$ that \[
    \forall n \geq 0.\
      \crel{\fix{x}{\subst{M}{\gamma}}}
           {\fixn{x}{n}{\subst{\complexity{M}}{\delta}}}
           {\ct{B}}
  \] By Fixed Point Induction (Lemma \ref{lemma:recbounds}(iii)), this
  allows us to conclude that \[
      \subst{(\fix{x}{M})}{\gamma}
    \aequiv
      \crel{\fix{x}{\subst{M}{\gamma}}}
          {\fix{x}{\subst{\complexity{M}}{\delta}}}
          {\ct{B}}
    \aequiv
        \subst{\complexity{\fix{x}{M}}}{\delta}
  \] If $n = 0$, then we obtain
  $\crel{\fix{x}{\subst{M}{\gamma}}}{\fix{x}{x}}{\ct{B}}$ from
  Lemma \ref{lemma:recbounds}(i). Assuming that the result is true
  for $n > 0$, we use the local IH to deduce that \[
      \vrel{\thunk{(\fix{x}{\subst{M}{\gamma}})}}
           {\fixn{x}{n}{\subst{\complexity{M}}{\delta}}}
           {U\ct{B}}
  \] and hence that \[
    \vrel{(\gamma, \thunk{(\fix{x}{\subst{M}{\gamma}})}/x)}
         {(\delta, \fixn{x}{n}{\subst{\complexity{M}}{\delta}/x})}
         {} : (\Gamma, x : U\ct{B})
  \] Then, the global IH for $M$ yields \[
    \crel{\subst{M}{\gamma, \thunk{(\fix{x}{\subst{M}{\gamma})}/x}}}
         {\subst{\complexity{M}}
                {\delta, \fixn{x}{n}{\subst{\complexity{M}}{\delta}}/x}}
         {\ct{B}}
  \] By head expansion, we get \[
    \crel{\fix{x}{\subst{M}{\gamma}}}
         {\subst{\complexity{M}}
                {\delta, \fixn{x}{n}{\subst{\complexity{M}}{\delta}}/x}}
         {\ct{B}}
  \] But then we can $\alpha$-convert the RHS to
  $\fixn{x}{n+1}{\subst{\complexity{M}}{\delta}}$.

  \end{indproof}
\end{proof}

\section{Inductive types}

Here we extend the definitions and proofs of the previous section
to handle lists. Figure~\ref{fig:ind-types} summarises the
relevant extensions. We will need an additional technical
extension: eliminative contexts of $\pcfc$ will also include
$\ectx{E} - \num 1$. We assume that compactness holds for this
type of eliminative context as well; note that subtraction by $1$
is strict, so this should not be too difficult to prove.

Lemmas~\ref{lemma:headexp} (head expansion/reduction),
\ref{lemma:unit} (charge),
and~\ref{lemma:bind-expansion} (bind expansion) are proved by
induction on computation types, so need no modification.  The same
goes for Lemma~\ref{lemma:recbounds}(i)(ii) (bounds for
recursion).


\begin{proof}[Proof of Lemma~\ref{lemma:bound} (Bound Weakening)]\hfill

\begin{indproof}
  \indcase{$\cbpvlist A$}
    We induct on $V$.
    \begin{indproof}
      \indcase{$\cbpvnil$}
        
        We have that $\num 0 \costleq E \costleq E'$, so the result
        follows by $(\hyperlink{trans}{\textsf{trans}})$.

      \indcase{$\cbpvcons(W, X)$} 

        By the IH we have $\vrel{X}{E - 1}{\cbpvlist A}$, but by
        $(\hyperlink{ctx}{\textsf{ctx}})$ we get $E - 1 \costleq E'
        - 1$, so we establish the result.
    \end{indproof}
\end{indproof}
\end{proof}

\begin{proof}[Proof of Lemma~\ref{lemma:recbounds}(iii) (Fixed point induction)]

  We establish the stronger results stated there by induction,
  taking care to include $\ectx{E} - \num 1$ as an eliminative
  context. All the proofs adapt, so we only show the new case.
  \begin{indproof}
    \indcase{$\cbpvlist A$}
      By induction on $V$. 

      \begin{indproof}
        \indcase{$\cbpvnil$}

        By the assumption we have that $\num 0 \costleq
        \ectx{E}[\fixn{x}{n}{E}]$ for all $n \geq 0$, so by
        $(\hyperlink{cpind}{\textsf{cpind}})$ we obtain the
        result.

        \indcase{$\cbpvcons(W, X)$}

        By the assumption we have for all $n \geq 0$ that \[
          \vrel{X}{\ectx{E}[\fixn{x}{n}{E}] - \num 1}{\cbpvlist A}
            \aequiv
          (\ectx{E} - \num 1)[\fixn{x}{n}{E}]
        \] We thus apply the IH with $\ectx{E} - \num 1$ to obtain \[
          \vrel{X}{(\ectx{E} - \num 1)[\fix{x}{E}]}{\cbpvlist A}
            \aequiv
          \ectx{E}[\fix{x}{E}] - \num 1
        \] from which we obtain the result.

      \end{indproof}
    \end{indproof}
\end{proof}

We will need the following additional axioms on the size order: \[
  \begin{prooftree}
      \Gamma \vdash E : \natt
    \justifies
      \Gamma \vdash E \costleq (E + \num 1) - \num 1 : \natt
    \using
      (\hypertarget{pone}{\textsf{+1}})
  \end{prooftree}
    \qquad
  \begin{prooftree}
      \phantom{\Gamma \vdash E : \natt}
    \justifies
      \Gamma \vdash \num 0 \costleq \num 1 : \natt
    \using
      (\hypertarget{unit}{\textsf{unit}})
  \end{prooftree}
    \quad
  \begin{prooftree}
      \Gamma, z : A \vdash \ectx{E}[z] : A
    \justifies
      \Gamma \vdash \fix x x \costleq \ectx{E}[\fix x x] : A
    \using
      (\hypertarget{efix}{\textsf{efix}})
  \end{prooftree}
\] It is easy to see that these axioms are true in the standard
semantics, and that they satisfy the desideratum expressed by
Lemma \ref{lemma:bounded}, viz. bounded terms are still a lower
set.

We will also need the following new lemmata.

\begin{lemma}
  \label{lemma:infinity-value-top}
  For any $V : A$ and $\ectx{E}$ we have $\vrel{V}{\ectx{E}[\fix x x]}{A}$.
\end{lemma}

\begin{proof} By induction on $A$. The inductive cases follow from
absorbing eliminators into $\ectx{E}$, or using preservation of
infinity and Lemma~\ref{lemma:recbounds}(i) in the case of
$U\ct{B}$. For the base case of $\natt$, notice that $\num n
\costleq \fix x x \costleq \ectx{E}[\fix x x]$ by the fact $\fix x
x$ is a top element and $(\hyperlink{efix}{\textsf{efix}})$, so we
just use $(\hyperlink{trans}{\textsf{trans}})$.  This leaves the
case of $\cbpvlist A$. The case for $\cbpvnil$ follows as the case
of $\natt$. For $\cbpvcons$: we show $\num 1 \costleq
\ectx{E}[\fix x x]$ as in the case of $\natt$, and for the rest we
use the IH for $\ectx{E} - 1$.

\end{proof}

\begin{lemma}
  \label{lemma:lower-bound-list}
  If $\vrel{V}{E}{\cbpvlist A}$ then $\num{0} \costleq E$.
\end{lemma}

\begin{proof}
  By induction on the value $V$. If $V \aequiv \cbpvnil$, then the
  result follows by the definition of $\vrel{}{}{\cbpvlist A}$. If
  $V \aequiv \cbpvcons(W, X)$, then we have $\num 1 \costleq E$.
  Hence, $\num 0 \costleq \num 1 \costleq E$, so the result
  follows by $(\hyperlink{unit}{\textsf{unit}})$ and
  $(\hyperlink{trans}{\textsf{trans}})$.
\end{proof}

\begin{proof}[Theorem~\ref{thm:bounding-theorem} (Bounding Theorem)] \hfill

\begin{indproof}
  \indcase{$\Gamma\vturn \cbpvnil : \cbpvlist A$}
  
  We have that $
            \subst{\cbpvnil}{\gamma} 
    \aequiv \vrel{\cbpvnil}{\num 0}{\cbpvlist A} 
    \aequiv \potential{\cbpvnil} 
    \aequiv \subst{\potential{\cbpvnil}}{\delta}
  $, as $\num{0} \costleq \num{0}$ by
  $(\hyperlink{refl}{\textsf{refl}})$.

  \indcase{$\Gamma \vturn \cbpvcons(W, X) : \cbpvlist A$}

  We need to show that \[
    \subst {\cbpvcons(W, X)} \gamma
      \aequiv
    \vrel{\cbpvcons(\subst W \gamma , \subst X  \gamma)}
         {\subst {\potential{\cbpvcons(W, X)}} \delta}
         {\cbpvlist A}
      \aequiv
    \subst {\potential X} \delta + \num{1}
  \] By the IH we know that $\vrel{\subst X \gamma}{\subst
  {\potential X} \delta}{\cbpvlist A}$, so we use the
  new axiom $(\hyperlink{pone}{\textsf{+1}})$ and bound weakening
  to obtain $\vrel{\subst X \gamma}{\left(\subst {\potential X}
  \delta + \num 1\right) - \num 1}{\cbpvlist A}$. It remains to
  obtain the lower bound of $\num 1$, which we get by \[
    \num 1
    \costleq \num 0 + \num 1
    \costleq \subst {\potential X} \delta + \num 1
  \] which follows by $(\hyperlink{betanum}{\textsf{num}_\beta})$
  and Lemma \ref{lemma:lower-bound-list} along with
  $(\hyperlink{ctx}{\textsf{ctx}})$.

\indcase{$\Gamma \cturn \cbpvlcase V {M_\cbpvnil} x {xs} {M_\cbpvcons} : \ct B$}
  We have that $
  \subst {\complexity{\cbpvlcase V {M_\cbpvnil} x {xs}
  {M_\cbpvcons}}} \delta
  $ is $\alpha$-equivalent to \begin{equation}
    \label{eq:case-RHS}
    \pif {\subst {\potential V} \delta}
         {\subst {\complexity{M_\cbpvnil}} \delta}
         {\subst {\complexity{M_\cbpvcons}} 
                 {\delta, \fix x x/x, \subst {\potential V} \delta - \num{1}/xs}}
  \end{equation}

  There are now two cases, depending on the structure of the
  closed term $\subst V \gamma$.

  \begin{indproof}

  \indcase{$\subst V \gamma \aequiv \cbpvnil$}

    By the IH we have $\cbpvnil \aequiv \vrel{\subst V
    \gamma}{\subst {\potential V} \delta}{\cbpvlist A}$, so
    \begin{equation}
      \label{eq:case-potential-1}
      \num{0} \costleq \subst {\potential V} \delta
    \end{equation} The IH also gives us that \[
      \crel{\subst {M_\cbpvnil} \gamma}
           {\subst {\complexity M_\cbpvnil} \delta}
           {\ct{B}}
    \] By $(\hyperlink{iftrue}{\textsf{if}_\textsf{tt}})$ and
    bound weakening, we may weaken the RHS to \[
      \crel{\subst {M_\cbpvnil} \gamma}
           {\pif {\num{0}}
                 {\subst {\complexity {M_\cbpvnil}} \delta}
                 {\subst {\complexity{M_\cbpvcons}} 
                         {\delta, \fix x x/x, \subst {\potential V} \delta - \num{1}/xs}}
           }
           {\ct{B}}
    \] By Lemma \ref{lemma:lower-bound-list} we know that $\num 0
    \costleq \subst{\potential{V}}{\delta}$, so we may use
    $(\textsf{\hyperlink{ctx}{\textsf{ctx}}})$,
    \eqref{eq:case-potential-1} to further weaken the RHS to \[
      \crel{\subst {M_\cbpvnil} \gamma}
           {\pif {\subst {\potential V} \delta}
                 {\subst {\complexity {M_\cbpvnil}} \delta}
                 {\subst {\complexity{M_\cbpvcons}} 
                         {\delta, \fix x x/x, \subst {\potential V} \delta - \num{1}/xs}}
           }
           {\ct{B}}
    \] so that the new RHS is just \eqref{eq:case-RHS}.  By Lemma
    \ref{lemma:headexp} (head expansion), we may replace the LHS
    by \[
      \cbpvlcase {\subst V {\gamma}} {\subst {M_\cbpvnil} \gamma} x {xs} {\subst {M_\cbpvcons} \gamma}
      \aequiv \subst {\cbpvlcase V {M_\cbpvnil} x {xs}
      {M_\cbpvcons}} \gamma
    \] 

  \indcase{$\subst V \gamma \aequiv \cbpvcons(W, X)$}

  By the IH for $V$, we have that \[
    \cbpvcons(W, X)
      \aequiv
    \vrel{\subst V \gamma}
         {\subst {\potential V} \delta}
         {\cbpvlist A}
  \] Thus, by the definition of $\vrel{}{}{\cbpvlist A}$, we get
  that \begin{gather}
      \label{eq:cons-lower-bound}
      \num 1 \costleq \subst {\potential V} \delta \\
      \label{eq:cons-IH}
      \vrel{X}{\subst {\potential V}{\delta} - 1}{\cbpvlist A}
  \end{gather} As Lemma \ref{lemma:infinity-value-top} implies
  that $\vrel{W}{\fix x x}{A}$, we may plug this along with
  \eqref{eq:cons-IH} into the IH to obtain \[
    \crel {\subst {M_\cbpvcons} {\gamma, W/x, X/xs}}
          {\subst {\complexity{M_\cbpvcons}} 
                  {\delta, \fix x x/x, \subst {\potential V} \delta - 1/xs}}
          {\ct{B}}
  \] Thus, by Lemma \ref{lemma:headexp} (head expansion), we
  get \[
    \crel {\subst {\cbpvlcase {\cbpvcons(W, X)} {M_\cbpvnil} x {xs} {M_\cbpvcons}} \gamma}
          {\subst {\complexity{M_\cbpvcons}} 
                  {\delta, \fix x x/x, \subst {\potential V} \delta - 1/xs}}
          {\ct{B}}
  \] We have thus obtained the desired term on the LHS. 

  We may then use $(\hyperlink{iffalse}{\textsf{if}_\textsf{ff}})$
  (with $n = 0$) and then \eqref{eq:cons-lower-bound} and
  $(\hyperlink{ctx}{\textsf{ctx}})$ to weaken the bound of the RHS
  to \[
    \pif {\subst {\potential V} \delta}
         {\subst {\complexity {M_\cbpvnil}} \delta}
         {\subst {\complexity{M_\cbpvcons}} 
                 {\delta, \fix x x/x, \subst {\potential V} \delta - 1/xs}}
  \] This is $\alpha$-equivalent to \[
    \subst
      {\left( \pif {\potential V}
                   {\complexity {M_\cbpvnil}}
                  {\subst {\complexity {M_\cbpvcons}} {\fix x x/x, \potential V - 1/xs}}
       \right)}
      \delta
  \] which is the desired RHS.

  \end{indproof}

\end{indproof}

\end{proof}

\section{Sized domains}

\begin{definition}[Complete partial order]
  Let $(D, \sqsubseteq)$ be a partial order.
  \begin{enumerate}
    \item
      An \emph{$\omega$-chain} $(x_i)_{i \in \omega}$ on $D$ is an
      increasing sequence of elements \[
        x_1 \sqsubseteq x_2 \sqsubseteq \dots
      \] 
    \item
      $(D, \sqsubseteq)$ is a \emph{complete partial order} (cpo)
      just if it has a least element $\bot \in D$, and a
      least upper bound \[
        \bigsqcup_{i \in \omega} x_i
      \] for every $\omega$-chain $(x_i)_{i \in \omega}$.
    \item
      Let $(D, \sqsubseteq_D)$ and $(E, \sqsubseteq_E)$ be cpos. A
      function $f : D \rightarrow E$ is \emph{Scott-continuous}
      just if \[
        f\left(\bigsqcup_{i \in \omega} x_i\right)
          =
        \bigsqcup_{i \in \omega} f(x_i)
      \] for all $\omega$-chains $(x_i)_{i \in \omega}$.
  \end{enumerate}
\end{definition}

Evidently, complete partial orders and Scott-continuous functions
are immediately seen to form a category $\textbf{Cpo}$.

\begin{remark}
  When our domains have countable bases, the above definition of
  cpo is equivalent to the usual one, which requires upper bounds
  for all \emph{directed sets} (assuming the axiom of choice). See
  \cite[\S 2.2.4]{Abramsky1994}.
\end{remark}

\begin{theorem}[Fixed Point Theorem]
  Every Scott-continuous function $f : D \rightarrow D$ on a cpo
  $(D, \sqsubseteq)$ has a least fixed point, namely \[
    \textbf{fix}_D(f) \defeq \bigsqcup_{i \in \omega} f^i(\bot_D)
  \] Moreover, the function \begin{align*}
    \textbf{fix}_D    : [D \rightarrow D] \rightarrow D \\
    \textbf{fix}_D(f) \defeq \bigsqcup_{i \in \omega} f^i(\bot_D)
  \end{align*} is Scott-continuous.
\end{theorem}

We define the category $\mathbf{SizeDom}$ to consist of
\begin{description}
  \item[objects] sized domains
  \item[morphisms] 
    a morphism \[
      f : (D_1, \sqsubseteq_1, \semleq_1, \lor_1, \bot_1, 0_1, 1_1) 
        \rightarrow 
          (D_2, \sqsubseteq_2, \semleq_2, \lor_2, \bot_2, 0_2, 1_2)
    \] is a function $f : D_1 \rightarrow D_2$ that is
    \emph{Scott-continuous} with respect to the $\omega$-cpos
    $(D_1, \sqsubseteq_1)$ and $(D_2, \sqsubseteq_2)$
\end{description} It is clear that $\mathbf{SizeDom}$ is a full
subcategory of $\mathbf{Cpo}$: not all domains are sized, but
between any two sized domains we keep all the continuous
functions.

\subsection{The product of sized domains}

Given two sized domains, $(D_1, \sqsubseteq_1, \semleq_1, \lor_1,
\bot_1, 0_1, 1_1)$ and $(D_2, \sqsubseteq_2, \semleq_2, \lor_2,
\bot_2, 0_2, 1_2)$ we define the \emph{product sized domain}, \[
  (D_1 \times D_2, \sqsubseteq, \semleq, \lor, \bot, 0, 1)
\] componentwise. That is: \begin{align*}
  (x_1, x_2) \sqsubseteq (y_1, y_2)\
    &\defiff
      x_1 \sqsubseteq_1 y_1 \text{ and } x_2 \sqsubseteq_2 y_2 \\
  (x_1, x_2) \semleq (y_1, y_2)\
    &\defiff
      x_1 \semleq y_1 \text{ and } x_2 \semleq y_2 \\
  (x_1, x_2) \lor (y_1, y_2)
    &\defeq
      (x_1 \lor_1 y_1, x_2 \lor_2 y_2) \\
  \bot\
    &\defeq
      (\bot_1, \bot_2) \\
  0\
    &\defeq
      (0_1, 0_2) \\
  1\
    &\defeq
      (1_1, 1_2)
\end{align*} 

\begin{proposition}
  $(D_1 \times D_2, \sqsubseteq, \semleq, \lor, \bot, 0, 1)$ is a
  sized domain. Moreover, it is the product of $(D_1,
  \sqsubseteq_1, \semleq_1, \bot_1, 0_1, 1_1)$ and $(D_2,
  \sqsubseteq_2, \semleq_2, \bot_2, 0_2, 1_2)$ in the category
  $\mathbf{SizeDom}$.
\end{proposition}

\begin{proof}
  As both $\sqsubseteq$ and $\semleq$ are defined pointwise, it is
  evident that $(D_1 \times D_2, \sqsubseteq)$ is a cpo, and that
  $(D_1 \times D_2, \semleq)$ is a preorder with chosen binary
  joins which are pointwise continuous and hence continuous.
  
  It remains to check last two axioms. If $(x_1, x_2) \sqsubseteq
  (y_1, y_2)$, then $x_i \sqsubseteq_i y_i$ for $i \in \{1, 2\}$.
  Hence, $y_i \semleq x_i$, and we conclude that $(y_1, y_2)
  \semleq (x_1, x_2)$. An $\omega$-chain $(a_i, b_i)_{i \in
  \omega}$ gives rise to $\omega$-chains $(a_i)_{i \in \omega}$ in
  $D_1$ and $(b_i)_{i \in \omega}$ in $D_2$. Thus, \begin{align*}
    \forall i \in \omega.\ (z_1, z_2) \semleq (a_i, b_i)
      \Longrightarrow\
    &\forall i \in \omega.\ 
      z_1 \semleq a_i \text{ and } z_2 \semleq b_i \\
      \Longrightarrow\
    &z_1 \semleq \bigsqcup_{i \in \omega} a_i
      \text{ and }
     z_2 \semleq \bigsqcup_{i \in \omega} b_i \\
      \Longrightarrow\
    &(z_1, z_2) \semleq \bigsqcup_{i \in \omega} (a_i, b_i)
  \end{align*} That this is the product follows from the fact
  $(D_1 \times D_2, \sqsubseteq)$ is the product in the category
  $\mathbf{Cpo}$ of complete partial orders.
\end{proof}

\subsection{The exponential of sized domains}

Given two sized domains $(D_1, \sqsubseteq_1, \semleq_1, \lor_1,
\bot_1, 0_1, 1_1)$ and $(D_2, \sqsubseteq_2, \semleq_2, \lor_2,
\bot_2, 0_2, 1_2)$, let $[D_1 \rightarrow D_2]$ consist of the
continuous functions $f : D_1 \rightarrow D_2$.  We define the
\emph{exponential sized domain} $([D_1 \rightarrow D_2],
\sqsubseteq, \semleq, \lor, \bot, 0, 1)$ componentwise. That is:
\begin{align*}
  f \sqsubseteq g
    &\defiff
      \forall x \in D_1.\ f(x) \sqsubseteq_2 g(x) \\
  f \semleq g
    &\defiff
      \forall x \in D_1.\ f(x) \semleq_2 g(x) \\
  f \lor g
    &\defeq
      x \in D_1 \longmapsto f(x) \lor_2 g(x) \in D_2 \\
  \bot
    &\defiff
      x \in D_1 \longmapsto \bot_2 \in D_2 \\
  0
    &\defeq
      x \in D_1 \longmapsto 0_2 \in D_2 \\
  1
    &\defeq
      x \in D_1 \longmapsto 1_2 \in D_2 \\
\end{align*} Notice that the above definitions depend neither on
the orders $\sqsubseteq_1$ and $\semleq_1$ nor on the chosen joins
$\lor_1$.

\begin{proposition}
  $([D_1 \rightarrow D_2], \sqsubseteq, \semleq, \bot, 0, 1)$ is a
  sized domain. Moreover, it is the exponential of $(D_1,
  \sqsubseteq_1, \semleq_1, \lor_1, \bot_1, 0_1, 1_1)$ and $(D_2,
  \sqsubseteq_2, \semleq_2, \lor_2 \bot_2, 0_2, 1_2)$ in the
  category $\mathbf{SizeDom}$.
\end{proposition}

\begin{proof}
  
  $([D_1 \rightarrow D_2], \sqsubseteq)$ is the exponential in the
  category $\textbf{Cpo}$ of complete partial orders. As both
  orders are defined pointwise, it is evident that $\lor$ is
  continuous and picks chosen joins.  It thus suffices to check
  the last two axioms. 

  If $(f_i)_{i \in \omega}$ is an $\omega$-chain in $[D_1
  \rightarrow D_2]$ we have \begin{align*}
    \forall i \in \omega.\ g \semleq f_i 
      \Longleftrightarrow\
    & \forall i \in \omega.\ \forall x \in D_1.\
        g(x) \semleq f_i(x) \\
      \Longleftrightarrow\
    & \forall x \in D_1.\ \forall i \in \omega.\
        g(x) \semleq f_i(x) \\
      \Longrightarrow\
    & \forall x \in D_1.\ 
        g(x) \semleq \bigsqcup_{i \in \omega} f_i(x) \\
      \Longleftrightarrow\
    & \forall x \in D_1.\ 
        g(x) \semleq \left(\bigsqcup_{i \in \omega} f_i\right)(x) \\
      \Longleftrightarrow\
    & g \semleq \bigsqcup_{i \in \omega} f_i
  \end{align*} The only non-invertible step in this proof follows
  by axiom (iv) in the definition of sized domains
\end{proof}

\subsection{Interpreting PCF with costs into sized domains}

Recall that when $\vec{d} \in \sem{\Gamma}{}$ we define
\begin{align*}
  \sem{\Gamma \vdash \lambda x.\ M : A \rightarrow B}{}
    (\vec{d})
    &\defeq 
        \left(
          x
            \mapsto
          \sem{\Gamma \vdash, x : A \vdash M : B}{}(\vec{d}, x)
        \right) \\
  \sem{\Gamma \vdash \fix{x}{E} : A}{}(\vec{d})
    &\defeq
  \bigsqcup_{n \in \omega} f_{\vec{d}}^n(\bot_A)\
  \text{ where } \begin{cases}
    f_{\vec{d}} : \sem{A}{} \rightarrow \sem{A}{} \\
    f_{\vec{d}}(x) \defeq \sem{\Gamma, x : A \vdash E : A}{}(\vec{d}, x)
  \end{cases} 
\end{align*} 

\subsubsection{Soundness}

\begin{lemma}[Substitution]
  \label{lemma:substitution}
  If $\Gamma \vdash M_i : A_i$ for $i = 1, \dots, n$, and $\Delta
  \vdash N : B$ where $\Delta \equiv x_1 : A_1, \dots, x_n : A_n$,
  then \[
    \forall \vec{d} \in \sem{\Gamma}.\quad
    \sem{\Gamma \vdash M[\vec{N}/\vec{x}] : B}{}(\vec{d})
      =
    \sem{\Delta \vdash M : B}{}\left(
      \sem{\Gamma \vdash M_1 : A_1}{}(\vec{d}),
        \dots
      \sem{\Gamma \vdash M_n : A_n}{}(\vec{d})
    \right)
  \] 
\end{lemma}

\begin{proof}
  By induction on the term $M$. The proof of this lemma is
  completely standard---see e.g. \cite[Lemma
  3.15]{Streicher2006}---so we only need to check the unusual
  clause for the conditional. Recall that \[
    \sem{\Gamma \vdash 
      \pif{M[\vec{N}/\vec{x}]}{P[\vec{N}/\vec{x}]}{Q[\vec{N}/\vec{x}]} : A}(\vec{d})
    \defeq
      \iflor_A(\sem{M}(\vec{d}), \sem{P}(\vec{d}), \sem{Q}(\vec{d}))
  \] where $\iflor_A$ is the continuous function \begin{align*}
    &\iflor_A : \mathbb{N}_\bot \times \sem{A} \times \sem{A} \rightarrow \sem{A} \\
    &\iflor_A(n, p, q) \defeq
      \begin{cases}
        \bot_A & \text{ if } n = \bot \\
        p & \text{ if } n = 0 \\
        p \lor_A q & \text{ otherwise}
      \end{cases}
  \end{align*} Letting $e_i \defeq \sem{\Delta \vdash M_i : A_i}$,
  we calculate that \begin{derivation}
      \sem{\Delta \vdash \pif{M}{P}{Q} : B}(e_1, \dots, e_n)
    \since[=]{definition of semantic map}
      \iflor_A\left(
        \sem{M}(e_1, \dots e_n),
        \sem{P}(e_1, \dots, e_n),
        \sem{Q}(e_1, \dots, e_n)
      \right)
    \since[=]{IH}
      \iflor_A\left(
        \sem{M[\vec{N}/\vec{x}]}(\vec{d}),
        \sem{P[\vec{N}/\vec{x}]}(\vec{d}),
        \sem{Q[\vec{N}/\vec{x}]}(\vec{d})
      \right)
    \since[=]{definition of semantic map}
      \sem{\Gamma \vdash \pif{M[\vec{N}/\vec{x}]}{P[\vec{N}/\vec{x}]}{Q[\vec{N}/\vec{x}]} : A}(\vec{d})
  \end{derivation}
\end{proof}

\begin{proposition}[Fixpoint Approximation]
  \label{prop:fixpoint-approx}
  If $\Gamma, x : A \vdash E : A$, then for all $\vec{d} \in
  \sem{\Gamma}{}$ \[
    \sem{\Gamma \vdash \fix{x}{E} : A}{}(\vec{d})
      =
    \bigsqcup_{n \in \omega}
      \sem{\Gamma \vdash \fixn{x}{n}{E} : A}{}(\vec{d})
  \]
\end{proposition}
\begin{proof}
  Recall the definition of $f_{\vec{d}}$ from before.  It follows
  easily by induction and the substitution lemma (Lemma
  \ref{lemma:substitution}) that \[
    \sem{\Gamma \vdash \fixn{x}{n}{E} : A}{}(\vec{d})
      =
    f_{\vec{d}}^n(\bot_A)
  \] for all $n \in \omega$. Hence \[
    \sem{\Gamma \vdash \fix{x}{E} : A}{}(\vec{d})
      =
    \bigsqcup_{n \in \omega}
      f_{\vec{d}}^n(\bot_A)
      =
    \bigsqcup_{n \in \omega}
      \sem{\Gamma \vdash \fixn{x}{n}{E} : A}{}(\vec{d})
  \]
\end{proof}

\begin{theorem}[Soundness]
  $
  \Gamma \vdash M \costleq N : A\
    \Longrightarrow\
  \forall \vec{d} \in \sem{\Gamma}{}.\
  \sem{M}{}(\vec{d}) \semleq_A \sem{N}{}(\vec{d})
  $
\end{theorem}
\begin{proof}
  Most inequalities hold up to equality in sized domains, so the
  standard proof---as found in e.g. \citep[\S
  3.2]{Streicher2006}---covers most cases. For $(\textsf{ctx})$,
  we proceed by induction on monotone contexts: it it suffices to
  notice that they are chosen exactly so that either (a) preserve
  inequalities, say at natural numbers, or (b)  so that they
  mirror the pointwise definitions of product and function types.

  For the two fixpoint rules, recall from the proof of Proposition
  \ref{prop:fixpoint-approx} that \[
    \sem{\Gamma \vdash \fixn{x}{n}{E} : A}{}(\vec{d})
      =
    f_{\vec{d}}^n(\bot_A)
  \] In the case of  $(\textsf{rat})$, we have that
  that for all $\vec{d} \in \sem{\Gamma}{}$ and $i \in \omega$, \[
    f_{\vec{d}}^i(\bot_A) \sqsubseteq_A f_{\vec{d}}^{i+1}(\bot_A)
  \] Hence, by the axioms of sized domains, for $
  \vec{d} \in \sem{\Gamma}$ and $i \in \omega$, \[
    f_{\vec{d}}^{i+1}(\bot_A) \semleq_A  f_{\vec{d}}^i(\bot_A)
  \] 
  
  For $(\textsf{cpind})$: if $\vec{d} \in \sem{\Gamma}{}$, let
  \begin{align*}
    &h : \sem{A}{} \rightarrow \sem{B}{} \\
    &h(x) \defeq 
      \sem{\Gamma, z : A \vdash \ectx{E}[z] : B}{}(\vec{d}, x)
  \end{align*} Recall that $h$ is continuous with respect to
  $\sqsubseteq$. Suppose that \[
    \sem{\Gamma \vdash M : B}{}(\vec{d})
      \semleq_B
    \sem{\Gamma \vdash \ectx{E}[\fixn{x}{n}{E}] : B}{}(\vec{d})
      =
    h\left(f_{\vec{d}}^i(\bot_A)\right)
  \] for all $i \in \omega$. But, as $h$ is continuous,
  $\left(h\left(f_{\vec{d}}^i(\bot_A)\right)\right)_{i \in
  \omega}$ is an $\omega$-chain wrt to $\sqsubseteq_B$ that is an
  upper bound for $\sem{M}{}(\vec{d})$, for all $i \in \omega$.
  Hence \begin{align*}
    \sem{\Gamma \vdash M : B}{}(\vec{d})
      \semleq_B\
    &\bigsqcup_{i \in \omega} 
        h\left(f_{\vec{d}}^i(\bot_A)\right) \\
      =\
    &h\left(\bigsqcup_{i \in \omega} 
        f_{\vec{d}}^i(\bot_A)
      \right) \\
      =\
    &h\left(\sem{\Gamma \vdash \fix{x}{E} : A}{}(\vec{d})\right) \\
      =\
    &\sem{\Gamma \vdash \ectx{E}[\fix{x}{E}] : B}{}(\vec{d})
  \end{align*} by the continuity of $h$.
\end{proof}

\subsubsection{Adequacy}

To prove adequacy, we follow a logical relations technique, as
demonstrated in \citet[\S 4]{Streicher2006}.

We define the relation \[
  d \rel{A} M
\] between elements $d \in \sem{A}{}$ of a sized domain and terms
$M : A$ of PCF, by the clauses \begin{alignat*}{3}
  & d \rel{\natt} M
    &&\defiff\ &&d \neq \bot
        \Longrightarrow 
      \exists n.\ \evalpcf{M}{\num{n}}{}\ \land\ n \leq_\mathbb{N} d\\
  & d \rel{\cost} M
    &&\defiff\ &&d \neq \bot
        \Longrightarrow 
      \exists n.\ \evalpcf{M}{\numc{n}}{}\ \land\ n \leq_\mathbb{N} d\\
  & (d_1, d_2) \rel{A_1 \times A_2} M
    &&\defiff\ &&d_1 \rel{A_1} \pi_1(M) \text{ and } d_2 \rel{A_2} \pi_2(M) \\
  & f \rel{A \rightarrow B} M
    &&\defiff\ &&\forall x \rel{A} N.\ f(x) \rel{B} M\,N \\
\end{alignat*} We read $d \rel{A} M$ as ``$d$ over-approximates $M$.''

We also introduce a notion of head reduction for PCF:
\begin{center}
\begin{tabular}{c}

  \\

\begin{prooftree}
  \justifies
    (\lambda x.\, M)N \wh{\mathrm{PCF}} M[N/x]
\end{prooftree}

\quad

\begin{prooftree}
  \justifies
    \pi_i\left(\tuple{M_1}{M_2}\right) \wh{\mathrm{PCF}} M_i
\end{prooftree}

\quad

\begin{prooftree}
    \phantom{M \wh{1} N}
  \justifies
    \fix{x}{M} \wh{\mathrm{PCF}} M[\fix{x}{M}/x]
\end{prooftree}

\\

\begin{prooftree}
    M \wh{\mathrm{PCF}} N
  \justifies
    \ectx{E}[M] \wh{\mathrm{PCF}} \ectx{E}[N]
\end{prooftree}

\quad

\begin{prooftree}
    \phantom{nothing}
  \justifies
    \pif{\num{0}}{P}{Q} \wh{\mathrm{PCF}} P
\end{prooftree}

\quad

\begin{prooftree}
    \phantom{nothing}
  \justifies
    \pif{\num{n+1}}{P}{Q} \wh{\mathrm{PCF}} Q
\end{prooftree}

\\

\begin{prooftree}
    \phantom{nothing}
  \justifies
    (\pif{N}{P}{Q})X \wh{\mathrm{PCF}} \pif{N}{PX}{QX}
\end{prooftree}

\\

\begin{prooftree}
    \phantom{nothing}
  \justifies
    \pi_i(\pif{N}{P}{Q}) \wh{\mathrm{PCF}} \pif{N}{\pi_i(P)}{\pi_i(Q)}
\end{prooftree}

\\

\end{tabular}
\end{center} where $\ectx{E}$ is an eliminative PCF context, as
defined in Figure \ref{fig:pcf-inequality}.

\begin{lemma} \hfill
  \label{lemma:pcf-banquet}
  \begin{enumerate}[label=(\roman*)]
    \item
      $\bot_A \rel{A} M$ for all $M : A$.
    \item
      If $(d_i)_{i \in \omega}$ is a $\omega$-chain in
      $\sem{A}{}$, \[
        \left(\forall i \in \omega.\ d_i \rel{A} M\right)
          \Longrightarrow
        \left(\bigsqcup_{i \in \omega} d_i\right) \rel{A} M
      \]
    \item
      $d \rel{A} M'\ \land\ M \wh{\mathrm{PCF}} M'\ 
        \Longrightarrow
      d \rel{A} M$
  \end{enumerate}
\end{lemma}
\begin{proof} \hfill
  \begin{enumerate}[label=(\roman*)]
    \item
      By induction on $A$.
    \item
      By induction on $A$.
    \item
      By induction on $M \wh{\mathrm{PCF}} M'$.
  \end{enumerate}
\end{proof}

\begin{lemma}[Conditional upper bound]
  \label{lemma:iflor}
  If $d \rel{A} P$, $e \rel{A} Q$, and $m \rel{\natt} M$,
  then \[
    \iflor_A(m, d, e) 
      \rel{A} 
    \pif{M}{P}{Q}
  \]
\end{lemma}

\begin{proof}
  If $m = \bot$, then $\iflor_A(m, d, e) = \bot_A$ which is
  related to all terms by Lemma \ref{lemma:pcf-banquet}(i), so
  there is nothing to prove.

  If $m = 0$, then as $m \rel{\natt} M$, it is easy to see that
  $\evalpcf{M}{\num{0}}{}$. Thus, $d \rel{A} P$ implies that $d
  \rel{A} \pif{M}{P}{Q}$.

  Finally, if $m \geq 0$ we proceed by induction on $A$.

  \begin{indproof}
    \indcase{$\natt$}

    We have $\iflor_A(m, d, e) \defeq d \lor_\natt e$. If $d
    \lor_\natt e \in \mathbb{N}$, then by assumption
    \begin{align*}
      \exists a.\ \evalpcf{P}{\num{a}}{}\ \land\ a \leq d \leq d \lor_\natt e \\
      \exists b.\ \evalpcf{Q}{\num{b}}{}\ \land\ b \leq e \leq d \lor_\natt e
    \end{align*} As $m \rel{\natt} M M$, we know that there exists
    a $c \in \mathbb{N}$ such that $\evalpcf{M}{c}{}$ and $c
    \costleq m$. Irrespective of whether this $c$ is zero or not,
    we know that $\pif{N}{P}{Q}$ converges to something bounded
    above by $d \lor_\natt e$, so the result holds.

    \indcase{$\cost$}

    Similar.

    \indcase{$A \rightarrow B$}

    We have $ \iflor_{A \rightarrow B}(m, d, e) \defeq d \lor_{A
    \rightarrow B} e$. We need to show that for all $x \rel{A} X$
    it is the case that \[
      (d \lor_{A \rightarrow B} e)(x) 
        \defeq
      d(x) \lor_B e(x)
        \rel{B}
      \left(\pif{M}{P}{Q}\right)X
    \] By assumption, we have \[
      d(x) \rel{B} P\,X\
        \text{ and }\
      e(x) \rel{B} Q\,X
    \] By the IH and along with $m \rel{\natt} M$ these imply that \[
      d(x) \lor_B e(x) \rel{B} \pif{M}{P\,X}{Q\,X}
    \] whence, by Lemma \ref{lemma:pcf-banquet}(iii), $d(x) \lor_B
    e(x) \rel{B} \left(\pif{M}{P}{Q}\right)X$.
    
    \indcase{$A_1 \times A_2$}

    Similar.

  \end{indproof}

\end{proof}

\noindent We can then show that

\begin{proposition}
  \label{prop:pcf-adeq-diagonal}
  Suppose $x_1 : A_1, \dots, x_n : A_n \vdash M : A$.  Then if
  $d_i \rel{A_i} M_i$ for $i = 1, \dots, n$, we have that \[
      \sem{\Gamma \vdash M : A}{}(\vec{d})
        \rel{A}
      M[\vec{N}/\vec{x}]
  \]
\end{proposition}
\begin{proof}
  By induction on $\Gamma \vdash M : A$.
  \begin{indproof}
    \indcase{$x_1 : A_1, \dots, x_n : A_n \vdash x_i : A_i$} Immediate.

    \indcase{$\Gamma \vdash \num{n} : \natt$}

    $\sem{\num{n}}(\vec{d}) = n \in \mathbb{N}$, and
    $\num{n}[\vec{N}/\vec{x}] \aequiv
    \evalpcf{\num{n}}{\num{n}}{}$ with $n \leq_\mathbb{N} n$.

    \indcase{$\Gamma \vdash \zeroc : \cost$}

    Similarly.

    \indcase{$\Gamma \vdash \unitc : \cost$}

    Similarly.

    \indcase{$\Gamma \vdash \pif{M}{P}{Q} : A$}

    By the IH we have \begin{align*}
      \sem{M}(\vec{d}) &\rel{\natt} M[\vec{N}/\vec{x}] \\
      \sem{P}(\vec{d}) &\rel{A} P[\vec{N}/\vec{x}] \\
      \sem{Q}(\vec{d}) &\rel{A} Q[\vec{N}/\vec{x}]
    \end{align*} Hence, by Lemma \ref{lemma:iflor}, we obtain 
    \begin{derivation}
        \sem{\pif{M}{P}{Q}}(\vec{d})
      \since[=]{definition}
        \iflor_A\left(
          \sem{M}(\vec{d}),
          \sem{P}(\vec{d}),
          \sem{Q}(\vec{d})
        \right)
      \since[\rel{A}]{lemma}
        \pif{M[\vec{N}/\vec{x}]}{Q[\vec{N}/\vec{x}]}{Q[\vec{N}/\vec{x}]}
    \end{derivation}

    \indcase{$\Gamma \vdash M \pcfopm N : \natt$}

    If $\sem{M}(\vec{d}) = \bot$ or $\sem{N}(\vec{d})$ then
    $\sem{M + N}(\vec{d}) = \bot$ as well, so there is nothing to
    prove.

    If $\sem{M}(\vec{d}) = m \in \mathbb{N}$, and
       $\sem{N}(\vec{d}) = n \in \mathbb{N}$,
    then $\sem{M + N}(\vec{d}) = m \pcfopm n$. By the IH
    \begin{align*}
      &\exists a.\ \evalpcf{M[\vec{N}/\vec{x}]}{\num{a}}{} \ \land \ a \leq_\mathbb{N} m \\
      &\exists b.\ \evalpcf{N[\vec{N}/\vec{x}]}{\num{b}}{} \ \land \ b \leq_\mathbb{N} n
    \end{align*} Hence $\evalpcf{M[\vec{N}/\vec{x}] +
    N[\vec{N}/\vec{x}]}{\num{a \pcfopm b}}{}$ with $a \pcfopm b
    \leq_\mathbb{N} m \pcfopm n$ whence the result.

    \indcase{$\Gamma \vdash M \pcfopa \num{n} : \natt$}

    If $\sem{M}(\vec{d}) = \bot$ then $\sem{M + \num{n}}(\vec{d})
    = \bot$ as well, so there is nothing to prove.

    If $\sem{M}(\vec{d}) = m \in \mathbb{N}$ then $\sem{M \pcfopa
    \num{n}} = m \pcfopa n$. By the IH \[
      \exists a.\ \evalpcf{M[\vec{N}/\vec{x}]}{\num{a}}{} \ \land \ a \leq_\mathbb{N} m
    \] Hence $\evalpcf{M[\vec{N}/\vec{x}] \pcfopa
    N[\vec{N}/\vec{x}]}{\num{a \pcfopa n}}{}$ with $a \pcfopa n
    \leq_\mathbb{N} m \pcfopa n$, whence the result.

    \indcase{$\Gamma \vdash M \pcfopa N : \natt$}

    If $\sem{M}(\vec{d}) = \bot$ then $\sem{M + N}(\vec{d}) =
    \bot$ as well, so there is nothing to prove.

    If $\sem{M}(\vec{d}) = m \in \mathbb{N}$
    then $\sem{M \pcfopa N} = m$. By the IH, \[
      \exists a.\ \evalpcf{M[\vec{N}/\vec{x}]}{\num{a}}{} \ \land \ a \leq_\mathbb{N} m
    \] Hence $\evalpcf{M[\vec{N}/\vec{x}] \pcfopa
    N[\vec{N}/\vec{x}]}{\num{b}}{}$ with $b \leq_\mathbb{N} a
    \leq_\mathbb{N} m$, whence the result.
    
    \indcase{$\Gamma \vdash M \plusc N : \cost$}

    Similar.

    \indcase{$\Gamma \vdash \tuple{M_1}{M_2} : A_1 \times A_2$}

    Let $e_i \defeq \sem{\Gamma \vdash M_i : A_i}(\vec{d})$. Then
    by the IH \begin{align*}
      e_1 &\rel{A_1} M_1[\vec{N}/\vec{x}] \\
      e_2 &\rel{A_1} M_2[\vec{N}/\vec{x}]
    \end{align*} By Lemma \ref{lemma:pcf-banquet}(iii) we get \[
      e_i 
      \rel{A_i}
        \pi_i\left(\tuple{M_1[\vec{N}/\vec{x}]}
                         {M_2[\vec{N}/\vec{x}]}\right)
      \aequiv
        \pi_i\left(\tuple{M_1}{M_2}[\vec{N}/\vec{x}]\right)
    \] and hence $\sem{\tuple{M_1}{M_2}}(\vec{d}) \defeq (e_1,
    e_2) \rel{A_1 \times A_2} \tuple{M_1}{M_2}[\vec{N}/\vec{x}]$.

    \indcase{$\Gamma \vdash \lambda x. M : A \rightarrow B$}

    Then by the IH we have for every $x \rel{A} N$
    that \[
      \sem{\Gamma, x : A \vdash M : B}(\vec{d}, x) 
        \rel{B}
      M[\vec{N}/\vec{x}, N/x]
    \] Hence, by Lemma \ref{lemma:pcf-banquet}(iii), \[
      \sem{\Gamma, x : A \vdash M : B}(\vec{d}, x) 
        \rel{B}
      (\lambda x. M[\vec{N}/\vec{x}])N
    \] But the LHS by definition is equal to \[
      \sem{\Gamma \vdash \lambda x. M : A \rightarrow B}(\vec{d})(x)
    \] and hence \[
      \sem{\Gamma \vdash \lambda x. M : A \rightarrow B}(\vec{d})
        \rel{A \rightarrow B}
      M[\vec{N}/\vec{x}]
    \]

    \indcase{$\Gamma \vdash \pi_i(M) : A_i$}

    Let $(e_1, e_2) \defeq \sem{\Gamma \vdash M : A_1 \times
    A_2}(\vec{d})$.  By the IH \[
      (e_1, e_2) \rel{A_1 \times A_2} M[\vec{N}/\vec{x}]
    \] Hence \[
      e_i \rel{A_i} \pi_i(M[\vec{N}/\vec{x}])
            \aequiv \pi_i(M)[\vec{N}/\vec{x}]
    \] But $\sem{\Gamma \vdash \pi_i(M) : A_i}(\vec{d}) \defeq e_i$.

    \indcase{$\Gamma \vdash M\,N : B$}

    By the IH \begin{alignat*}{3}
      &\sem{\Gamma \vdash M : A \rightarrow B}(\vec{d})\
        &&\rel{A \rightarrow B}\
      &&M[\vec{N}/\vec{x}] \\
      &\sem{\Gamma \vdash N : A}(\vec{d})\
        &&\rel{A}\
      &&N[\vec{N}/\vec{x}]
    \end{alignat*} Hence \[
      \sem{M\,N}(\vec{d})
        \defeq
      \sem{M}(\vec{d})(\sem{N}(\vec{d}))
        \rel{B}
      M[\vec{N}/\vec{x}] N[\vec{N}/\vec{x}]
        \aequiv
      (M\,N)[\vec{N}/\vec{x}]
    \]

    \indcase{$\Gamma \vdash \fix{x}{M} : A$}

    Recall that, by Proposition \ref{prop:fixpoint-approx}, \[
      \sem{\Gamma \vdash \fix{x}{M} : A}{}(\vec{d})
        =
      \bigsqcup_{n \in \omega}
        \sem{\Gamma \vdash \fixn{x}{n}{M} : A}{}(\vec{d})
    \] Thus, by Lemma \ref{lemma:pcf-banquet}(ii), it suffices to
    show that \[
        \sem{\Gamma \vdash \fixn{x}{n}{M} : A}{}(\vec{d})
          \rel{A}
        \fix{x}{M[\vec{N}/\vec{x}]}
    \] which we show by induction on $n$.

    \begin{indproof}
      \indcase{$0$} 
      
      Then $\sem{\fixn{x}{0}{M[\vec{N}/\vec{x}]}}(\vec{d}) =
      \bot_A \rel{A} \fix{x}{M[\vec{N}/\vec{x}]}$ by Lemma
      \ref{lemma:pcf-banquet}(i).

      \indcase{$n+1$}

      Then \begin{align*}
          \sem{\fixn{x}{n+1}{M}}(\vec{d})
        &\defeq\
          \sem{M[\fixn{x}{n}{M}/x]}(\vec{d}) \\
        &=\
          \sem{M}(\vec{d}, \sem{\fixn{x}{n}{M}}(\vec{d})) \\
        &\rel{A}\
          M[\vec{N}/\vec{x}, 
            (\fix{x}{M[\vec{N}/\vec{x}]})/x]
      \end{align*} by the global IH, as $d_i \rel{A_i} N_i$ by
      assumption, and $\sem{\fixn{x}{n}{M}}(\vec{d}) \rel{A}
      \fix{x}{M[\vec{N}/\vec{x}]}$ by the local IH. But the RHS is
     $\alpha$-equivalent to \[
        \fixn{x}{n+1}{M[\vec{N}/\vec{x}]}
      \]
    \end{indproof}
  \end{indproof}
\end{proof}

\begin{theorem}[Adequacy, Theorem~\ref{theorem:pcf-adequacy}] \hfill
  \begin{enumerate}
    \item
      If $M : \natt$, then $\sem{M}{} = m$ implies
      $\evalpcf{M}{\num{k}}{}$ for some $k \leq_\mathbb{N} m$.
    \item
      If $M : \cost$, then $\sem{M}{} = m$ implies
      $\evalpcf{M}{\numc{k}}{}$ for some $k \leq_\mathbb{N} m$.
  \end{enumerate}
\end{theorem}

\begin{proof}
  By Proposition \ref{prop:pcf-adeq-diagonal}, $\sem{M} R_{\natt}
  M$, which amounts to the result. The case for $\cost$ is
  identical.
\end{proof}

%
%



%% file: wh.tex
\renewcommand{\arraystretch}{3}

\begin{small}
\begin{tabular}{c}

    CBPV eliminative contexts: 

    $
    \ectx{E}\
      ::=\
             []
        \mid \pi_i(\ectx{E})
        \mid \ectx{E}\,V
    $

  \\

  $
  \begin{prooftree}
      \phantom{M \wh{0} N}
    \justifies
      (\lambda x.\, M)V \wh{0} M[V/x]
  \end{prooftree}
  $

  \quad

  $
  \begin{prooftree}
      \phantom{M \wh{0} N}
    \justifies
      \pi_i\left(\tuple{M_1}{M_2}\right) \wh{0} M_i
  \end{prooftree}
  $

  \quad

  $
  \begin{prooftree}
      \phantom{M \wh{0} N}
    \justifies
      \fix{x}{M} \wh{0} M[\thunk{(\fix{x}{M})}/x]
  \end{prooftree}
  $

  \\

  $
  \begin{prooftree}
      \phantom{yeah}
    \justifies
      \force{(\thunk{M})} \wh{0} M
  \end{prooftree}
  $

  \quad

  $
  \begin{prooftree}
    \phantom{M \wh{} N}
    \justifies
      \bind{x}{\return{V}}{N} \wh{0} N[V/x]
  \end{prooftree}
  $

  \quad

  $
  \begin{prooftree}
      \phantom{yeah}
    \justifies
      \ectx{E}[\charge{M}] \wh{0} \charge{\ectx{E}[M]}
  \end{prooftree}
  $

  \\

  $
  \begin{prooftree}
      \phantom{yeah}
    \justifies
      \ifz{\numl{0}}{P}{Q} \wh{0} P
  \end{prooftree}
  $

  \quad

  $
  \begin{prooftree}
      \phantom{yeah}
    \justifies
      \ifz{\numl{n+1}}{P}{Q} \wh{0} Q
  \end{prooftree}
  $

  \quad

  $
  \begin{prooftree}
      \phantom{yeah}
    \justifies
      \splitprod{(V_1, V_2)}{x_1}{y_2}{Q} \wh{0} Q[V_1/x_1, V_2/x_2]
  \end{prooftree}
  $

  \\

  $
  \begin{prooftree}
      M \wh{0} N
    \justifies
      \ectx{E}[M] \wh{0} \ectx{E}[N]
  \end{prooftree}
  $

\end{tabular}
\end{small}